\documentclass[a4paper,fleqn,usenatbib]{mnras}

\usepackage[T1]{fontenc}
\usepackage{ae,aecompl}

\usepackage{graphicx}	
\usepackage[caption = false]{subfig}
\usepackage{amsmath}	
\usepackage{amssymb}	
\usepackage{multirow}
\usepackage{pdflscape}
\usepackage{deluxetable}

\title[Mass models of gas-rich void dwarf galaxies]{Mass models of gas-rich void dwarf galaxies}

\author[Kurapati et al.]{
Sushma Kurapati$^{1}$\thanks{E-mail: sushma@ncra.tifr.res.in},
Jayaram N. Chengalur $^{1}$, Peter Kamphuis$^{2}$, Simon Pustilnik$^{3}$  \\
$^{1}$ National Centre for Radio Astrophysics, Tata Institute of Fundamental Research, PO Box 3, Pune 411007, India \\
$^{2}$ Astronomisches Institut Ruhr-Universit\"{a}t Bochum, Germany \\
$^{3}$ Special Astrophysical Observatory, Russian Academy of Sciences, Nizhnii Arkhyz, 369167 Russia 
}

\date{Accepted XXX. Received YYY; in original form ZZZ}

\pubyear{2017}

\begin{document}
\label{firstpage}
\pagerange{\pageref{firstpage}--\pageref{lastpage}}
\maketitle

\newcommand{\MB}{\ensuremath{\rm M_B}}
\newcommand{\mb}{\ensuremath{\rm M_b}}
\newcommand{\mhi}{\ensuremath{\rm M_{HI}}}
\newcommand{\mg}{\ensuremath{\rm M_{gas}}}
\newcommand{\msun}{\ensuremath{\rm M_{\odot}}}
\newcommand{\fatm}{\ensuremath{\rm f_{atm}}}
\newcommand{\jb}{\ensuremath{\rm j_b}}
\newcommand{\jg}{\ensuremath{\rm j_{gas}}}
\newcommand{\kms}{\ensuremath{\rm km \ s^{-1}}}
\newcommand{\kpc}{\ensuremath{\rm kpc}}
\newcommand{\LCDM}{\ensuremath{\Lambda{\rm CDM}}}
\newcommand{\aips}{{\sc aips }}
\newcommand{\gipsy}{{\sc gipsy }}
\newcommand{\fat}{{\sc fat }}
\begin{abstract}
We construct mass models of eight gas rich dwarf galaxies that lie in the Lynx-Cancer void. From NFW fits to the dark matter halo profile, we find that the concentration parameters of halos of void dwarf galaxies are similar to those of dwarf galaxies in normal density regions. We also measure the slope of the central dark matter density profiles, obtained by converting the rotation curves derived using 3-D (\fat) and 2-D (ROTCUR) tilted ring fitting routines, into mass densities. We find that the average slope ($\alpha = -1.39 \pm 0.19$), obtained from 3-D fitting is consistent with that expected from an NFW profile. On the other hand, the average slope measured using the 2-D approach is closer to what would be expected for an isothermal profile. This suggests that systematic effects in velocity field analysis have a significant effect on the slope of the central dark matter density profiles. Given the modest number of galaxies we use for our analysis, it is important to check these results using a larger sample.
\end{abstract}




\section{Introduction}

The standard $\Lambda$ Cold Dark Matter (\LCDM) paradigm has been remarkably successful at large scales, with predictions from \LCDM\ numerical simulations providing an excellent match to observational data \citep[e.g.][]{ade14, springel06, clowe06, dawson13, baur16}. However, discrepancies between \LCDM\ predictions and the observations remain at smaller scales, \citep[see][for a review]{bullock17}. One of the earliest noticed, as well as most studied, of these is the so called ``cusp-core problem'', where observations of isolated dwarf and low surface brightness galaxies (i.e. galaxies for which observations are expected to be least affected by systematic errors arising from non circular motions and uncertainties in the stellar mass to light ratio) typically find that the dark matter halo has a constant density core towards the centre, while \LCDM\ models predict a cuspy density profile \citep[e.g.][]{deblok10,oh11,ogiya14,ogiyamori14,ogiya15}.

In detail, \LCDM\ simulations which do not include baryonic physics predict that dark matter halos of all masses have a universal density profile with the density distribution in the inner regions following a power law $\rho \sim$ r$^{\alpha}$ with slope $\alpha $ = -1 \citep{navarro96, navarro97}.  Steeper slopes $\alpha \approx -1.5$ were found by \citet{moore98, moore99}, and more recent  simulations find that the inner slope shows some variation and mass dependence with  $\alpha$ lying in the range -(0.8 -- 1.4) \citep{ricotti03, ricotti07, delpopolo10, delpopolo12, dicintio14}. For smaller galaxies the slope is found to  decrease towards the center reaching $\alpha $= -0.8 at $\sim 100$~pc from the centre \citep{stadel09, navarro10, delpopolo11}. In contrast however, observations of dwarf and low-surface brightness (LSB) galaxies indicate a flat ($\alpha \approx -0.2$) dark matter density core \citep[e.g.][]{deBlok01, deBosma02, oh11,oh15}. Some observational studies \citep{simon05, adams14} find intermediate slopes, i.e. steeper than would be expected from a constant density core, but still significantly shallower than the slopes predicted by simulations.

Various solutions have been proposed to resolve the cusp-core problem; these broadly fall into three categories. The first set of solutions propose that the dark matter could be warm or weakly self interacting, this would erase the cusps that arise in pure \LCDM\ \citep[e.g.][]{spergel00, rocha13, elbert15, kaplinghat16, schneider17}. The second category of solutions invokes baryonic processes to generate cores from an originally cuspy distribution. Basically, repeated gas outflows resulting from supernova explosions from star formation concentrated at the galaxy centre result in a re-distribution of the baryonic as well as dark matter \citep[see e.g.][]{pontzen12, governato12, pontzen14, read16}. However, some simulations find that the density profile of the dark matter is consistent with cuspy profiles even after including baryonic outflows \citep{ceverino09, marianacci14,schaller15}. The third category of solutions suggest that there are residual systematic problems in determining and modelling the rotation curves of galaxies, and these problems could result in a mis-identification of cores as cusps \citep[e.g.][]{vandenbosch00, swaters03, hayashi04,oman17,pineda17}. These residual systematic effects include smoothing of the rotation curve because of the finite resolution of the observations \citep{vandenbosch00}, incorrectly measured inclination angles \citep[e.g.][]{rhee04, read16b}, improperly modelled pressure support \citep[e.g.][]{rhee04, valenzuela07, pineda17} or unmodelled non-circular motions \citep[e.g.][]{rhee04, valenzuela07, oman17}. All of these can lower the inner rotation velocities and mask the cuspy distributions.
More recently, \citet{pineda17} studied systematic effects in observational studies using high-resolution hydrodynamic simulations of dwarf galaxies and they find that the cored isothermal halos are favoured in spite of the fact that their simulations contain NFW halos. 
 
Traditionally, the H{\sc i} rotation curves used in mass modelling were derived from the 2-D velocity fields \citep[e.g.][]{deBlok02,oh11,oh15}. Recently there have been several software packages developed that determine the rotation curves by directly modelling the full 3D data cube,  including modelling of instrumental effects such as beam smearing. Rotation curves derived in this way would be expected not to suffer from the flattening associated with beam smearing, and as such better trace the underlying circular velocities. In this work, we derive the rotation curves by using both the 3D and 2D tilted ring fitting routines. We compare the properties of dark matter halos that were obtained using the rotation curves from both the approaches. 
 
The galaxies that we have in our sample all lie in voids. As such the dark matter distribution in these galaxies is also of interest by itself. There are at least two possible reasons why void galaxies may have different dark matter halo properties than galaxies in higher density regions. The first is that the large scale environment is expected to correlate with the properties of individual galaxy halos. Simulations find that, for halo masses $<$ 5 $\times$ 10$^{11}$ $h^{-1}$ M$_{\odot}$, halos in cluster regions are on average $\sim$ 30 -- 40 $\%$ more concentrated and have $\sim$ 2 times higher central densities than halos in voids \citep{avila05}. In models where the distribution of dark matter in central regions of the galaxy is driven by stellar feedback processes \citep[e.g.][]{pontzen12, governato12}, void galaxies, with their typically higher star formation rates \citep{moorman16} could be more affected by such baryonic processes as compared to galaxies in high density regions. 
 
This paper is organized as follows. In \S \ref{obs}, we describe the sample and the procedures used for the derivation of rotation curves. The construction of mass models is described \S \ref{massmodels}. In \S \ref{results}, we discuss the dark matter density profiles derived from the 2-D and 3-D rotation curves and finally, we summarize the main results in \S \ref{summary}.

\section{Kinematic analysis}
\label{obs}

\subsection{Rotation curves}
\label{rot}

Our sample galaxies are gas rich dwarfs, lying in nearby voids, and for which we have reasonably well resolved H{\sc i} data cubes. The sample selection and the data reduction is discussed in detail in \citet[][hereafter Paper~I]{kurapati18}, and the interested reader is referred to that for more details. The derivation of rotation curves is also presented in detail in Paper~I, and is briefly summarized below. As discussed earlier, rotation curves can be derived by fitting a tilted ring model either to the data cube (i.e. a ``3-D'' approach) or to the velocity field (i.e. a ``2-D'' approach). We use both of these approaches to derive rotation curves for our sample galaxies. The velocity field was determined using the `MOMNT' task in \aips; the moment method is a commonly used method for determining H{\sc i} velocity fields.  
For comparison, we also derive the velocity field using Gauss-Hermite fits to the individual spectra, a comparison of the results obtained using the moment method and the Gauss-Hermite fits is presented in Sec.~\ref{ssec:gauss-hermite}. 
In the ``2-D'' approach, we derive the rotation curves by fitting the tilted ring model to this velocity field using the `ROTCUR' task from the \gipsy\ software package \citep{vanderhulst92}. In the ``3-D'' approach we derive the rotation curves by fitting the tilted ring model directly to the H{\sc i} data cube using the \fat pipeline \citep[v5.0.2,][]{kamphuis15}. We fit flat discs, where all parameters except the surface brightness and the rotational velocity of each ring have the same value at all radii. Our data do not show obvious warping signatures. The \fat and ROTCUR derived rotation curves \citep[shown in Fig. A1, A2, and A3 of][]{kurapati18} are in broad agreement for 8 of the 11 galaxies. The main difference is that the inner parts of the rotation curves derived using \fat are steeper than those derived using ROTCUR. This is likely due to the fact that \fat operates directly on data cubes instead of velocity fields. It is therefore expected to be less affected by projection effects such as ``beam smearing". For the remaining 3 galaxies, the rotation curves derived using the \fat pipeline were unreliable in the central regions probably due to their clumpy H{\sc I} distribution. Hence, we exclude these 3 galaxies for the construction of the mass models. 

The ``3-D'' approach works well for galaxies with inclinations upto 90$^{\circ}$ whereas the ``2-D'' approach is expected to be reliable for the galaxies with inclinations upto $\sim$ 70$^{\circ}$ \citep{begeman89}. In our sample, the galaxy UGC4148 is above this limit of  70$^{\circ}$ and the galaxies J0929+1155 and J0926+3343 have inclinations of $\sim$ 70$^{\circ}$. The optical and kinematic parameters (as derived from the {\sc fat}) for the selected 8 galaxies are listed in Table \ref{tab:sample}. The columns are as follows. (1): galaxy name, (2): distance to the galaxy in Mpc, (3): absolute B-band magnitude (corrected for Galactic extinction using extinction values from NED), (4): H{\sc i} mass ($\times $ 10$^{7}$ M$_{\odot}$), and (5): inclination angle in degrees,  (6): velocity dispersion in \kms\ and  (7): maximum rotation velocity in \kms\ as derived from the {\sc fat}.

\subsection{Correction for pressure support}
\label{pressure}

\begin{table}
\begin{footnotesize}
\caption{Parameters of galaxies selected for this study}
\label{tab:sample}
\begin{tabular}{ p{1.4cm} p{0.6cm} p{0.8cm} p{1.0cm} p{0.4cm} p{0.7cm} p{0.5cm}}
\\
\hline
 Name  		& $d$ (Mpc) & \MB & $\mhi$ (10$^{7}$M$_{\odot}$) & i$^{ a}$ & $\sigma^{ b}$  & V$_{\rm max}^{ b}$\\ 
\hline
 KK246$^{c}$	    & 6.85 & -13.69 & 9.0  & 62 & 10.9 & 42.0 \\ 
 UGC4115	    & 7.73 & -14.75 & 31.9 & 55 & 13.6 & 56.5 \\ 
 J0926+3343         & 10.6 & -12.90 & 5.2  & 69 & 11.5 & 31.5 \\
 UGC5288	    & 11.4 & -15.61 & 90.2 & 38 & 9.1  & 72.4 \\
 UGC4148	    & 13.5 & -15.18 & 78.4 & 83 & 8.4  & 63.9 \\
 J0630+23           & 22.9 & -15.89 & 135.1& 51 & 10.9 & 80.2 \\
 J0626+24           & 23.2 & -15.64 & 63.8 & 62 & 10.2 & 80.3 \\
 J0929+1155         & 24.3 & -14.69 & 36.6 & 72 & 9.1  & 62.3 \\
\hline
\multicolumn{7}{l}{ $^{ a}$ The inclination angle (i) is in degrees,  }\\
\multicolumn{7}{l}{$^{ b}$ Velocity dispersion ($\sigma$), maximum velocity (V$_{\rm max}$) are in km~s$^{-1}$ } \\
\multicolumn{7}{l}{$^{ c}$ The \MB\ and $d$ values for KK246 and all other galaxies are}\\
\multicolumn{7}{l}{$^{ c}$  from \citet{kreckel11} and \citet{pustilnik11}.}\\
\end{tabular}
\end{footnotesize}
\end{table}

The observed rotation velocities are usually smaller than the true circular velocities, as the random gas motions in the gaseous disk provide ``pressure support'' to the disk. This effect is particularly significant in dwarf galaxies whose velocity dispersions are comparable to the maximum rotation velocity. In order to construct the mass models, the observed rotation velocities have to be corrected for the pressure support \citep[e.g.][]{meurer96,begum04}. This correction is given by: \\

$v_{c}^{2} = v_{o}^{2} - r \times \sigma ^{2} \bigg[\frac{d}{dr}(ln \  \Sigma_{HI}) \ + \frac{d}{dr}(ln \  \sigma^{2}) \ - \frac{d}{dr}(ln \  2h_{z}) \bigg]$ \\

%
where, $v_{c}$ is the corrected velocity, $v_{o}$ is the observed rotation velocity, $\Sigma_{HI}$ is the H{\sc i} surface density, $\sigma$ is the velocity dispersion and $h_{z}$ is the scale height of the disc. We assume that the scale height doesn't vary with radius and that the velocity dispersion is constant across the galaxy (but have also checked that assuming a linear gradient in both velocity dispersion and scale height does not make a significant difference - see below). Under the assumption of a constant scale height and velocity dispersion the pressure correction simplifies to: 
 $v_{c}^{2} = v_{o}^{2} - r \times \sigma^{2} \bigg[\frac{d}{dr}(ln \  \Sigma_{HI}) \bigg]$ 


We use parametric fits to the de-projected radial surface density profiles (themselves obtained using task `ELLINT' in \gipsy software) to determine the pressure support correction. For most of the galaxies, either a Gaussian or double Gaussian profile fits best to the radial surface density distribution, An exponential profile fits best for the case of KK246. For the velocity dispersion, we use the values determined from the data cubes by the \fat pipeline.
 
The rotation curves after correcting for the ``pressure support'' are shown in Appendix A. For all the galaxies, we find that the corrections are usually small in the inner regions, but are significant in the outer regions except J0926+3343, which has the smallest angular diameter. Our measurement of the DM density profile slope $\alpha$ is most sensitive to the inner part of the rotation curve, and is hence not much affected by the pressure support correction. We have also computed the corrections if we assume  a linear gradient in both scale height and velocity dispersion (6--12 km s$^{-1}$), and find that the conclusion that the corrections are not significant at the inner radii remains unchanged. 

\begin{figure*}
\noindent
\captionsetup[subfigure]{aboveskip=-1pt,belowskip=-1pt}
\subfloat{\includegraphics[width = 3.25in]{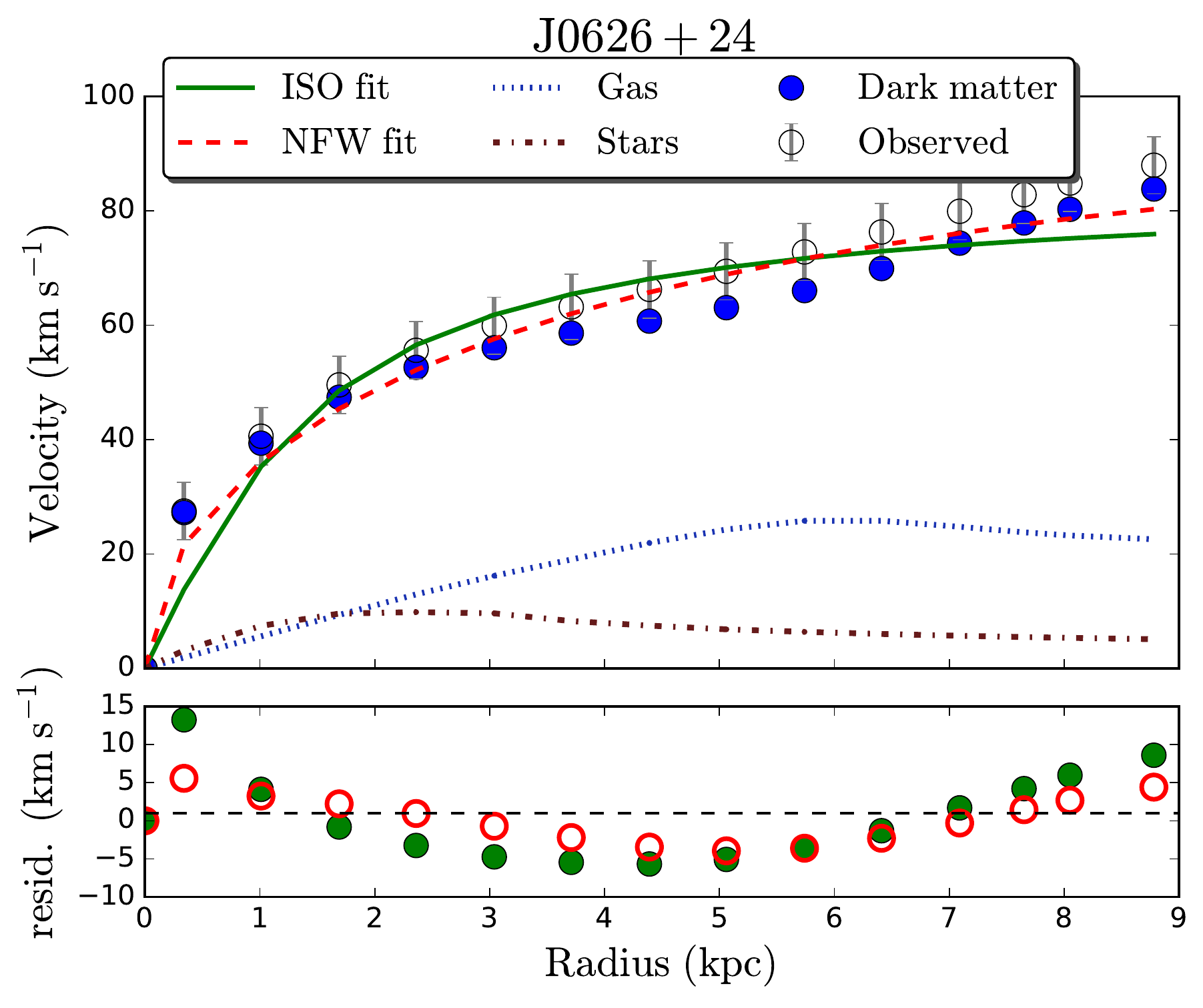}}
\subfloat{\includegraphics[width = 3.25in]{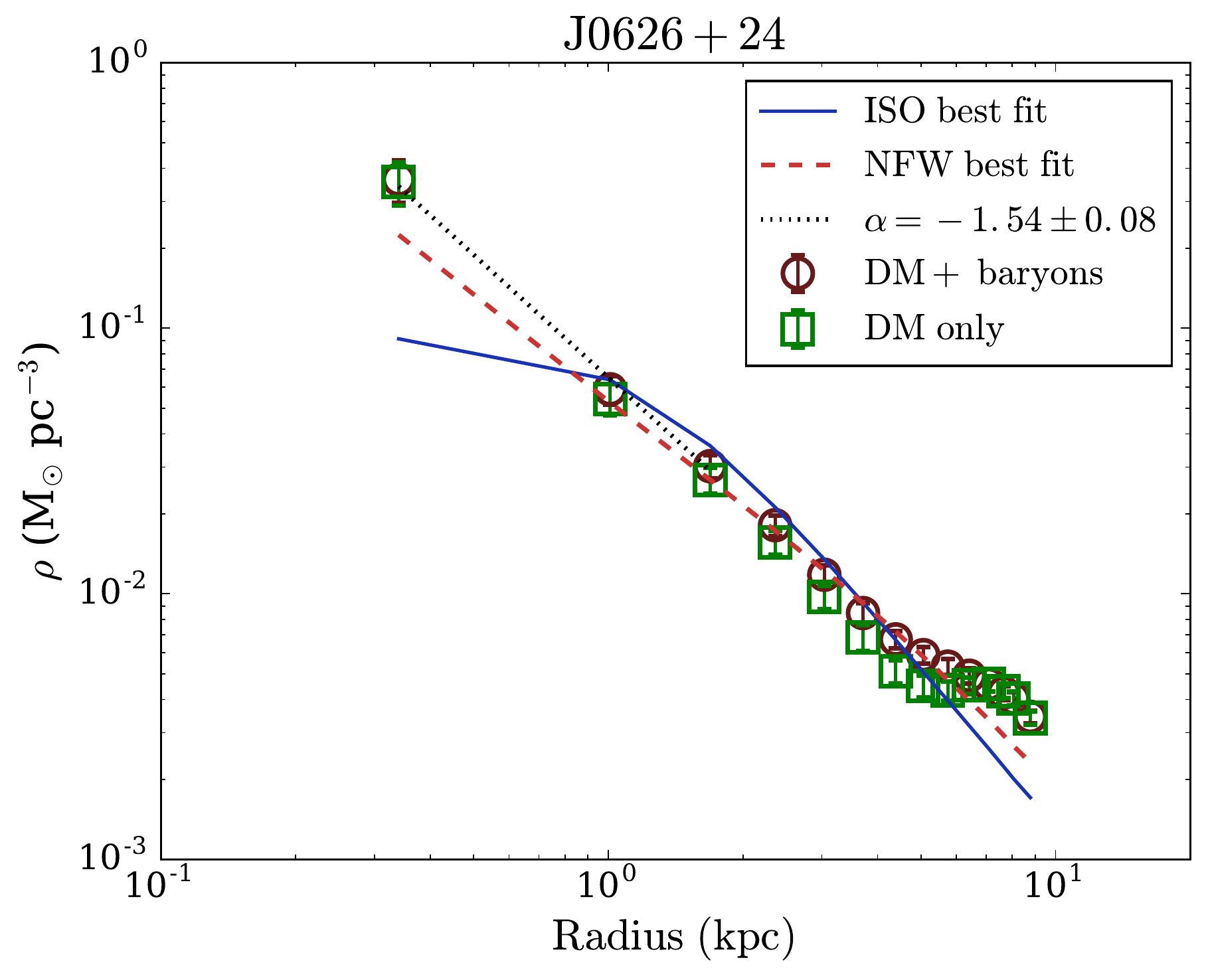}} \\
\subfloat{\includegraphics[width = 3.25in]{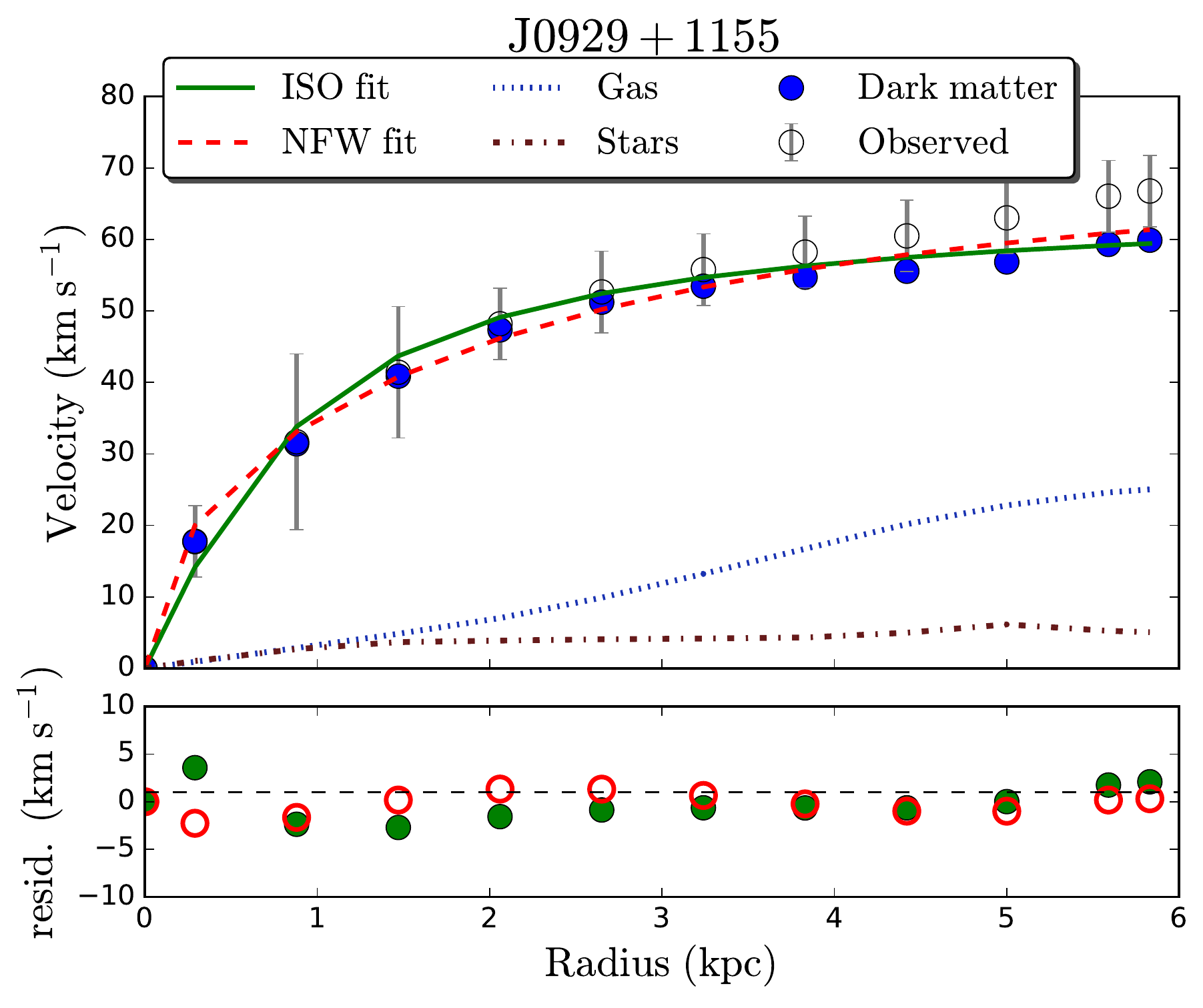}} 
\subfloat{\includegraphics[width = 3.25in]{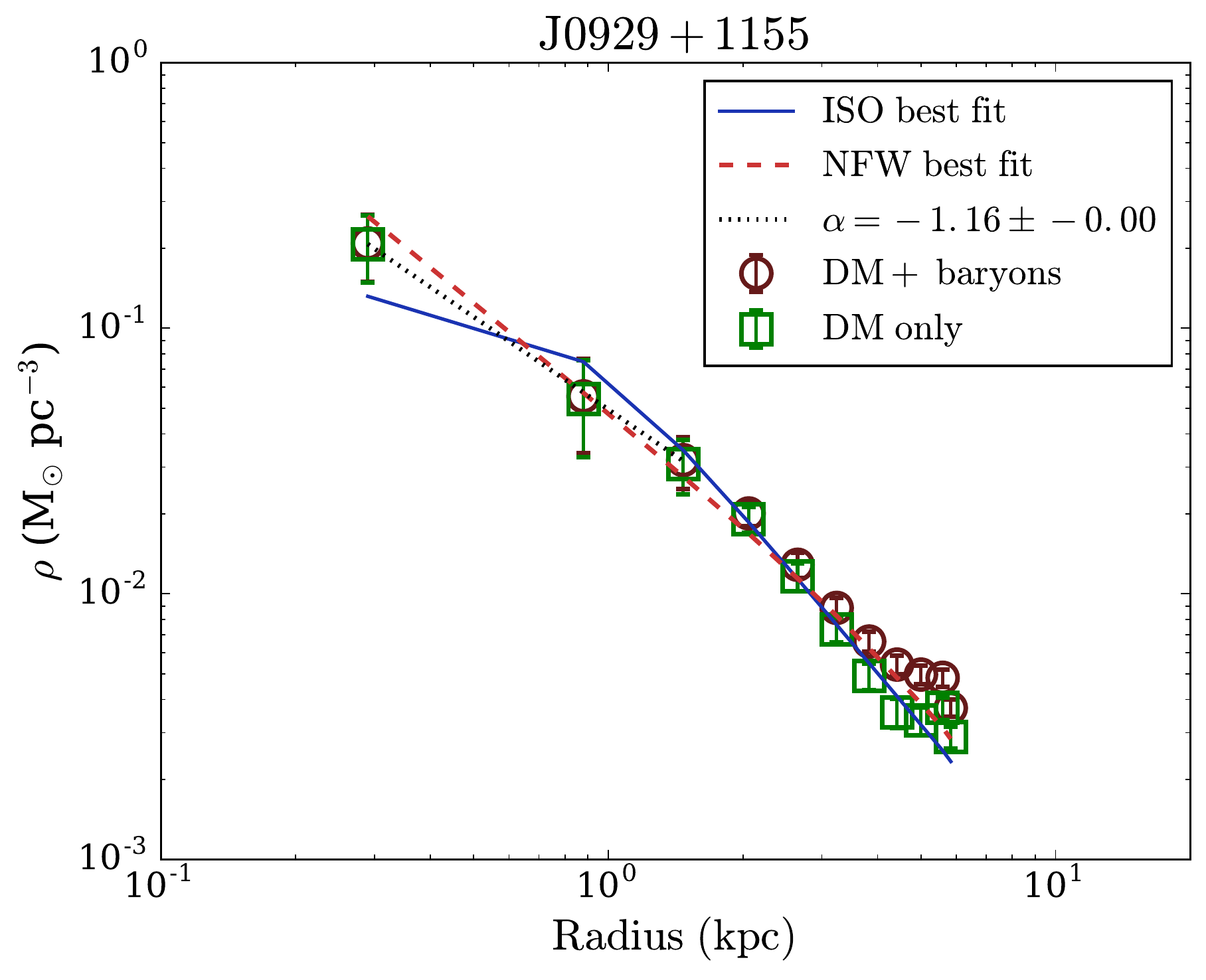}} \\
\subfloat{\includegraphics[width = 3.25in]{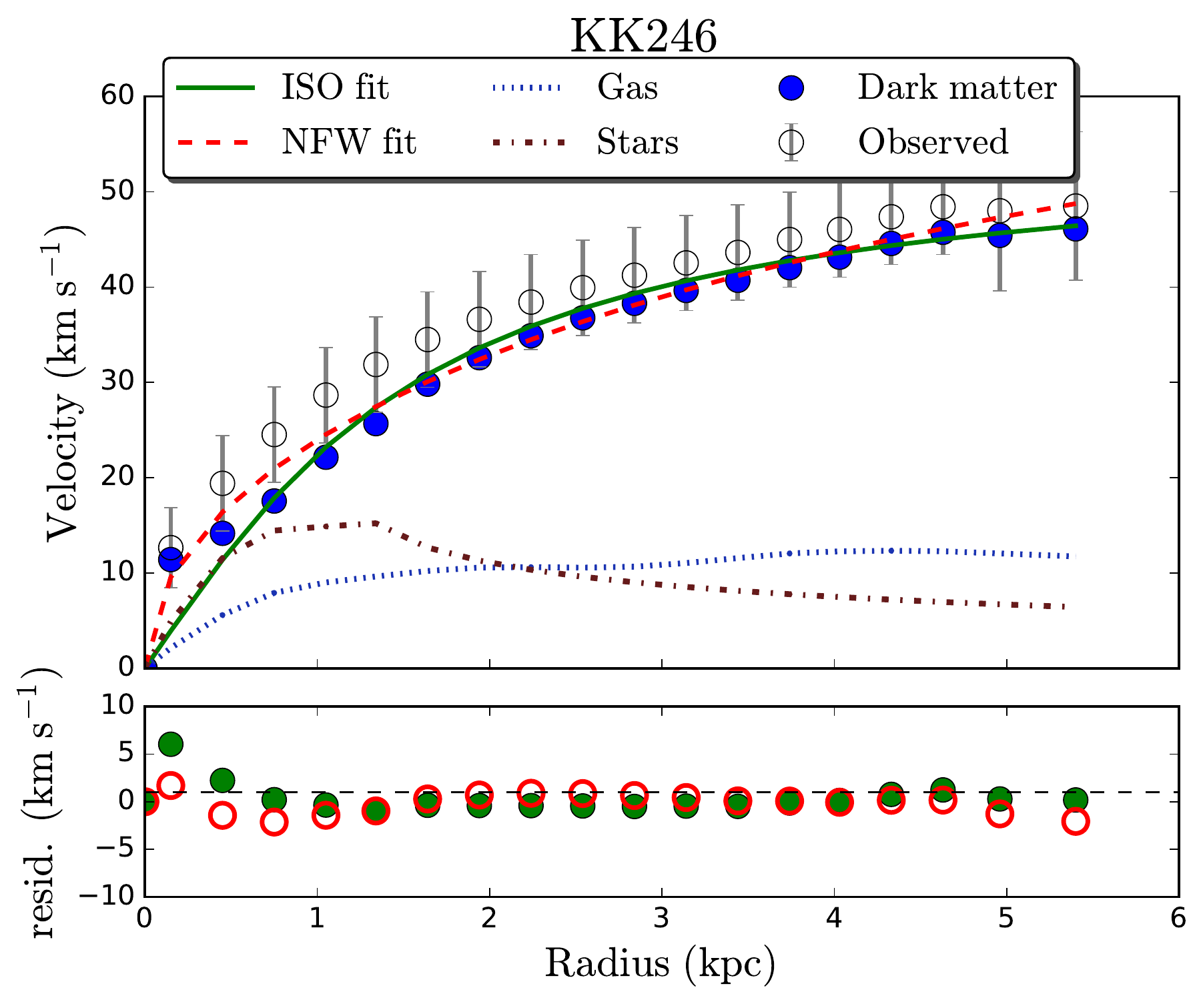}}
\subfloat{\includegraphics[width = 3.25in]{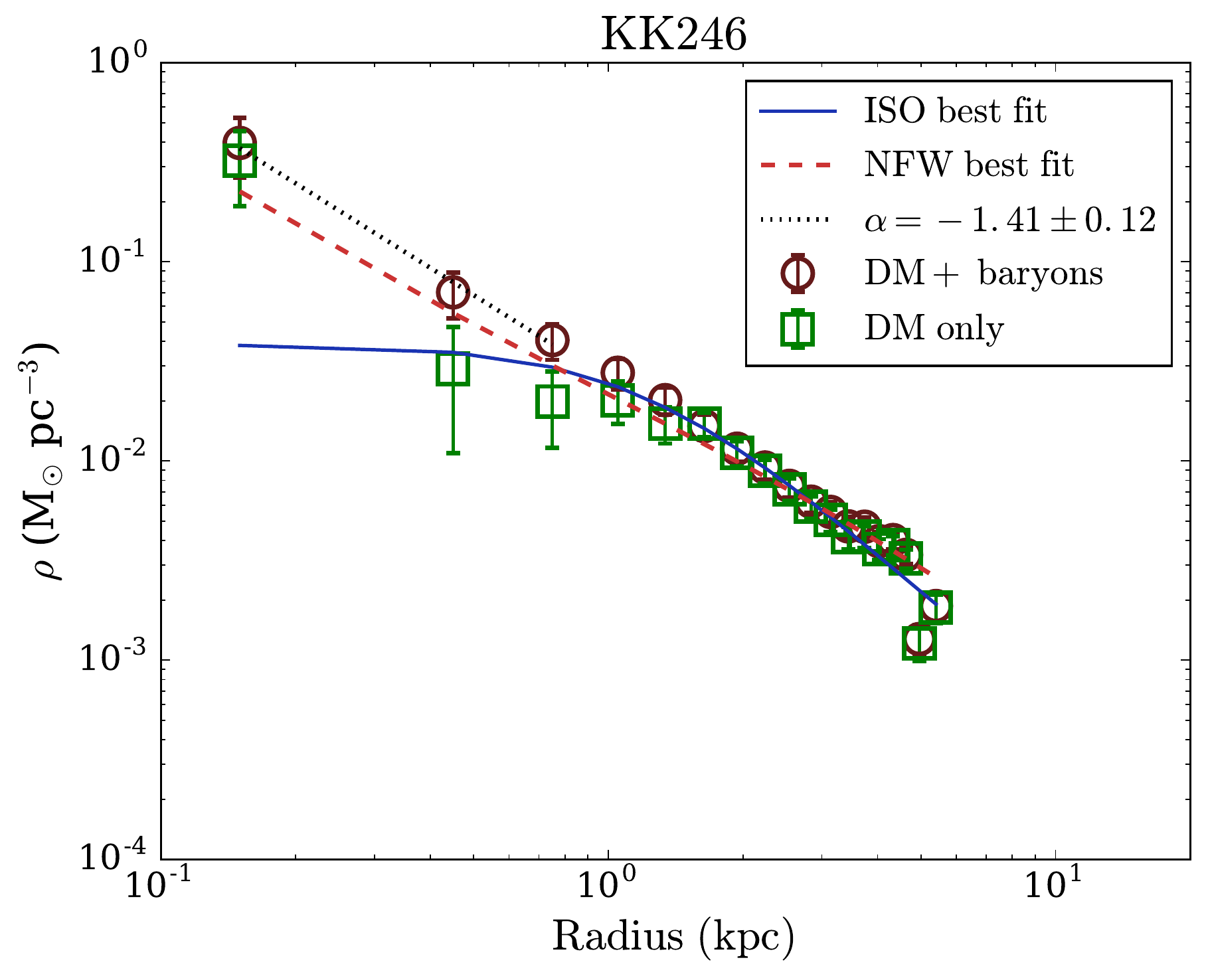}}
\caption{Left panels: Mass models for J0626+24, J0929+1155 and KK246. The open circles indicate the total observed velocities after the correction for pressure support. The blue filled circles indicate the velocity contribution of the dark matter halo (after subtracting the contribution of baryons). The red dashed line and the green solid line represent the best fit NFW model and the best fit ISO model. The blue dotted line and the brown dashed line represent the gas and stars respectively. The residual velocities from best-fit ISO and NFW model are shown by the green filled circles and the red open circles. Right panels: The mass density profiles of J0626+24, J0929+1155 and KK246. The black dotted line indicates the data points used for the estimation of inner density slope.}
\label{fig:dens1}
\end{figure*}

\begin{figure*}
\noindent
\captionsetup[subfigure]{aboveskip=-1pt,belowskip=-1pt}
\subfloat{\includegraphics[width = 3.25in]{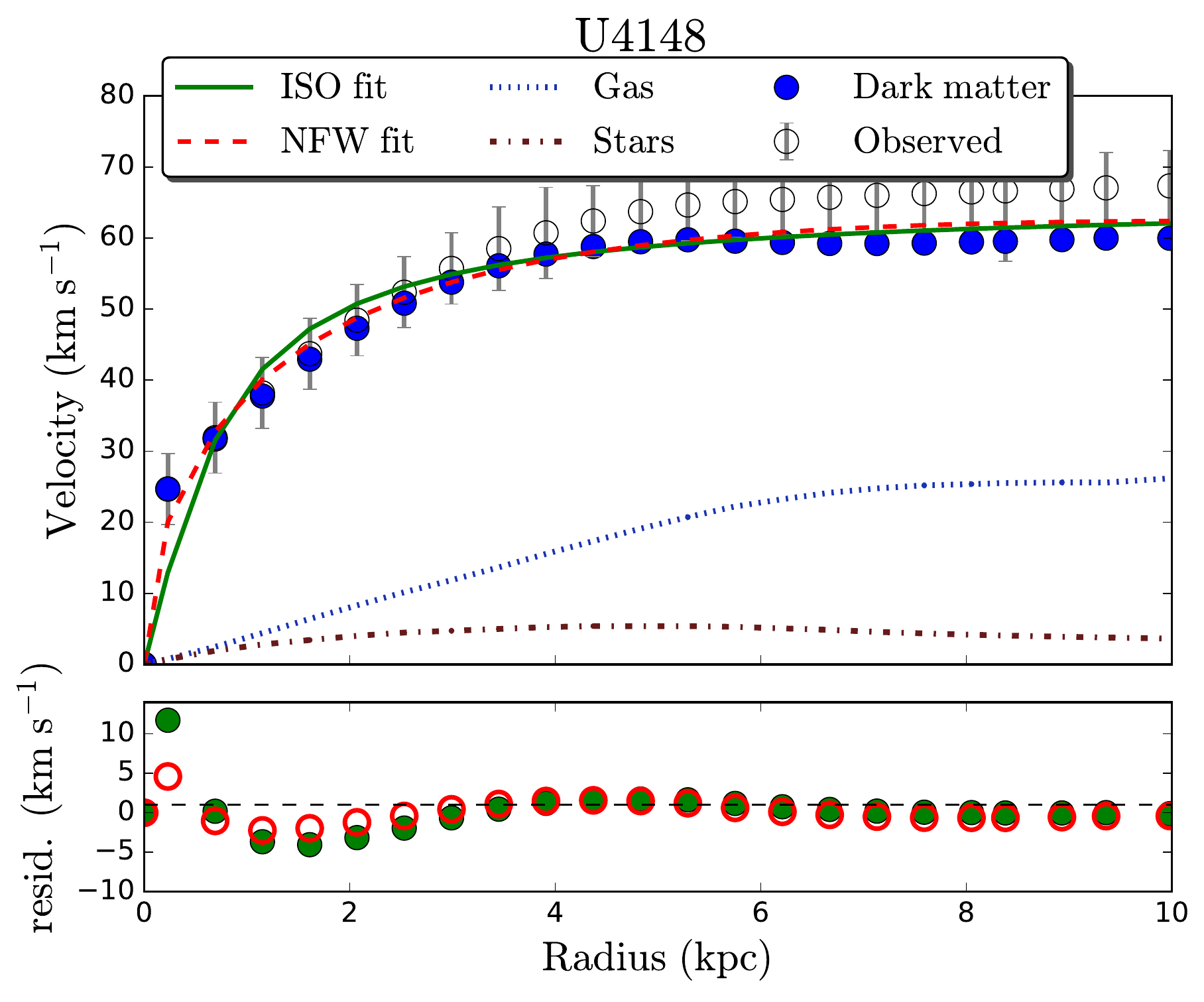}}
\subfloat{\includegraphics[width = 3.25in]{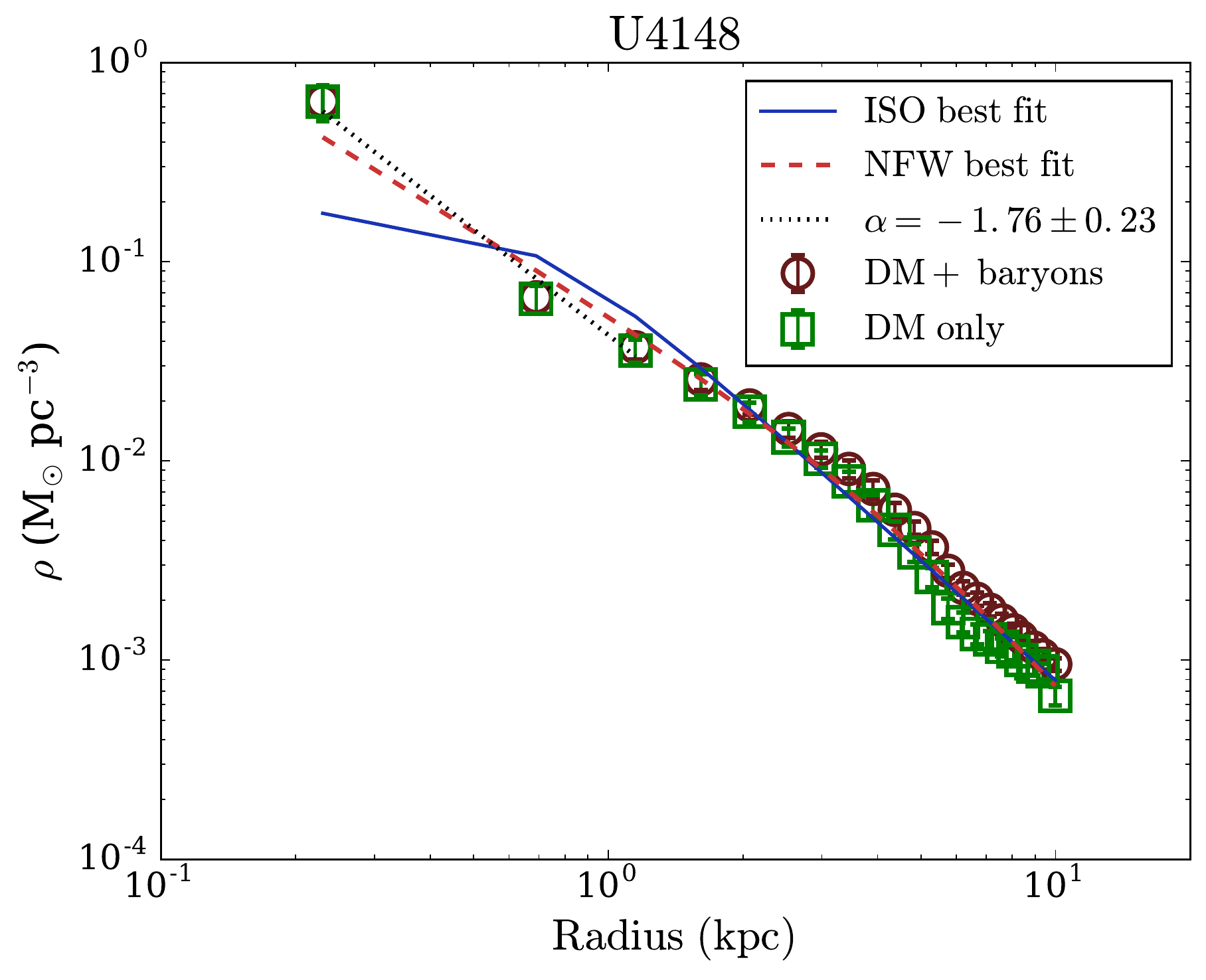}} \\
\subfloat{\includegraphics[width = 3.25in]{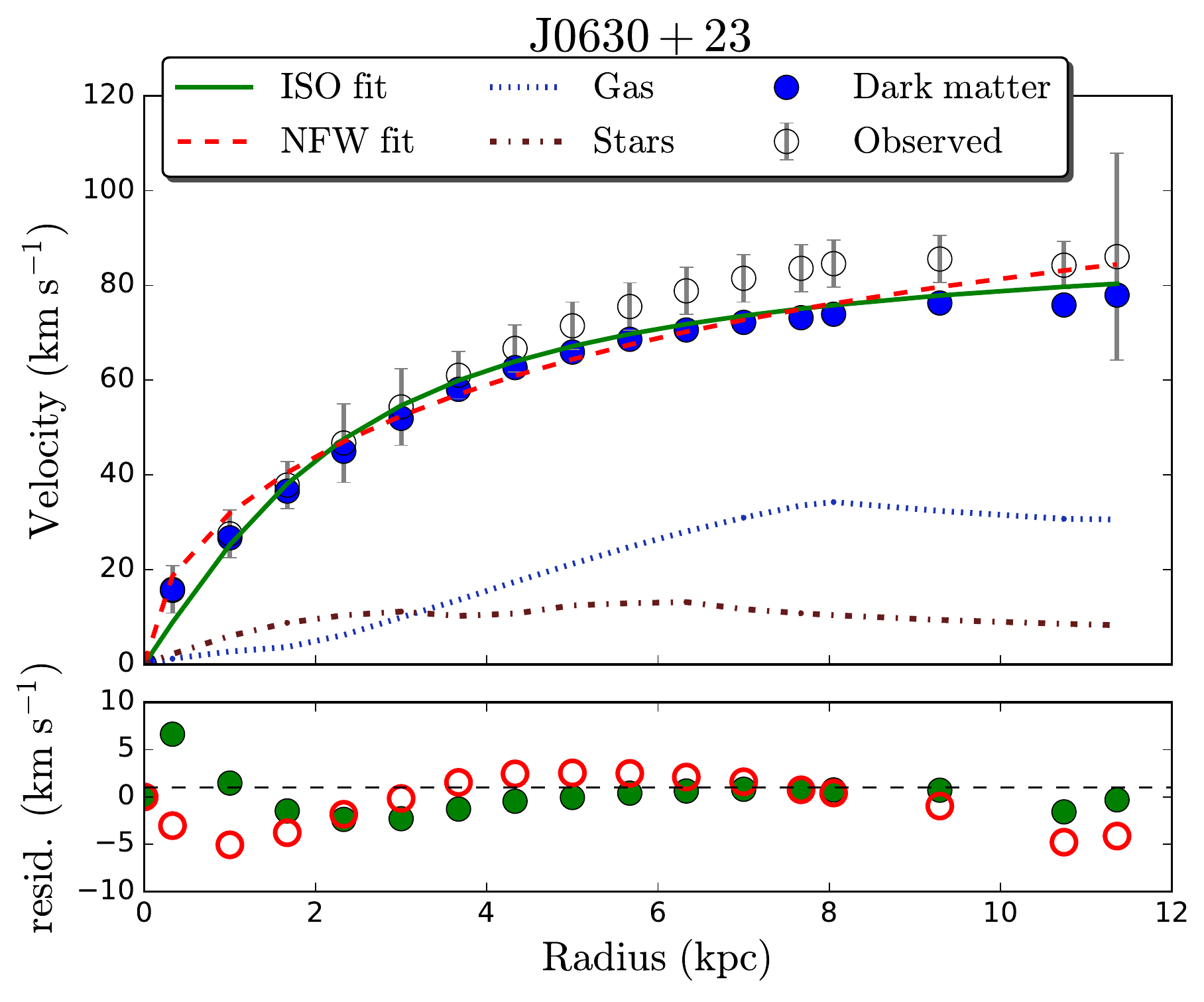}}
\subfloat{\includegraphics[width = 3.25in]{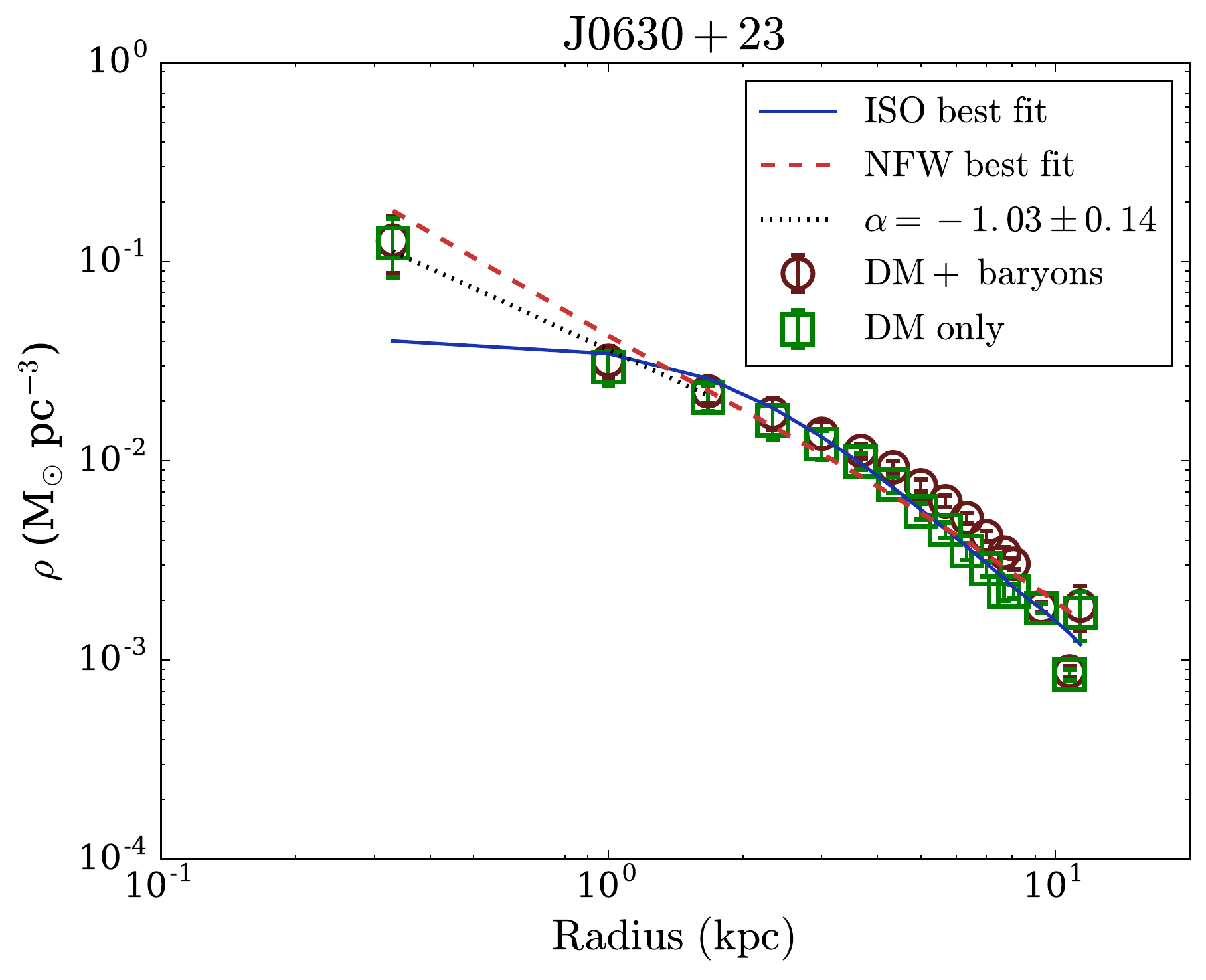}} \\
\subfloat{\includegraphics[width = 3.25in]{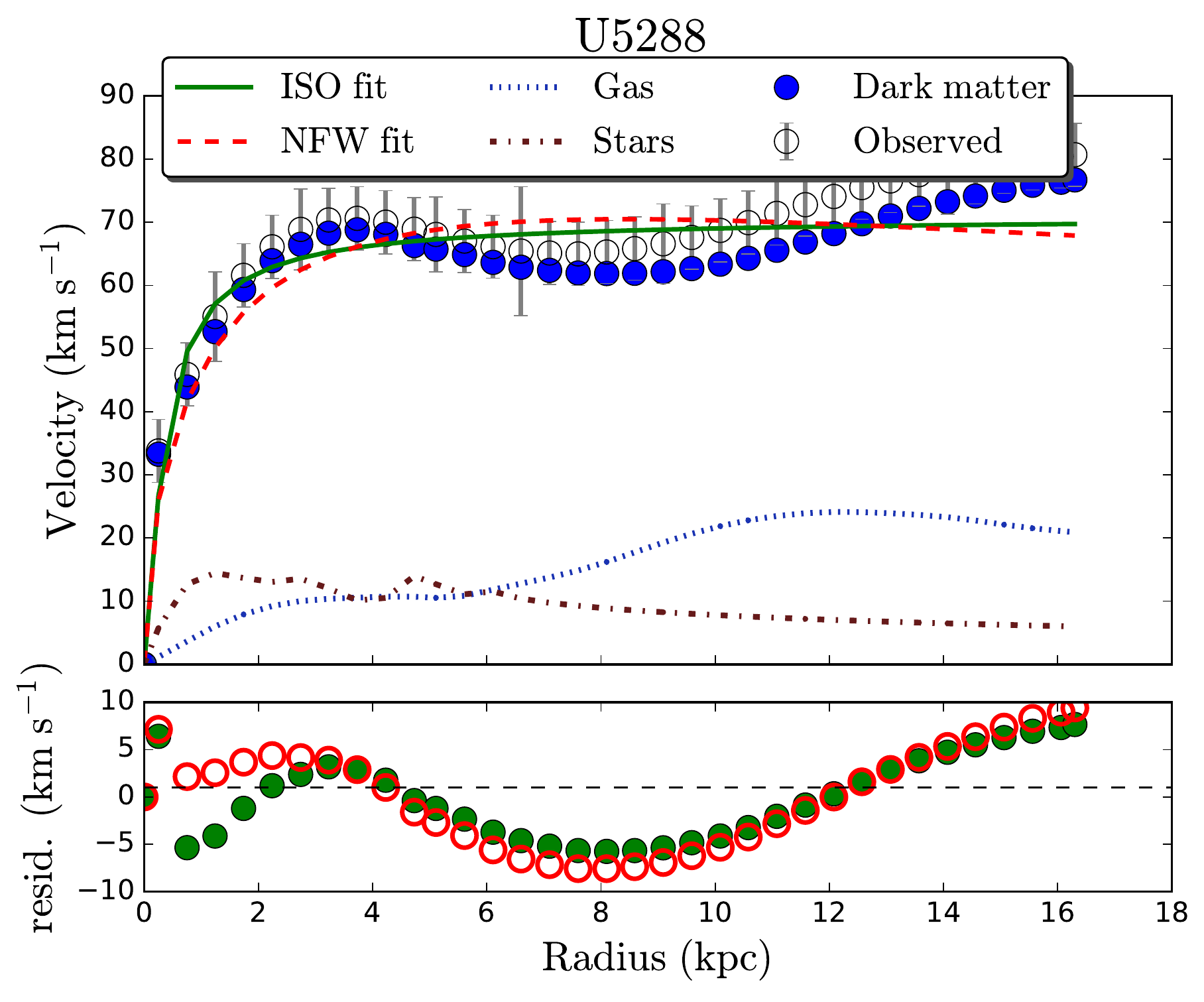}} 
\subfloat{\includegraphics[width = 3.25in]{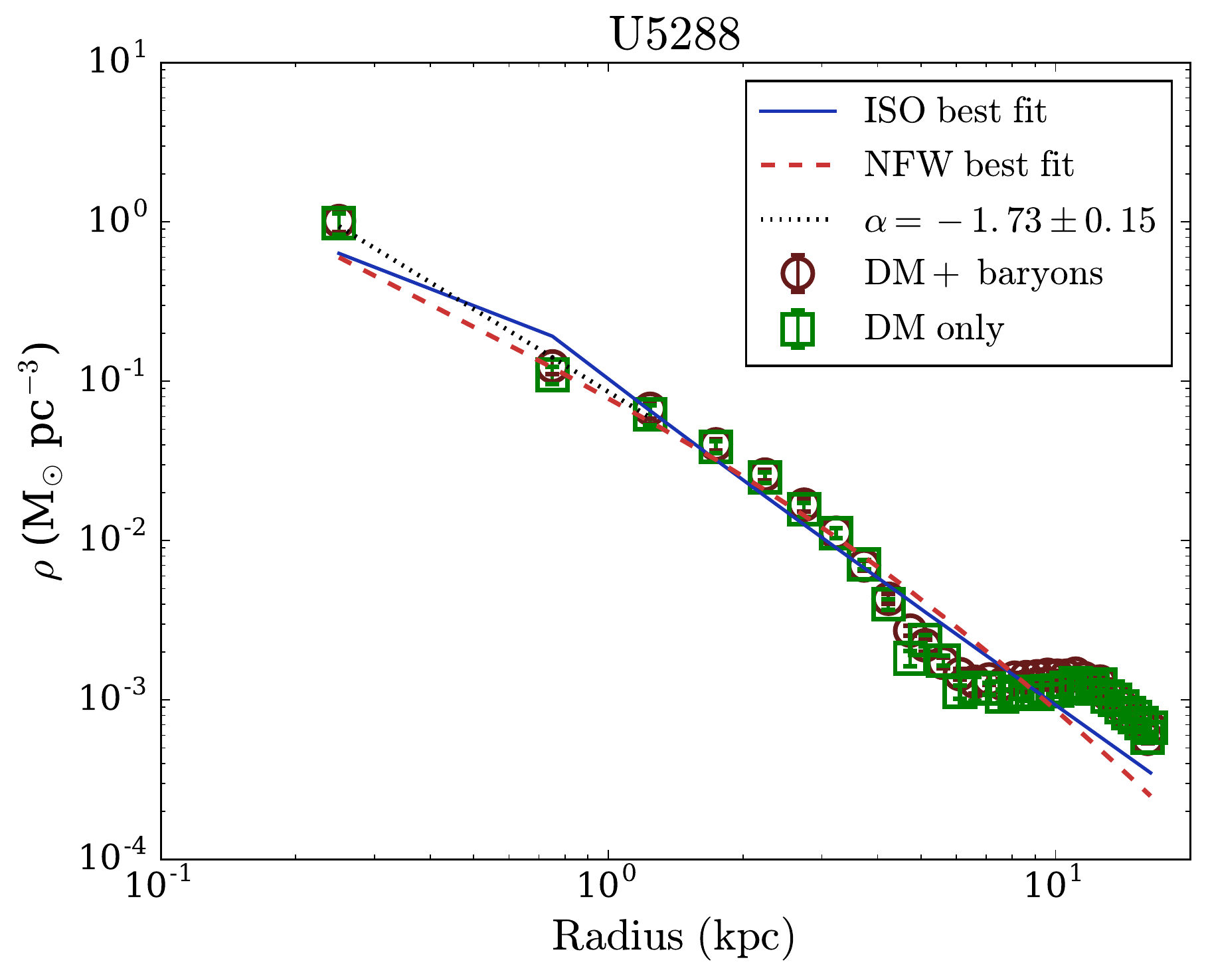}}
\caption{Left panels: Mass models for UGC4148, J0630+23 and UGC5288. The open circles indicate the total observed velocities after the correction for pressure support. The blue filled circles indicate the velocity contribution of the dark matter halo (after subtracting the contribution of baryons). The red dashed line and the green solid line represent the best fit NFW model and the best fit ISO model. The blue dotted line and the brown dashed line represent the gas and stars respectively. The residual velocities from best-fit ISO and NFW model are shown by the green filled circles and the red open circles. Right panels: The mass density profiles of UGC4148, J0630+23 and UGC5288. The black dotted line indicates the data points used for the estimation of inner density slope.}
\label{fig:dens2}
\end{figure*}

\begin{figure*}
\noindent
\captionsetup[subfigure]{aboveskip=-1pt,belowskip=-1pt}
\subfloat{\includegraphics[width = 3.25in]{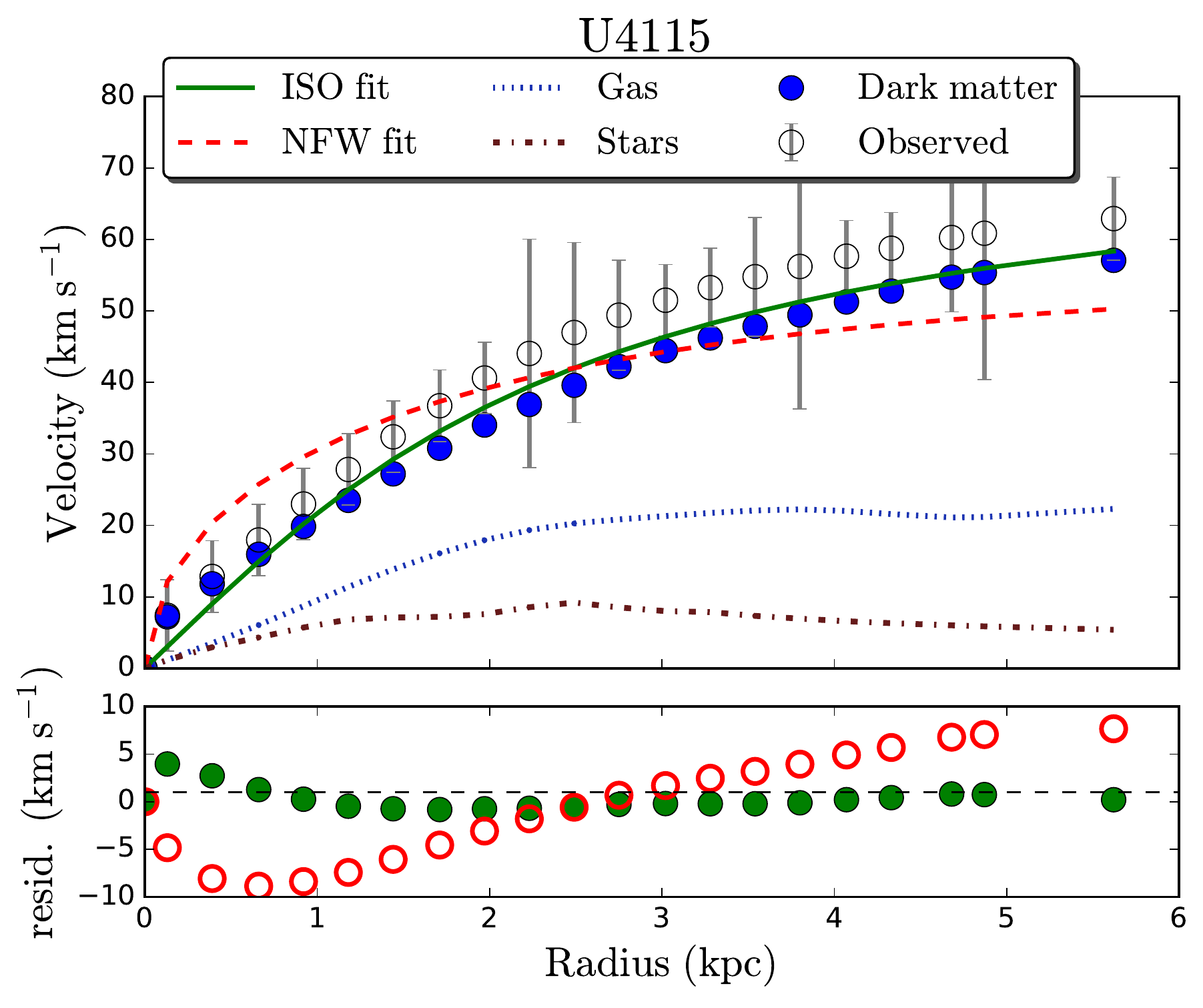}} 
\subfloat{\includegraphics[width = 3.25in]{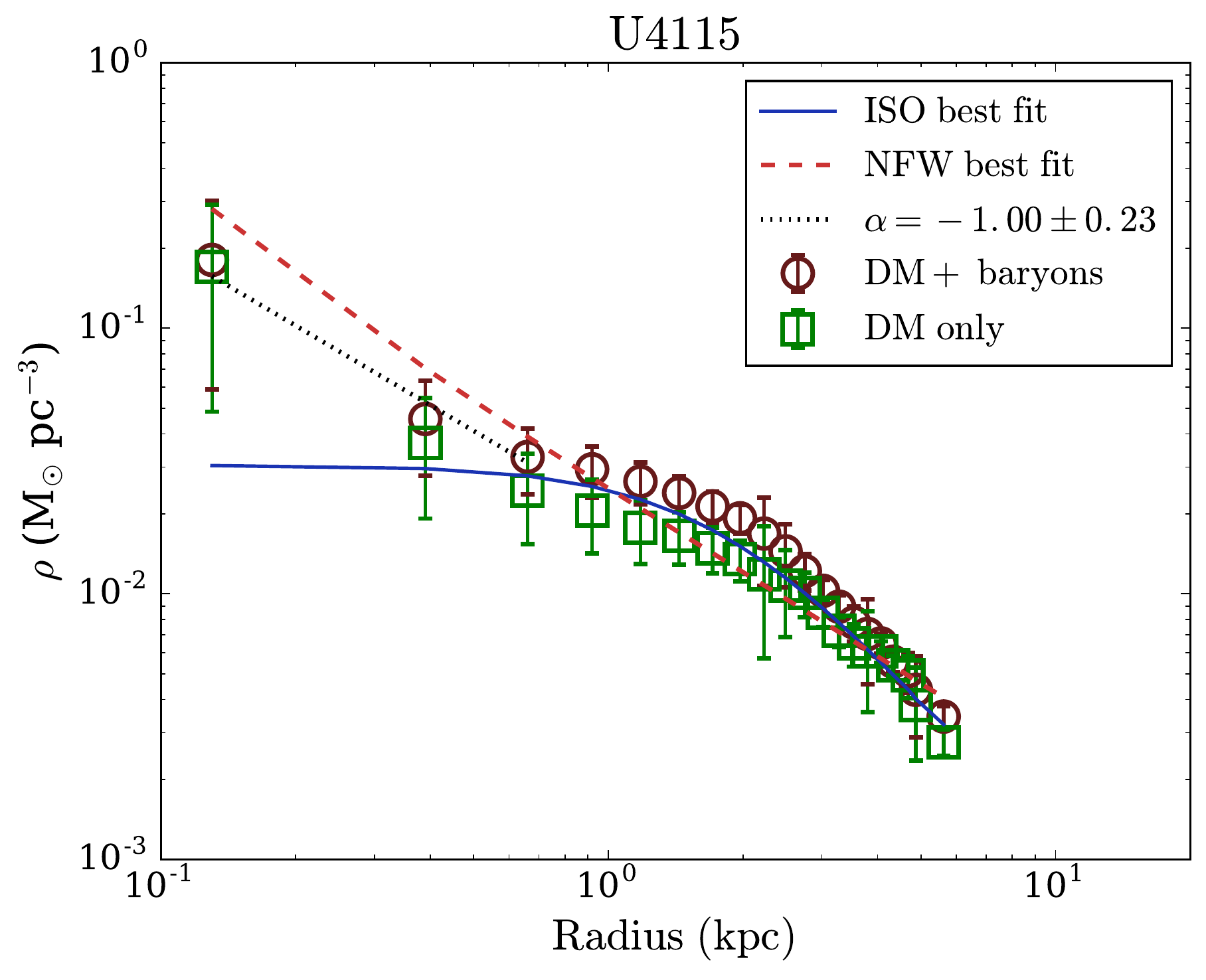}} \\
\subfloat{\includegraphics[width = 3.25in]{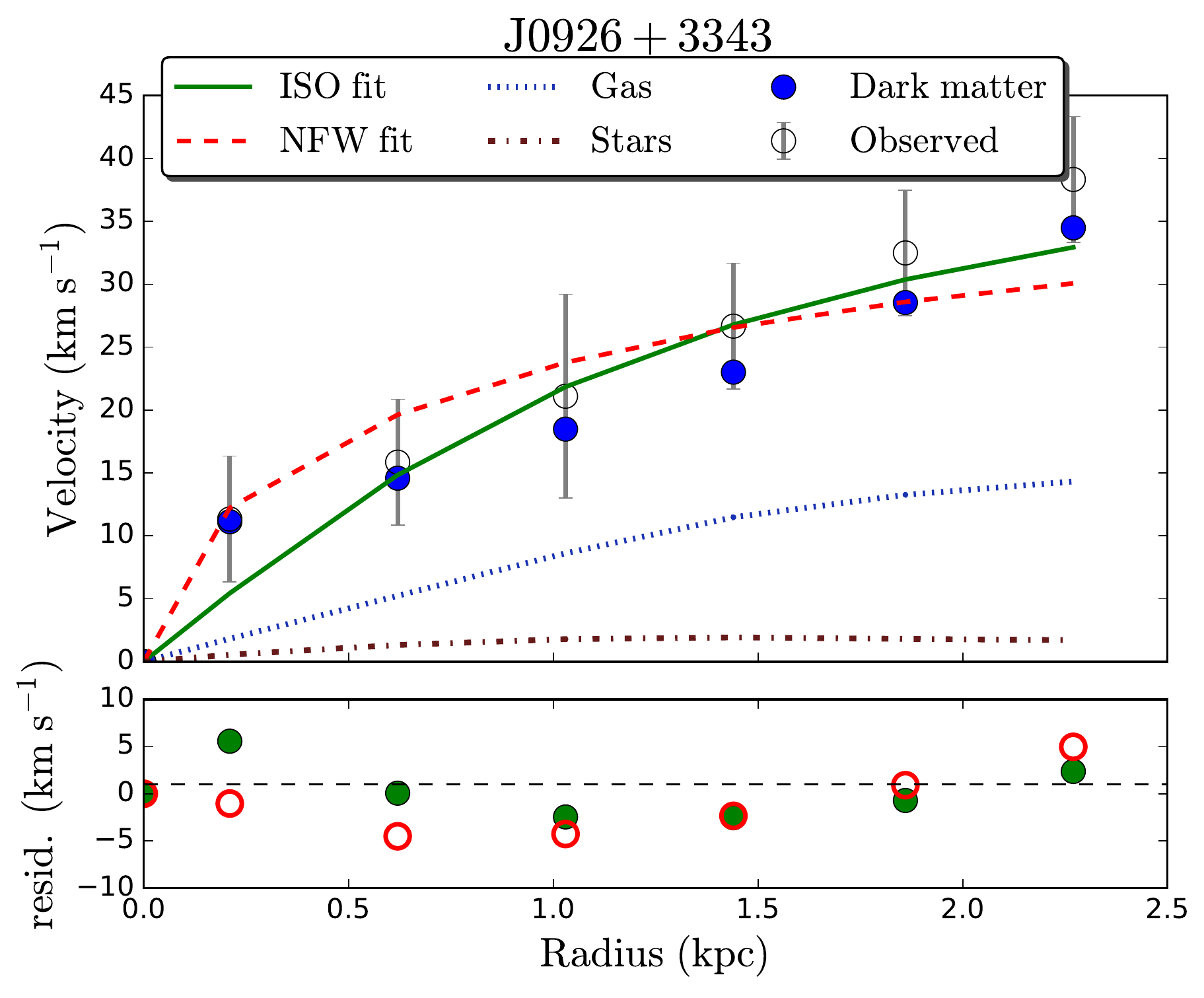}} 
\subfloat{\includegraphics[width = 3.25in]{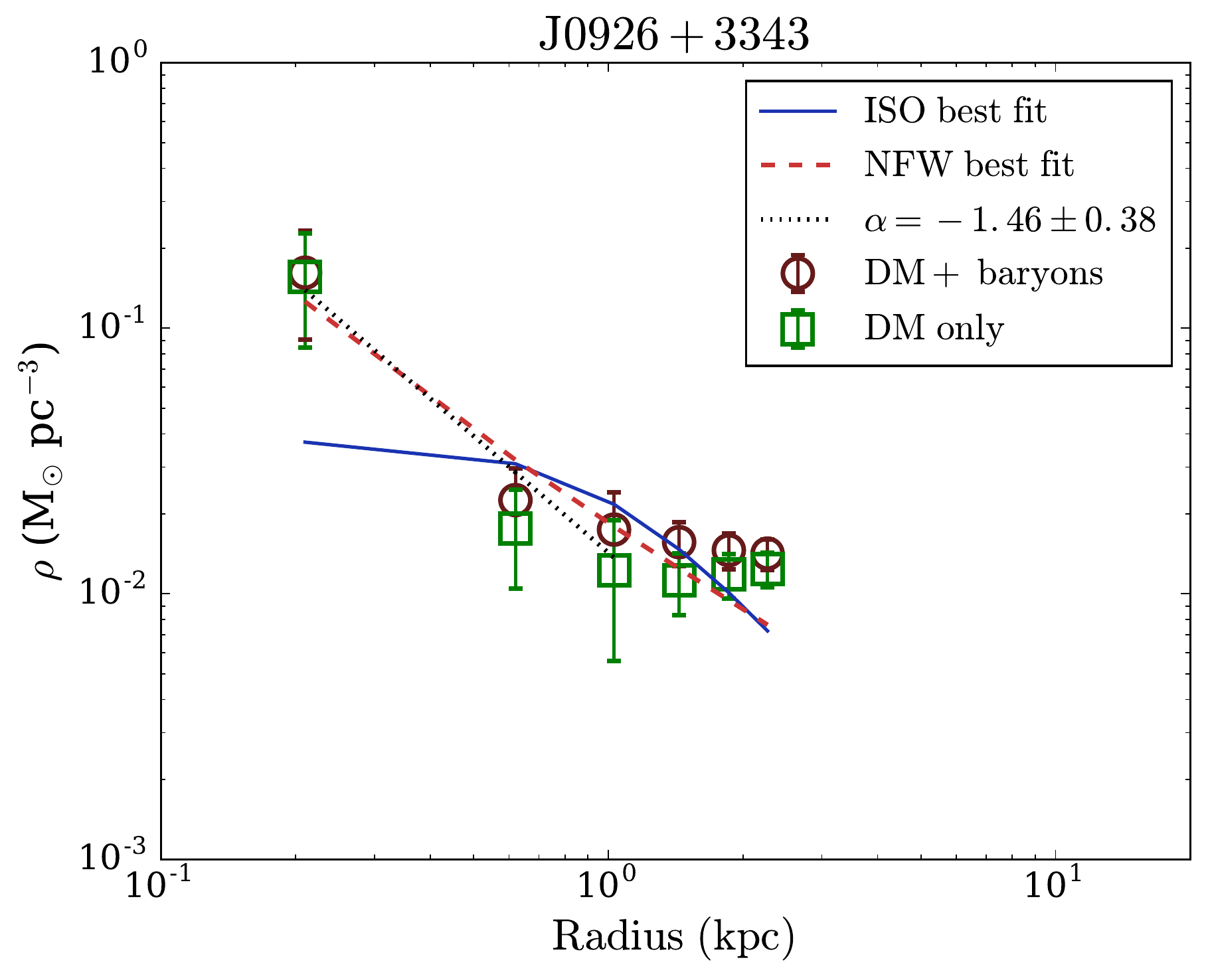}}
\caption{Left panels: Mass models for UGC4115 and J0926+3343. For UGC4115, we didn't yield physical results when both c and R$_{200}$ were allowed to vary. Hence, we fit the model by fixing the c value to the average concentration parameter for other galaxies ( $\sim$ 5.5 ). The open circles indicate the total observed velocities after the correction for pressure support. The blue filled circles indicate the velocity contribution of the dark matter halo (after subtracting the contribution of baryons). The red dashed line and the green solid line represent the best fit NFW model and the best fit ISO model. The blue dotted line and the brown dashed line represent the gas and stars respectively. The residual velocities from best-fit ISO and NFW model are shown by the green filled circles and the red open circles. Right panels: The mass density profiles of UGC4115 and J0926+3343. The black dotted line indicates the data points used for the estimation of inner density slope.}
\label{fig:dens3}
\end{figure*}

\section{Mass models}
\label{massmodels}

\begin{table*}
\begin{footnotesize}

\caption{Optical and dark matter properties }
\label{table:dm}
\begin{tabular}{ p{1.6cm} p{0.7cm}  p{0.8cm}  p{1.2cm} p{0.7 cm} p{1.4cm} p{1.9cm} p{0.7cm}  p{1.4cm} p{2.1cm} p{0.7cm}}
\\
\hline
 &  &  &  & & Isothermal &  &  & NFW &  &   \\
\hline
Galaxy		& D &	M$_{B}$ & V$_{rot}$  & $\gamma_{\ast}$ & r$_{c}$ & $\rho_{0}$  &$\chi^{2}_{r}$ &  	c            & r$_{200}$	& $\chi_{r}^{2}$ \\		
 & (Mpc) & & (km s$^{-1}$) & & (kpc) & (M$_{\odot}$ pc$^{-3}$) & & & (kpc) & \\
\hline
KK246		     & 6.85 &	-13.69  & 48.5 & 1.07 & 1.20$\pm$0.26 & 39$\pm$12   & 0.30 & 2.7$\pm$0.7 &	52$\pm$10 & 0.06 \\
U4115$^{ a}$     & 7.73 &	-14.75  & 62.9 & 0.29 & 1.87$\pm$0.20 & 31$\pm$4    & 0.09 & ..(5.5)   &.. (37$\pm$3)&..(0.78)\\		
J0926+3343       & 10.6 &	-12.90  & 38.3 & 0.06 & 0.34$\pm$0.19 & 163$\pm$143 & 0.52 & 1.9$\pm$6.1 & 60$\pm$161  & 0.15 \\ 
U5288		     & 11.4 &	-15.61  & 80.6 & 0.36 & 0.25$\pm$0.05 & 1341$\pm$507& 0.71 & 11.4$\pm$1.7& 41$\pm$2 & 1.21 \\	
U4148		     & 13.5 &	-15.18  & 67.3 & 0.17 & 0.50$\pm$0.11 & 289$\pm$118 & 0.69 & 8.4$\pm$0.6 & 39$\pm$1 & 0.14 \\	
J0630+23	     & 22.9 &	-15.89  & 86.0 & 0.24 & 1.92$\pm$0.24 & 42$\pm$8    & 0.34 & 3.4$\pm$0.8 & 71$\pm$10 & 0.39 \\	
J0626+24	     & 23.2 &	-15.64  & 88.0 & 0.27 & 0.89$\pm$0.33 & 144$\pm$90  & 2.14 & 5.1$\pm$1.1 & 60$\pm$8 & 0.52 \\	
J0929+1155	     & 24.3 &	-14.69  & 66.8 & 0.22 & 0.64$\pm$0.10 & 185$\pm$48  & 0.17 & 5.8$\pm$0.7 & 46$\pm$3 & 0.07 \\
\hline  
\multicolumn{11}{l}{$^{ a}$ For UGC4115, we didn't yield physical results when both c and R$_{200}$ were allowed to vary for the NFW halo model.  }\\
\multicolumn{11}{l}{Hence, we fix the c value to the average concentration parameter for other galaxies ( $\sim$ 5.5 ) to fit the NFW halo model. }
\end{tabular}

\end{footnotesize}

\end{table*}

The circular velocity reflects the gravitational potential of total matter content of galaxy which includes stars, gas  and dark matter. We use the H{\sc i} and optical de-projected radial surface brightness profiles and subtract the dynamical contribution of the atomic gas and stars in order to constrain the dark matter distribution. The derivation of the gas and stellar surface densities are described in \S \ref{ste} and \S \ref{gas}.  In \S \ref{ml}, we discuss the various different assumptions regarding the stellar mass to luminosity ratio, and the effect they have on the derived dark matter distribution. We assume that the contribution of the molecular gas is negligible since for faint dwarf galaxies such as those in our sample, the molecular fraction is believed to be low \citep[e.g.][]{taylor98, schruba12, cormier14}. The contribution of the ionized gas is also expected to be small and has been neglected. The H{\sc i} gas and stars hence form the dominant baryonic component of the galaxy. In \S \ref{dark} below, we describe the different dark matter models that were used for the construction of mass models. Once the form of contribution of each of these components was determined, mass models were fit to the pressure support corrected rotation curves using `ROTMAS' task in \gipsy. The modelling procedure consists of a $\chi^{2}$ minimization of 
$v_{c}^{2} -\gamma v_{*}^{2} -v_{g}^{2} -v_{h}^{2}  $, where $v_{c}$ is the rotation velocity after correcting for the pressure support, $\gamma$ is the mass to luminosity ratio, $v_{*}, v_{g} $ and $v_{h} $ are the rotation velocities contributed by stars, gas and dark matter halo respectively. 

\subsection{Stellar Component}
\label{ste}

We use the $g$-band optical images to calculate the contribution of the stellar disk to the rotation curve. The optical $g$-band images were either taken from  SDSS \citep{ahn12} or from PanSTARRS \citep{flewelling16} for those galaxies which lie outside the SDSS footprint. The de-projected luminosity profiles were derived using the task `ELLINT' in `\gipsy' using the parameters obtained from the tilted ring fits. We fit an exponential to the extracted luminosity profile to obtain the scale length (h). We derive the $g$-band mass to luminosity ratio from the $g-i$ colors  \citep[taken from][]{perepelitsyna14} and the stellar mass to light relations given in \citet{zibetti09}. We use the derived mass to luminosity ratio to convert the luminosity profiles into  stellar mass profiles. For the mass modelling, we assume a vertical sech$^{2}$(z) scale height distribution, with the ratio of scale length to scale height (h/z$_{\circ}$) to be 5.  We have also confirmed, that the choices of the vertical profile and the value of h/z$_{\circ}$ do not affect the mass models significantly. 

\subsection{Gas Component}
\label{gas}

We use the total integrated H{\sc i} intensity maps to derive the contribution of the gas to the observed rotation velocity. We apply the tilted ring parameters (\S  \ref{rot}) to derive the de-projected H{\sc i} radial surface density profiles using task `ELLINT' in {\sc gipsy}. For most galaxies, the H{\sc i} surface density profiles derived in this way match broadly with the profiles derived from using the \fat pipeline to directly fit to the data cube. In some cases however, we find that the \fat\ derived surface density tends to be slightly higher in the inner regions of the galaxy. The reason for this difference is unclear. We show in \S \ref{results} that our main results are not affected significantly if one uses the \fat derived surface brightness profiles instead of 'ELLINT' ones. We scale the derived H{\sc i} surface density profiles by a factor of 1.35 to take the contribution of Helium and metals to the gas mass. As discussed above, we assume that the contribution of molecular gas and ionized gas to the total gas mass is negligible.

\subsection{Dark matter halo}
\label{dark}

We use the two well-known models, viz. the Isothermal and the NFW models to parametrize the dark matter distribution. The parameters of these models are briefly summarized below.

\subsubsection{Isothermal halo model}

The density distribution of the observationally motivated pseudo-isothermal (ISO) halo model \citep[e.g.][]{begeman91} is:

\begin{equation}
\rho_{iso} (r) =\rho_{0}[1+(r/r_{c})^{2}]^{-1}
\end{equation}

where, $\rho_0$ is the core density and $r_c$  the core radius of the halo. The corresponding circular velocity is:

\begin{equation}
V_{iso} (r) = \sqrt{4 \pi \rho_{0} (r)  r_{c}^{2}\Big [1-\dfrac{r_{c}}{r} \tan^{-1}(\dfrac{r}{r_{c}})\Big ]}
\end{equation}

The inner slope ($\rho$ $\sim$ r$^{\alpha}$) of the mass density profile for the Isothermal halo model is $\alpha = 0$ (since for  $r<< r_{c}$, $\rho_{iso} \approx \rho_{\circ}$).

\subsubsection{NFW halo model}

The density profile of the NFW halo model \citep{navarro96} is given by: 

\begin{equation}
\rho_{NFW} (r) =\dfrac{\rho_{i}}{(r/r_{s})(1+r/r_{s})^{2}}
\end{equation} 

where $ r_{s} $ is a characteristic radius of the dark matter halo and $\rho_{i}$ is the initial density of the universe at the time of collapse of the halo.  

The corresponding circular velocity is:

\begin{equation}
V_{NFW} (r) = v_{200} \sqrt{\dfrac{ln(1+cx)-cx/(1+cx)}{x[ln(1+c)-c/(1+c)]}}
\end{equation}

where $ c=r_{200}/r_{s} $ is the concentration parameter and $ x=r/r_{200} $; $ r_{200} $ being the radius at which the mean density of the halo is equal to 200 times the critical density and $ v_{200} $ is the rotational velocity at $ r_{200}$. $ v_{200} $ is related to the $ r_{200}$ as $ v_{200} $ = h$ r_{200}$, where h is the dimensionless Hubble parameter. The inner slope of the density distribution is $\alpha$ $\sim$ -1  (at radii $r<< r_{s}$, $\rho_{NFW} \approx \rho_{i}$ \big($\frac{r_{s}}{r}$ \big)).

\subsection{Constructing the mass models}
\label{ml}

As described above mass models for the galaxies were fit using three components i.e. the stellar disk, the gas disk and the dark matter halo. A number of different fits were performed. Firstly for the dark matter halo we fit both the Isothermal halo and NFW halo models. For each of these we allowed four different possibilities for the contributions from the baryonic components, viz. (1) {\it Constant $\gamma_{\ast}$:}  In this case, we use a fixed mass to light ratio as derived from the stellar population synthesis (SPS) models. (2) {\it Maximum disc:}  This model assumes that the observed rotation curve in the inner regions is almost entirely due to the stellar component. In this case, we scale the rotation curve due to the stellar component to the maximum value for which the dark matter density is non-negative at all radii.  This model constrains the value of dark matter density to be minimum at all radii. (3) {\it Minimum disc:} In this case, we assume that the contribution of baryons to the observed rotation curve is zero. This assumption provides an upper limit to the dark matter density.  (4) {\it Minimum disc + gas:} the stellar disk contribution to the rotation velocity is assumed to be zero, however the contribution of the gas disk is fully included. (5) {\it Free $\gamma_{\ast}$: } In this model, we allow the mass-to-light ratio to be a free parameter. In some of the galaxies, this assumption did not yield physical results. The fit results of individual galaxies for various assumptions for $\gamma_{\ast}$ are presented in Table \ref{table:gamma} in the Appendix.

\section{Results}
\label{results}

 Fig \ref{fig:dens1}--\ref{fig:dens3} show the fit results for case(1), i.e. where the stellar mass to light ratio is fixed using the $g-i$ colors, using the rotation curves derived with \fat and the corresponding fit parameters are given in Table~\ref{table:dm}. We find that the NFW halos provide a better fit in most of the galaxies in terms of fit quality. However, the $\chi^{2}_{red}$ is less than 1 in almost all cases, indicating that the errors provided by the \fat package are likely to be overestimated. Although both fits are formally acceptable, the residual velocities for NFW model(shown by red open circles) are generally smaller than the residual velocities for the isothermal model (shown by green filled circles) in the central regions. For 2 galaxies, viz. UGC4115 and UGC5288, the isothermal halo gave a better fit compared to the NFW halo. We note that the galaxy UGC5288 has a strong bar. For the galaxy UGC4115, we did not get physical results when both $c$ and R$_{200}$ were allowed to freely vary during the fitting process. For this galaxy we hence fix the value of $c$ to the average concentration parameter for other galaxies ( $\sim$ 5.5 ).  

\subsection{Dark matter density profiles}

\begin{figure}
\centering
\includegraphics[width=1.0\linewidth]{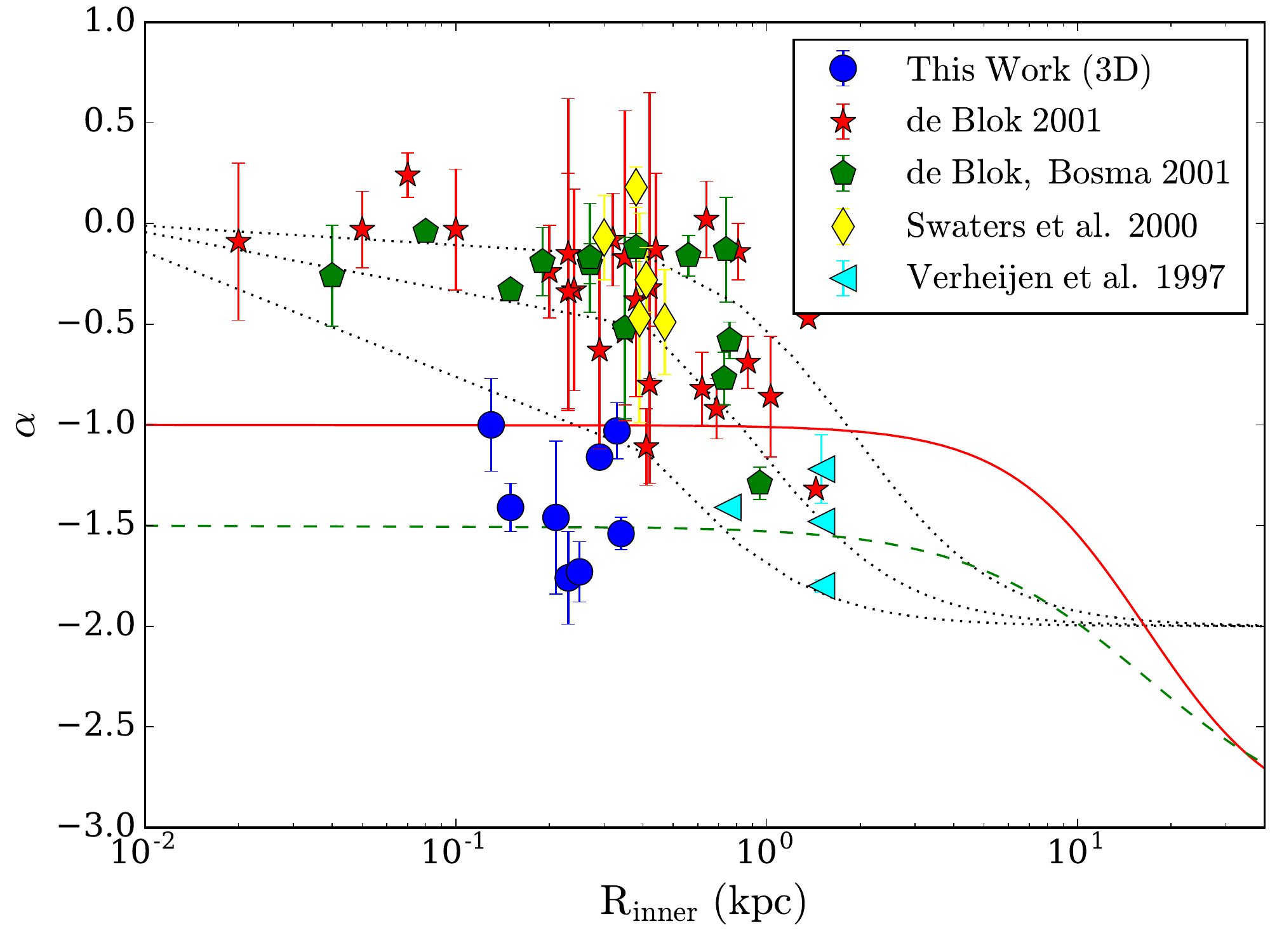}
\includegraphics[width=1.0\linewidth]{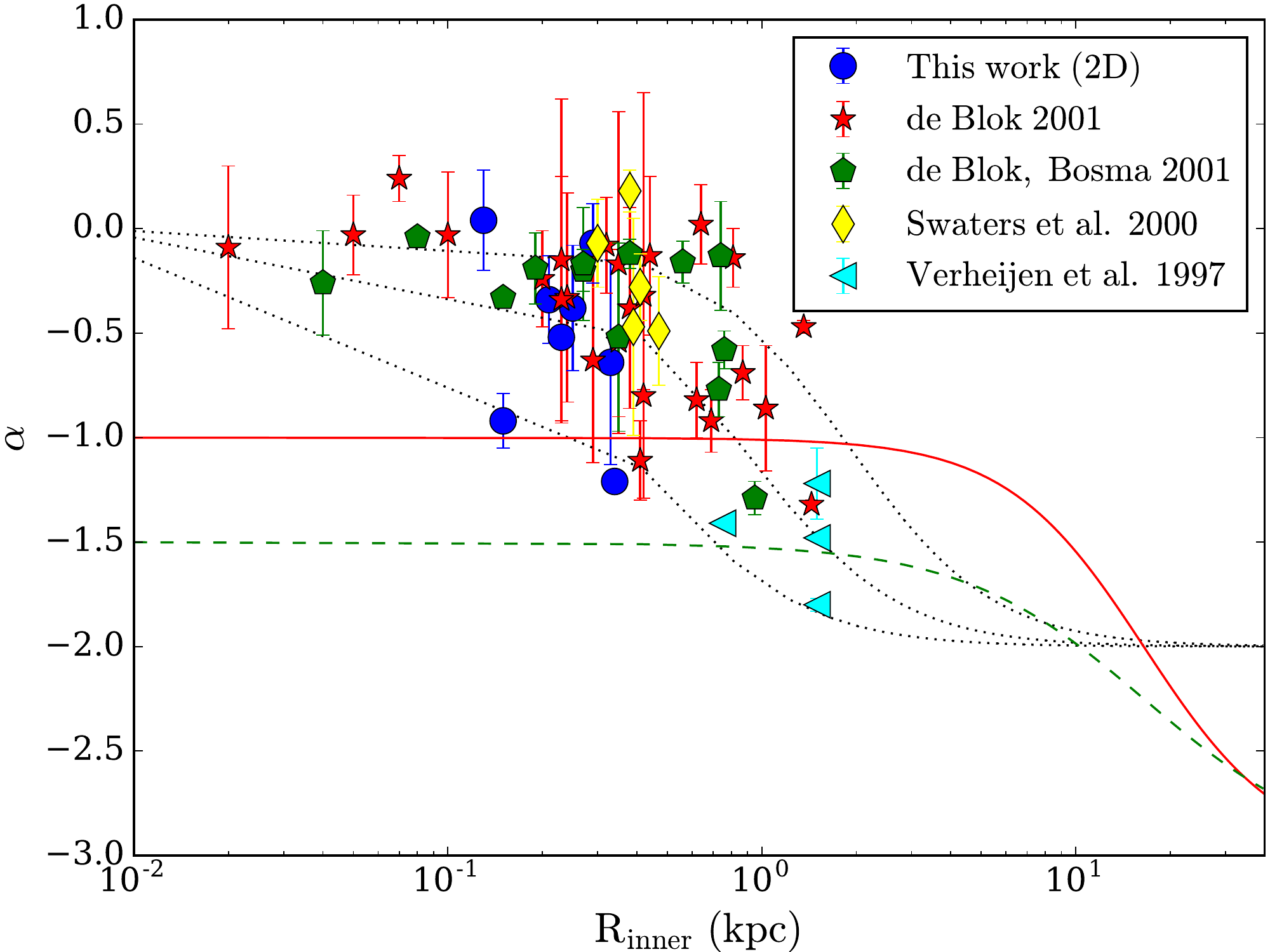}
\caption{The value of inner slope ($\alpha$) of the dark matter density distribution versus the radius of the innermost point. The data points from this work are shown by blue circles. The red stars are from \citet{deBlok01}, the green pentagons are from \citet{deBosma02}, the yellow diamonds are from \citet{swaters00}, and the cyan triangles are from \citet{verheijen97}. The black dotted lines are the theoretical slopes for the isothermal halo with core radii of 0.5 (left), 1 (middle) and 2(right) kpc. The red solid line is the NFW model (r$^{-1}$)  and the green dashed line is the CDM r$^{-1.5}$ model with c=8 and v$_{200}$ = 100 km s$^{-1}$. upper panel: slopes obtained using the rotation curves derived from \fat (3-D approach), lower panel: slopes obtained using the rotation curves derived from ROTCUR (2-D approach).}
\label{fig:alpha}
\end{figure}%

Apart from fitting mass models, one can also use the rotation curve to directly determine the inner slope of the density profile \citep[e.g.][]{oh11a}. We assume a spherical halo potential \citep{trachternach08} to convert the rotation curve to the dark matter density profile using the following equation:

\begin{equation}
\rho(R) = \frac{1}{4\pi G} \bigg[2\frac{V}{R} \frac{\partial V}{\partial R} + \bigg( \frac{V}{R} \bigg)^{2} \bigg]
\end{equation}

where G is the gravitational constant and $V(R)$ is the rotation velocity after correcting for pressure support at a radius R.  The right panels of Fig~\ref{fig:dens1}--\ref{fig:dens3} show the derived dark matter density profiles of the individual galaxies. We plot both the dark matter densities derived from the total rotation curve (brown circles) and from the rotation curve after subtracting the contribution of baryons (green squares). The best-fit dark matter density profiles of isothermal halo and NFW halo models for a fixed mass to light ratio are also overplotted. As can be seen, the central dark matter density profiles are steeper than expected for the isothermal halo model, but are a good match to what would be expected from the NFW model. 

In order to quantify the cuspiness of the dark matter distribution in the central regions of the galaxy, the logarithmic inner slope of the density profiles were measured following the method described in \citet{deBlok01} \citep[see also][]{deBlok02, oh11, oh15}. We first determine the ``break radius" in the central regions of the galaxy. The break radius is defined as the radius at which the variation of the logarithmic slope of the density profile is maximum. The inner density slope ($\alpha$) is then measured using a least-squares fit to the data points that lie within the break radius. The range over which the fit is performed is shown in the right panels of Fig~\ref{fig:dens1}--\ref{fig:dens3} by a black dotted line. The uncertainty in $\alpha$ is taken to be half of the difference between the slopes measured when one includes or excludes the first data point outside the break radius while doing the fit. This error estimate is probably more appropriate than that derived from the formal fit error. The inner density slopes were measured from both the observed rotation curve as well as the rotation curve obtained after subtracting the baryonic contribution. The former case corresponds to the ``minimum disc assumption", in the latter case the rotation velocity corresponds to the contribution of the dark matter alone. The measured slope $\alpha$ and the uncertainty in slope $\Delta \alpha$ of the galaxies for the dark matter only profiles  are shown in the right panels of Fig~\ref{fig:dens1}--\ref{fig:dens3}. We find that the mean values of the inner density slopes are $\alpha$ = -1.39 $\pm$0.19 and $\alpha$ = -1.48 $\pm$0.23 for the minimum disc assumption and the dark matter only profile respectively.
We note that the values overlap within the errorbars. The measured logarithmic inner density slopes for all the galaxies are shown in table \ref{table:alpha}. These values are in better agreement with the logarithmic inner slope ($\sim$ -1) of the NFW cuspy profiles, and inconsistent with the slope of $\sim 0$ expected for the ISO halo. We also note that for all the galaxies, the radius within which we measure the inner density slope (r$_{\rm d}$) is much less than the best fit characteristic radius (r$_{s}$) of the NFW halo, where r$_{\rm d}$ is typically within $\sim$ 10\%  of r$_{s}$ and within 30 $\%$ of r$_{\rm s}$ for all the cases. But, for isothermal haloes, we find that r$_{\rm d}$ is typically comparable to the best fit core radius (r$_{c}$) of the isothermal halo and is higher than r$_{c}$ in a few cases, which may lead to steeper DM density profiles (steeper than $\alpha \sim 0$).

\subsection{Comparison with earlier studies}
\begin{figure}
\centering
\includegraphics[width=1.0\linewidth]{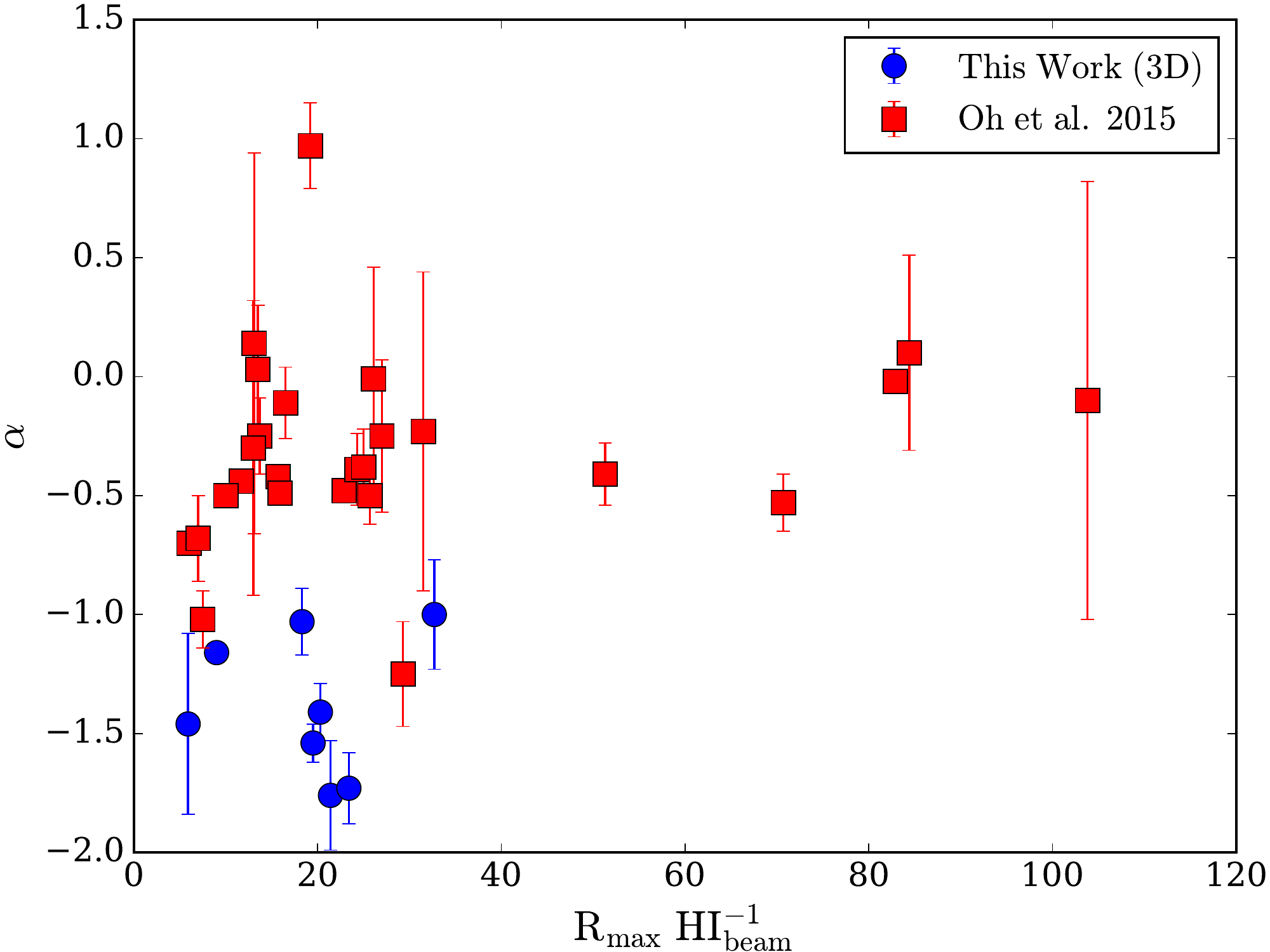}
\includegraphics[width=1.0\linewidth]{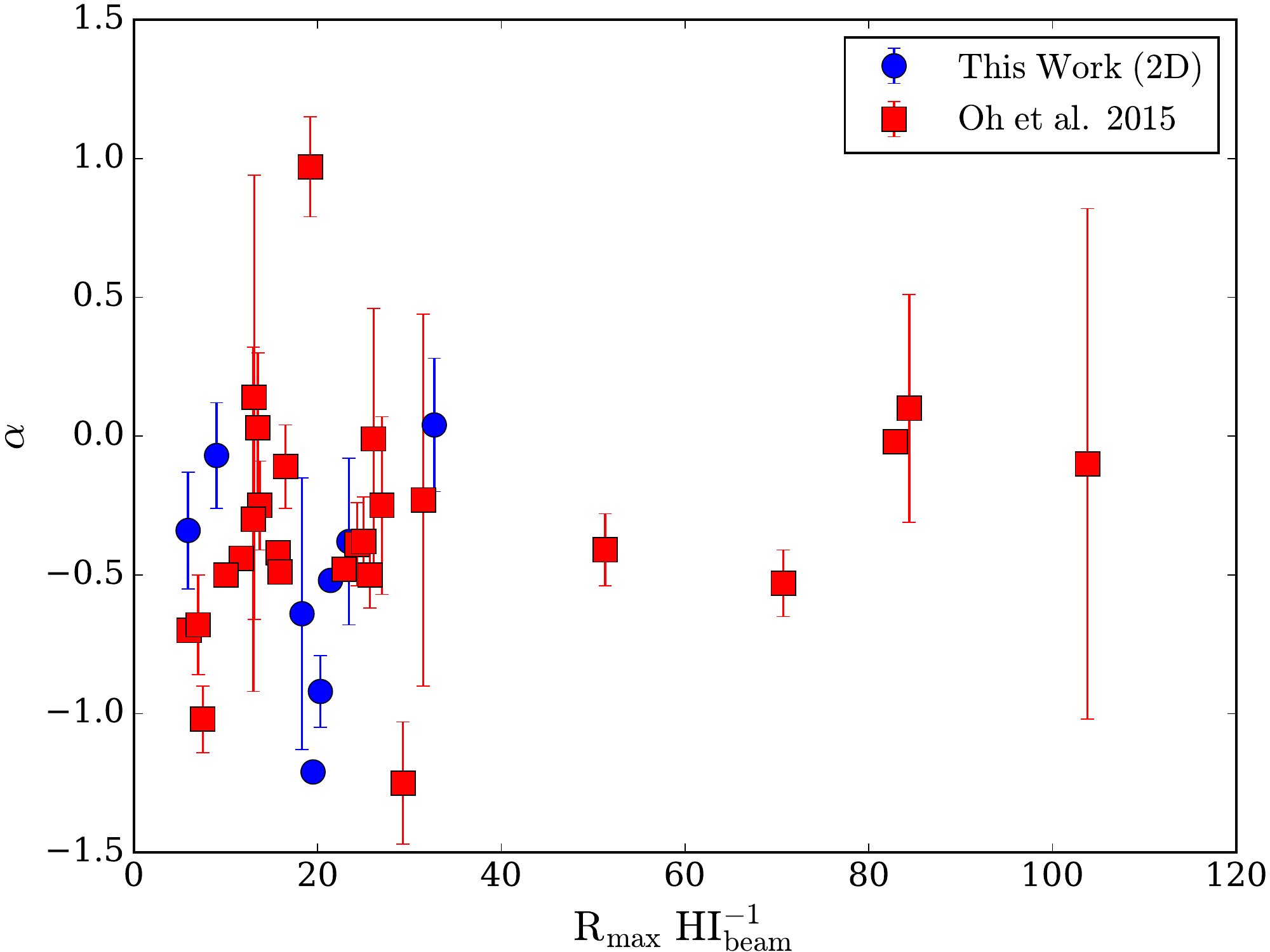}
\caption{The value of inner slope ($\alpha$) of the dark matter density distribution versus the number of resolution elements across the major axis of the galaxy. The data points from this work are shown by blue circles. The red squares are from \citet{oh15}. upper panel: slopes obtained using the rotation curves derived from \fat (3-D approach), lower panel: slopes obtained using the rotation curves derived from ROTCUR (2-D approach).}
\label{fig:alpha_rmax}
\end{figure}%

The steep inner slopes that we measure are in contradiction to several earlier measurements based on independent samples. For example, \citet{deBlok01} measure $\alpha$ to be $-0.2 \pm 0.2$, while  \citet{oh11} find $\alpha = -0.29 \pm 0.07$ and \citet{oh15} get $\alpha = -0.42 \pm 0.21$ using the minimum disc assumption. Some earlier studies \citep[e.g.][]{verheijen97} have however found steeper slopes, i.e. $\alpha \sim -1.8$. Insufficient sampling of the dark matter density profiles in the inner region could lead to an artificial steepening of the slope \citep{deBlok01}. This is because the logarithmic slope of the density profile becomes steeper towards the outer regions, steeper slopes would hence be obtained if data points from the outer region are included for the estimation of slope. To check whether this under sampling is the cause for the steeper slope that we measure, we compare the number of resolution elements across the galaxy for the galaxies in our sample and the literature. Fig \ref{fig:alpha_rmax} (upper panel) shows the plot of value of inner slope versus the number of resolution elements across the major axis of the galaxy. The blue circles indicate the galaxies from our sample and the red squares are the galaxies from earlier studies. As can be seen the galaxies in our sample have similar number of resolution elements as compared to the galaxies in the literature. This indicates that the obtained steeper slopes are not due to a different sampling in the inner regions of the dark matter halos. The other point of departure of this study is that we use rotation curves derived using a fit to the 3-D data cube, instead of a fit to the 2-D velocity field. To see what difference this makes, we show in Fig \ref{fig:alpha_rmax} (lower panel) the slopes measured for our sample using the 2-D velocity field and the \gipsy task ROTCUR. As can be seen the agreement between the slopes measured from our sample, and the slopes measured for earlier samples match quite well when we use rotation curves derived from the 2-D velocity field. Fig.~\ref{fig:alpha}, which is a plot of the value of inner slope ($\alpha$) versus the radius of the innermost point shows a similar result. Fig.~\ref{fig:alpha}~(a), shows slopes for our galaxies based on \fat rotation curves, while in Fig.~\ref{fig:alpha}~(b) shows the slopes measured using ROTCUR on the 2-D velocity fields. The data points from this work are shown by blue circles and other symbols are from measurements taken from other studies (see the caption for details). The black dotted line shows the slope for the isothermal model, the red solid line and the green dashed lines show the $\Lambda$CDM r$^{-1}$ \citep{navarro96} and r$^{-1.5}$ \citep{moore99} models respectively. The mean value of $\alpha$ for estimates from rotation curves derived from the 2-D velocity field, is  $\alpha = -0.49 \pm 0.24$, which is consistent with values obtained in the literature.

\begin{figure}
\centering
\includegraphics[width=1.0\linewidth]{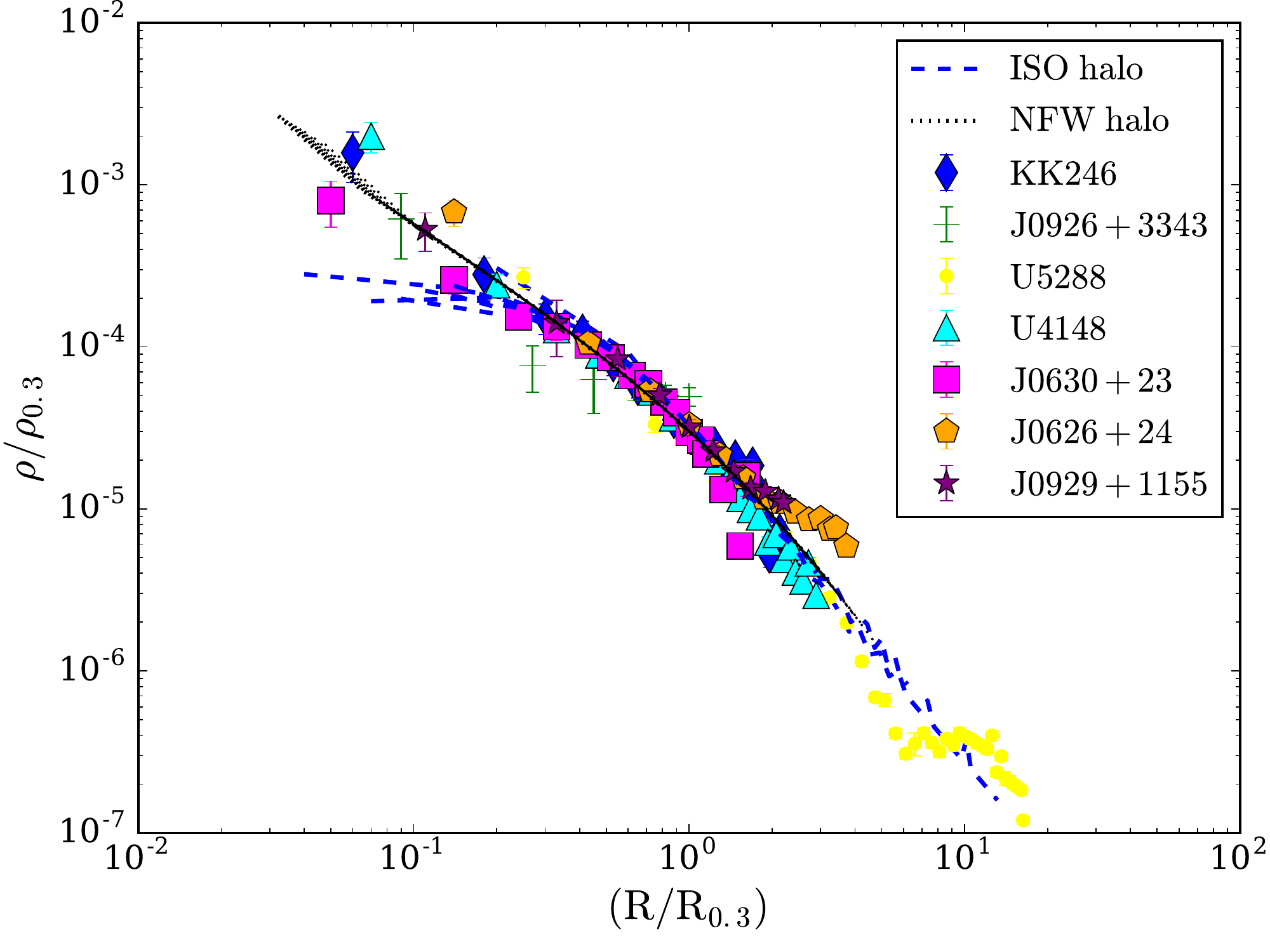}
\includegraphics[width=1.0\linewidth]{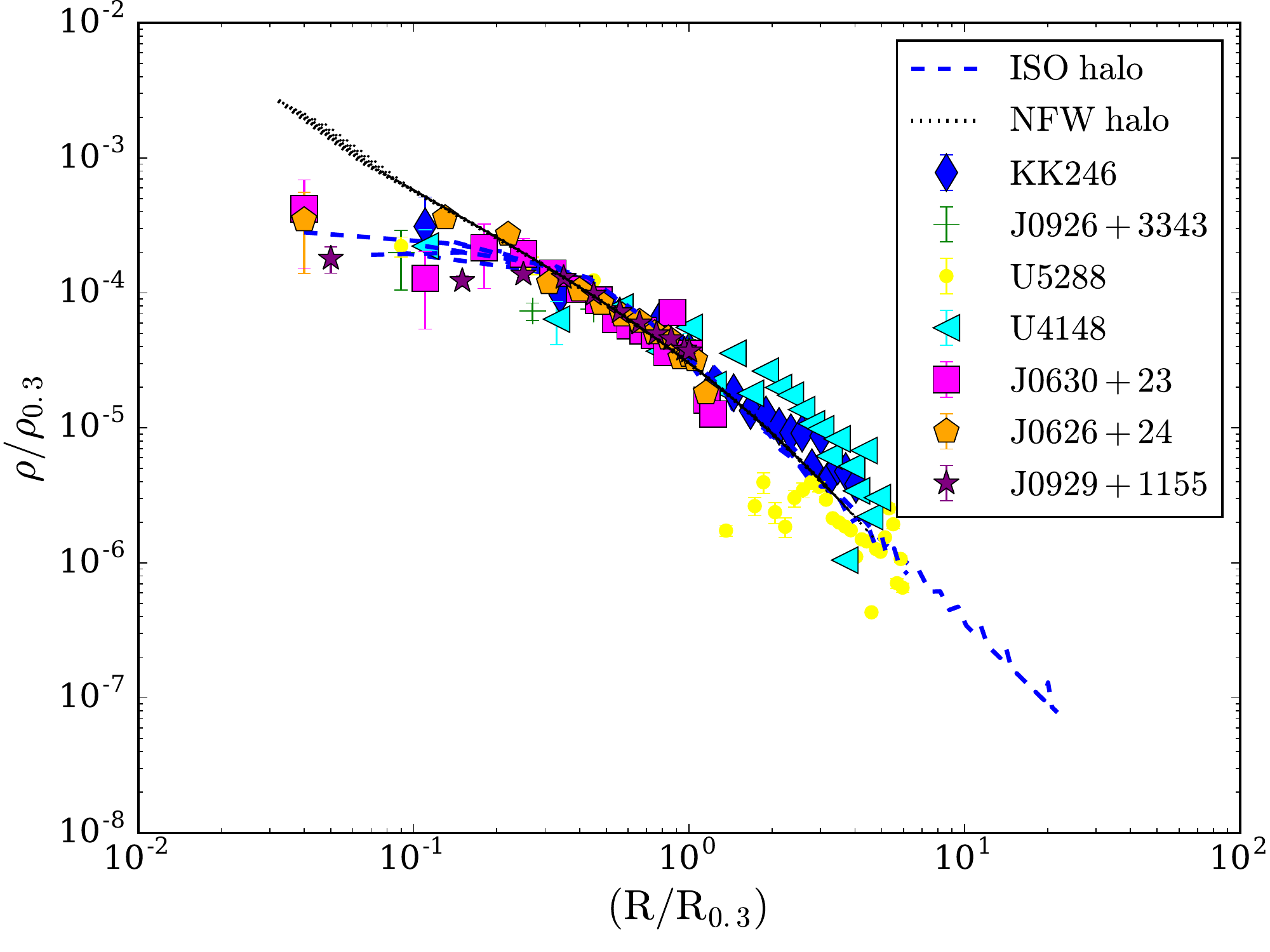}
\caption{upper panel: The density profiles derived using \fat (3-D approach) rotation curves, lower panel: The density profiles derived using ROTCUR (2-D approach) rotation curves. The black dotted line represents NFW halo  with ($\alpha$ $\sim$ -1) with V$_{200}$ 10--110 km s$^{-1}$ and the blue dashed line is the best-fit isothermal halo model.}
\label{fig:density}
\end{figure}%

\subsection{Scaled density profiles} 

So far we have been dealing with each individual galaxy separately, and combining the results from parametric fits to the individual rotation curve. A complementary approach would be to try and suitably scale each rotation curve, so that they can be combined together, and then compare this composite curve with theoretical models. To do this, we follow \citet[][]{hayashi06, oh11, oh15} and scale the pressure corrected rotation velocities to V$_{0.3}$ and the radius to R$_{0.3}$, where R$_{0.3}$ is the radius at which the logarithmic slope of the rotation curve ($d$ log$V$/$d$ log$R$) becomes 0.3 and V$_{0.3}$ is the velocity at this radius.  This scaling radius can generally be well determined for all the galaxies as it lies between the raising ($d$ logV/$d$ logR = 1) and the flat ($d$ logV/$d$ logR = 0) rotation curve. For our sample, we can determine R$_{0.3}$ for all the galaxies except for J0926+3343; in the case of J0926+3343 we take the outermost radius is taken as the scaling radius. In order to compare this composite rotation curve with  the NFW model, we combine the theoretical NFW curves with V$_{200}$ ranging from 10--110 km s$^{-1}$, concentration  parameter obtained using the empirical relation between  c--V$_{200}$ \citep{deBlok03, mcGaugh07} by also scaling them with respect to R$_{0.3}$ and V$_{0.3}$ Fig.~\ref{fig:density} shows the plot of the scaled density versus the scaled radius. The black dotted line represents NFW halo  with ($\alpha$ $\sim$ -1) with V$_{200}$ 10--110 km s$^{-1}$ and the blue dashed line is the best-fit isothermal halo model. Once again we see a clear difference between the rotation curve derived from the 2-D and 3-D rotation curve. The upper panel shows the density profile obtained using the \fat-derived rotation curve, which is consistent with the cuspy NFW profile. The lower panel shows the density profile derived using the ROTCUR-derived rotation curve. This is not consistent with the NFW profile but is in good agreement with best-fit isothermal model in contrast to the \fat derived rotation curve.

\subsection{Further comparison of the 2D and 3D approaches}
%


The difference between the rotation curves derived in the 2D and 3D method could, in principle arise from \fat incorrectly modelling the data cube. To test this, we derive moment maps from the best fit model data cube produced by \fat and derive rotation curves as well as surface brightness profiles from them using the 2D routines in GIPSY. We show rotation curves as well as surface brightness profiles for all the galaxies in Fig. \ref{fig:fatmodel1}, \ref{fig:fatmodel2}, and \ref{fig:fatmodel3}. For each of these quantities 3 sets of curves are shown, these are the curves derived by \fat (red squares) and GIPSY (black circles) when run on the original observed data, as well as the curves produced by the GIPSY tasks when run on the model produced by \fat (green diamonds). We find that in most cases, the rotation curves produced by GIPSY routines run on the original data very closely match those produced when the same routines are run on the \fat model, especially in the inner regions. This indicates that the \fat model is a good fit to the observed data, and that the differences that we see in the curves arise because of the systematic differences between the 2D and 3D approaches. There are significant differences in both the rotation curves as well as in the surface brightness profiles. The differences in the surface brightness profiles would lead to differences in the estimates of the contribution of the mass of the gas disk to the total rotation velocity as well as to differences in the estimated pressure support correction. We confirm that the differences to the pressure support correction are small in the inner regions (i.e. the regions that we use to determine the inner slope of the density profile). We remind the reader that we compare our results for the minimum disk approximation with results from the literature that also use the minimum disk approximation. As such this comparison is unaffected by the assumed surface density profile.

\subsection{Comparison with Gauss-Hermite velocity fields}
\label{ssec:gauss-hermite}
\begin{figure}
\centering
\includegraphics[width=1.0\linewidth]{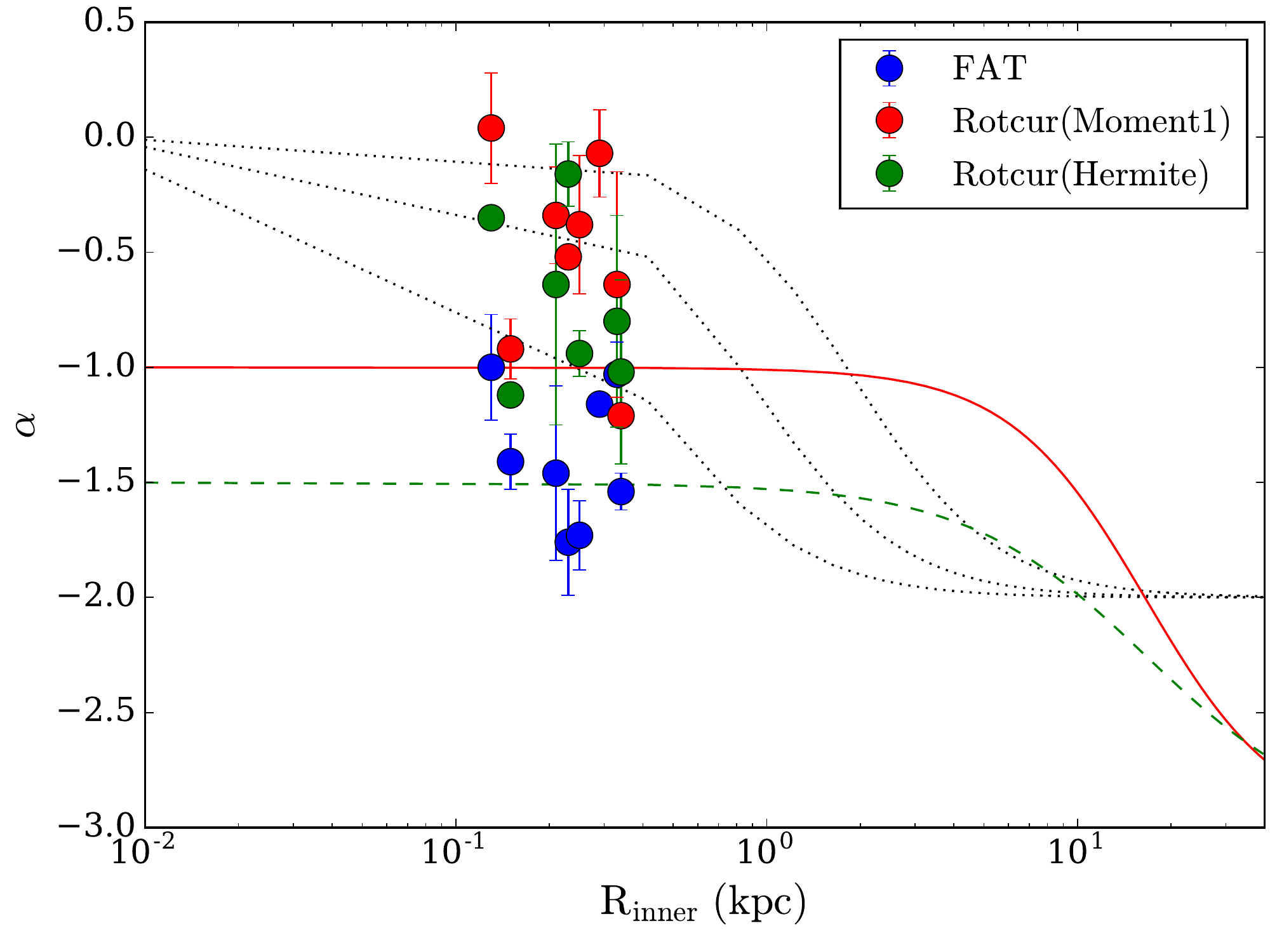}
\caption{The value of inner slope ($\alpha$) of the dark matter density distribution versus the radius of the innermost point. The slopes derived with the FAT rotation curves are shown by blue circles. The green and red circles indicate the slopes derived from the the Gauss-Hermite fit (and ROTCUR) and moment method (and `Rotcur') respectively.}
\label{fig:slope-compare}
\end{figure}

We also derive the velocity fields using Gauss-Hermite fits to the individual spectra using the task `XGAUFIT' in \gipsy\ . The Gauss-Hermite polynomial includes an $h_{3}$ (skewness) term to the velocity profiles, and hence takes into account asymmetries in the profiles. We could not derive a reliable Hermite velocity field for the galaxy J0929+1155, the galaxy with the lowest signal to noise. The difference in slopes from the Hermite velocity field and the first moment map is significant for the galaxy UGC5288, which is a barred galaxy. As before we used the velocity fields to calculate the rotation curves using the task ROTCUR, these were then used to calculate the logarithmic slope of the density profile in exactly the same manner as detailed above. These slopes are listed in table \ref{table:alpha}, and the logarithmic slopes derived using the three different estimates of rotation curve (i.e. from the moment method and ROTCUR, the Gauss-Hermite fit and ROTCUR, and FAT) are shown in Fig.~\ref{fig:slope-compare}. As can be seen slopes derived with Hermite velocity fields are somewhat steeper than the slopes derived with the first moment maps, but not as steep as the FAT derived curves. The average slope obtained using the Gauss Hermite fits is -0.71 $\pm$ 0.33. If we exclude the galaxy J0929+1155, the average slope obtained using moment maps and FAT is -0.56 $\pm$ 0.25 and -1.41 $\pm$ 0.21 respectively.

\subsection{Comparison with simulated galaxies}
\begin{figure}
\centering
\includegraphics[width=1.0\linewidth]{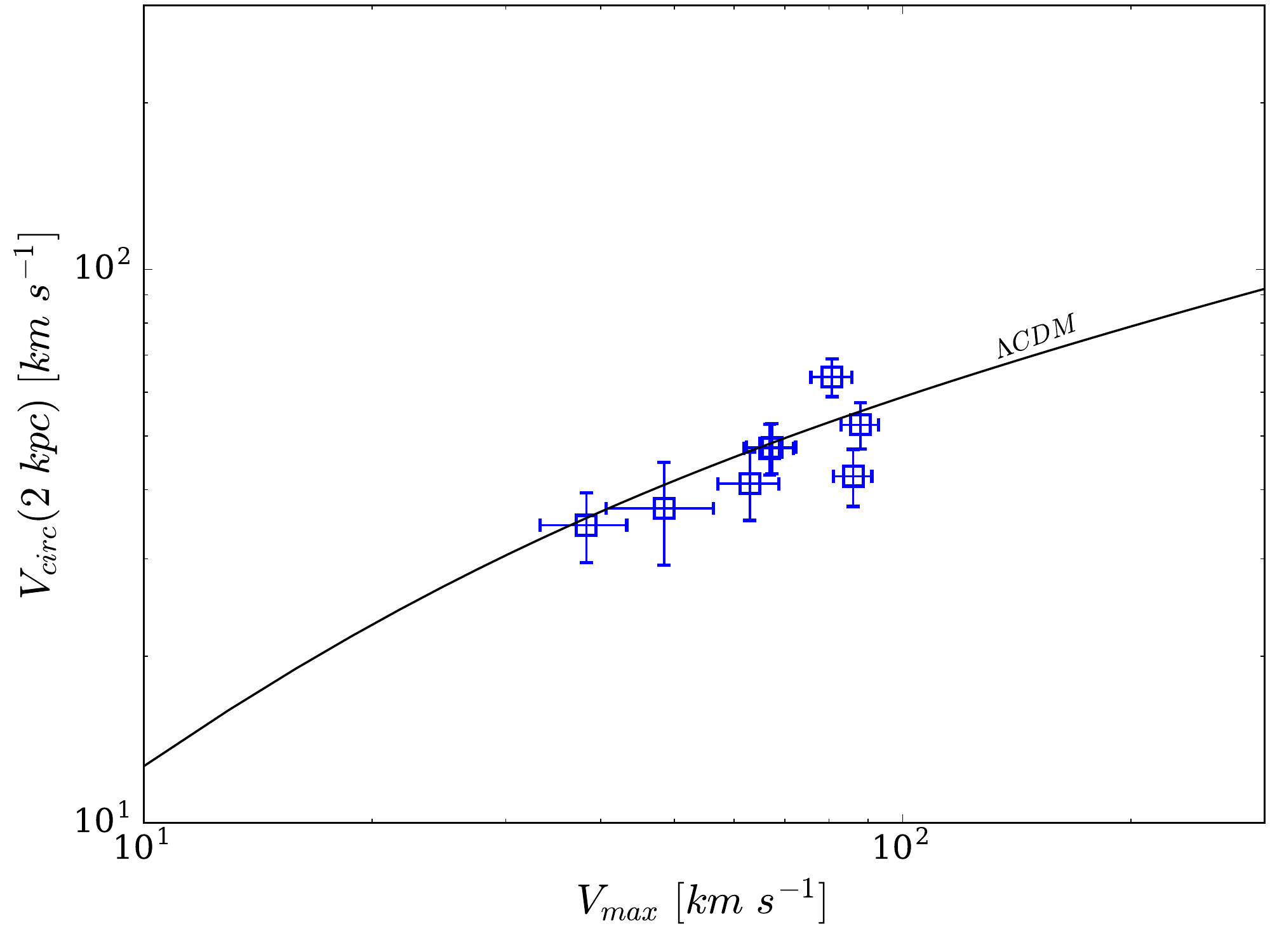}
\caption{Circular velocity at r = 2 kpc versus the maximum circular velocity, V$_{max}$, for 8 void dwarf galaxies. The black solid line represents the correlation expected for the NFW halo.}
\label{fig:vcirc}
\end{figure}

The role of systematic effects in observational studies to the cusp-core problem was investigated by \citet{pineda17} using hydrodynamical simulations of dwarf galaxies. They mimic realistic kinematic observations and fit mock rotation curves. Their model galaxies also suggest that the minimum disc approximation would cause one to infer that the DM profile is flatter than it actually is, which is in contrast with the widely accepted claim. Our results are in agreement with this, we find a flatter inner density slope with the minimum disc assumption ($\alpha$ = -1.39 $\pm$0.19) than for the dark matter only profile where we find a relatively steeper slope ($\alpha$ = -1.48 $\pm$0.23), although it matches within the error bars.
They also find that it is extremely challenging to fully correct for the pressure support even with the data available from the highest quality cusp-core studies and that even small errors of a few km s$^{-1}$ can cause dark matter cusps to be disguised as cores. In our particular case, we note that the pressure correction terms are small in the central part of the galaxy, and different assumptions do not make a significant difference to the final rotation curve (see \S \ref{pressure}).

\citet{oman15} find that the `cusp-core problem' is better characterized as an `inner mass deficit' problem, where they compare the inner circular velocities of observed galaxies with those of $\Lambda$CDM galaxies of same maximum velocity (V$_{\rm max}$). Following their prescription in Fig \ref{fig:vcirc}, we plot the circular velocity at 2 kpc  against the maximum measured rotation speed using the rotation curves derived using \fat. We interpolate linearly between nearby data points to get the velocity at exactly 2 kpc. The black solid line represents the correlation expected for the NFW halo. Dwarf galaxies in our sample are in good agreement with the NFW haloes.
  
\citet{oman17} measure H{\sc i} rotation curves using $^{\rm 3D}${\small BAROLO} tilted ring modelling tool for the galaxies simulated from the APOSTLE $\Lambda$CDM hydrodynamical simulations. Their results suggest that non-circular motions in the gas lead to underestimate the circular velocities in the central regions, which can be misinterpreted as evidence for cores in the dark matter. They suggest that the failure of tilted ring models when applied to galaxies with non-negligible non-circular motions could be a possible resolution to the cusp-core problem. In our case, we find no clear evidence for non-circular motions, except in the case of UGC~5288, where there appears to be a strong bar and we indeed find that NFW halo is not a good fit for this galaxy.

\subsection{Environmental dependence of DM halo properties}

\begin{figure}
\centering
\includegraphics[width=1.0\linewidth]{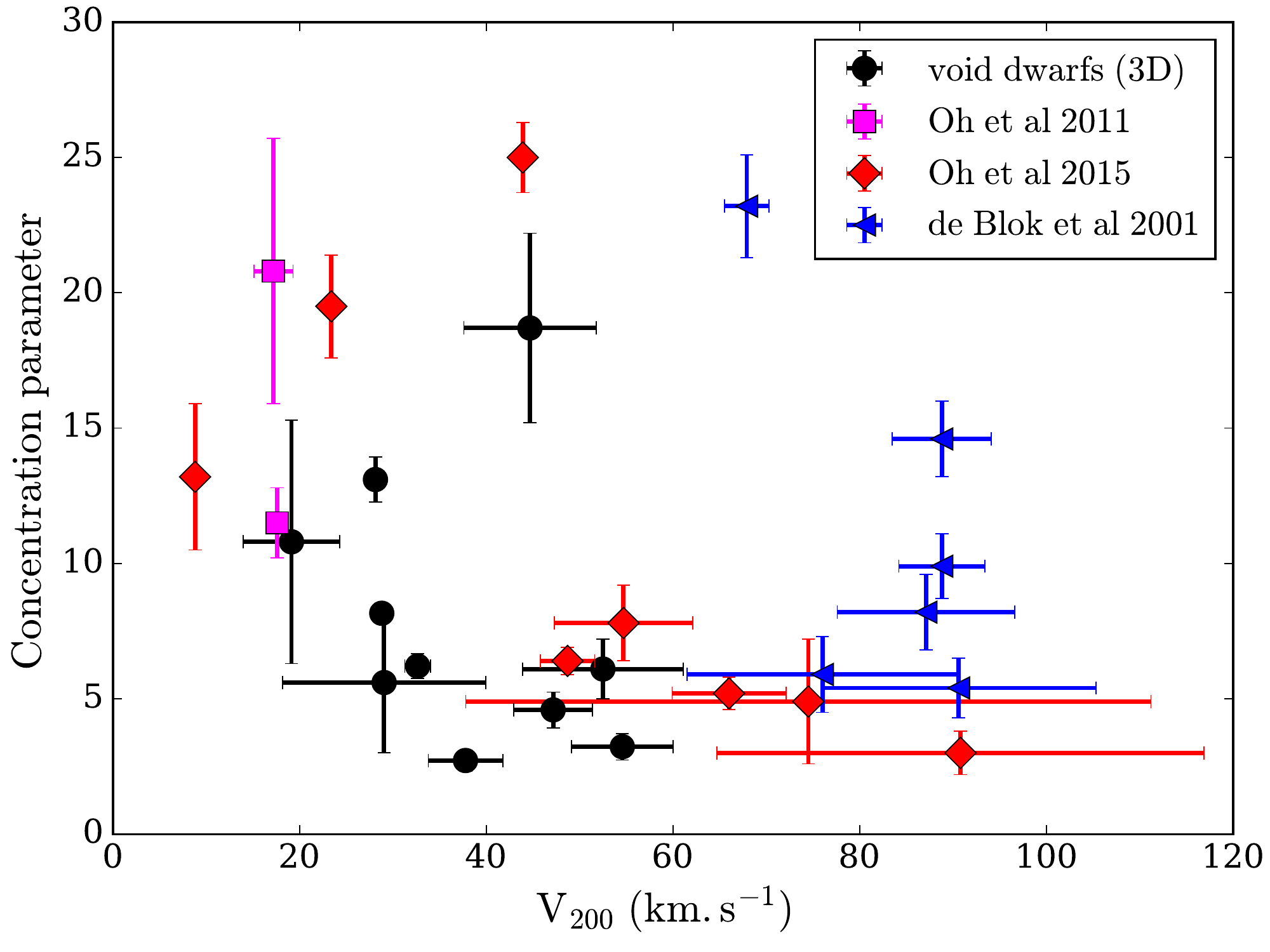}
\includegraphics[width=1.0\linewidth]{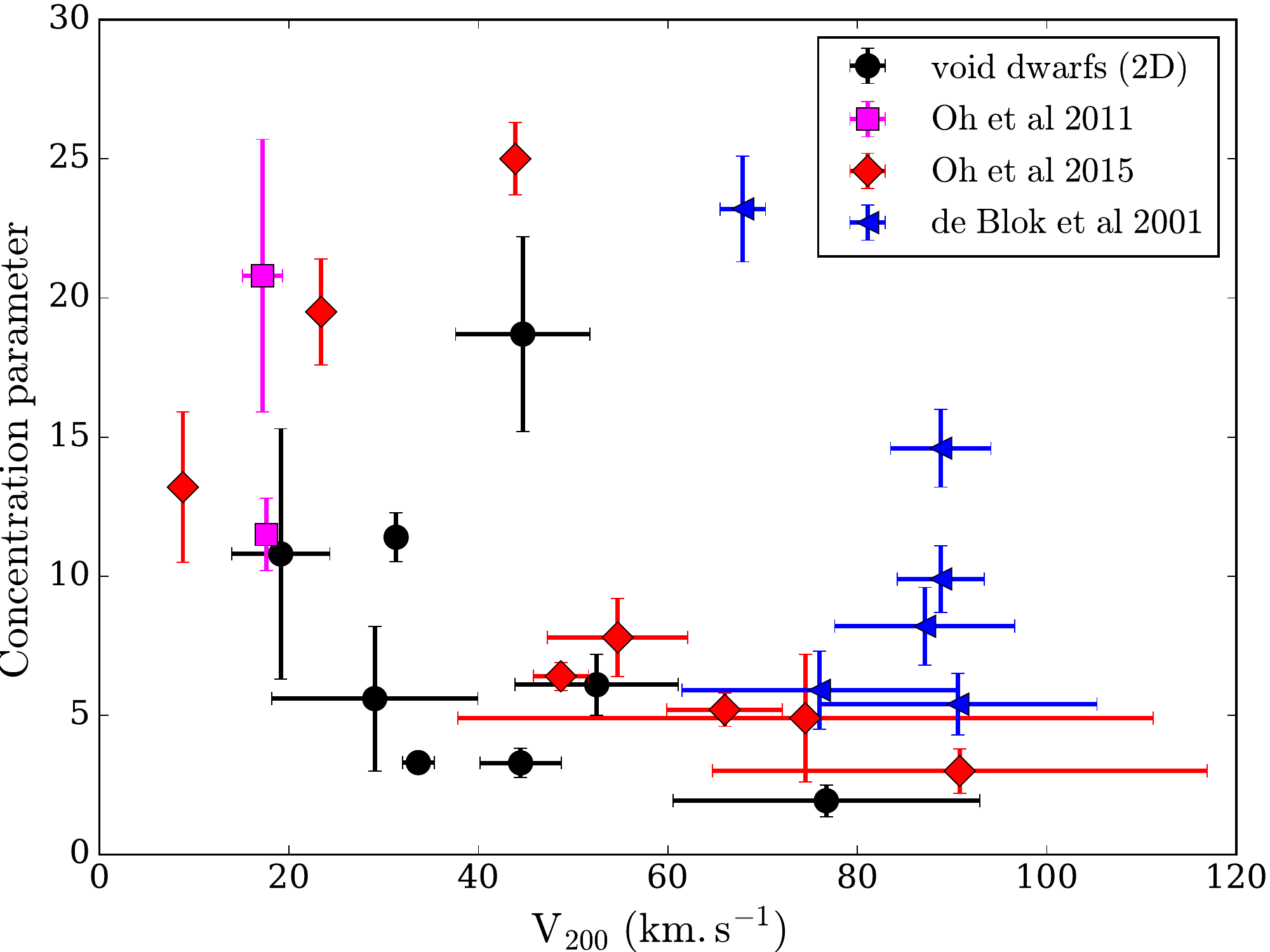}
\caption{The concentration parameters versus the V$_{200}$ in km s$^{-1}$.  upper panel: The parameters for the galaxies in our sample are using the rotation curves derived from \fat (3-D approach), lower panel: The parameters for the galaxies in our sample are using the rotation curves derived from ROTCUR (2-D approach). }
\label{fig:cv200}
\end{figure}%
One of the aims of this study was to check if the void galaxies have different DM halo properties as compared to the galaxies outside voids. Figure \ref{fig:cv200} shows the concentration parameters versus the V$_{200}$ (km s$^{-1}$). The black circles in the upper panel and lower panel represent the NFW halo parameters derived for the void dwarf galaxies using \fat and ROTCUR respectively. We were able to obtain physical values for the concentration parameters only for 6 rotation curves derived with \fat and 4 rotation curves derived with ROTCUR. The magenta squares \citep{oh11a}, red diamonds \citep{oh15}, blue triangles \citep{deBlok01} represent the NFW halo parameters of galaxies outside voids. 

As can be seen from the plot, the data for the void dwarfs from our sample and other dwarfs from the literature overlap. We perform a 2D two-sample Kolmogorov-Smirnov (KS) test to compare the distribution of dwarf galaxies in the voids and dwarf galaxies in the average density regions quantitatively. We do not include \citet{deBlok01} sample for the comparison as their galaxies lie outside the rotational velocity range of our galaxies. Since four galaxies (DDO43, DDO46, DDO52, and F564-V3) from the comparison sample happen to reside in voids \citep{pustilnik18}, we include them in the void galaxy sample. This test gives a probability of 0.11 for the void galaxies and the galaxies from average density regions being drawn from the same distribution. This p-value indicates that there is no clear statistical evidence for the two samples (i.e. dwarfs from voids and average density regions) being drawn from different populations. However, we require a larger sample to draw stronger conclusions. It would also be interesting to compare against a sample of dwarfs specifically chosen to lie in regions of higher than average density.

\section{Summary and Conclusions}
\label{summary}

We have derived rotation curves of 8 galaxies that lie in Lynx-Cancer void using 3-D and 2-D tilted ring fitting routines. We construct mass models and we find that both the isothermal and NFW halos are a good description of dark matter distribution of galaxies in terms of fit-quality (i.e. $\chi^{2}_{red}$). We convert the rotation curves derived using 2-D and 3-D approaches into density profiles. This allows us to examine the central dark matter density profiles and to distinguish between the cores and cusps at the center of galaxies.  We find that the dark matter halo density profiles derived using 3-D approach are consistent with the NFW profile and the measured inner slopes ($\alpha$ = -1.39 $\pm$0.19 ) are more steep than the values of the slopes from the literature ($\alpha$ = -0.2 $\pm$ 0.2  in \citet{deBlok01}, $\alpha$ = -0.29 $\pm$ 0.07  in \citet{oh11} and $\alpha$ = -0.42 $\pm$ 0.21  in \citet{oh15}). Since the mass models for the galaxies in the literature were constructed using the 2-D approaches, we use the rotation curves derived using 2D approach to estimate the inner density slope ($\alpha$) values. The value of $\alpha$ $\sim$ -0.5-0.7 we get using  the 2D approach is consistent with the slopes obtained in literature. This suggests that the fundamental differences in 3-D and 2-D tilted ring fitting routines affect the slope of the central dark matter density profiles. Since our sample size is modest, it is important to check the results using larger samples.

\section*{Acknowledgements}

 This paper is based in part on observations taken with the GMRT. We thank the staff of the
GMRT who made these observations possible. The GMRT is run by the National Centre for Radio Astrophysics of the Tata Institute of Fundamental Research. PK is partially supported by BMBF project 05A17PC2 for D-MeerKAT. The work of SAP on this project was supported by RSCF grant No. 14-12-00965.

\appendix

\section{Data}

\begin{figure*}
\centering
\includegraphics[width = 4.75in]{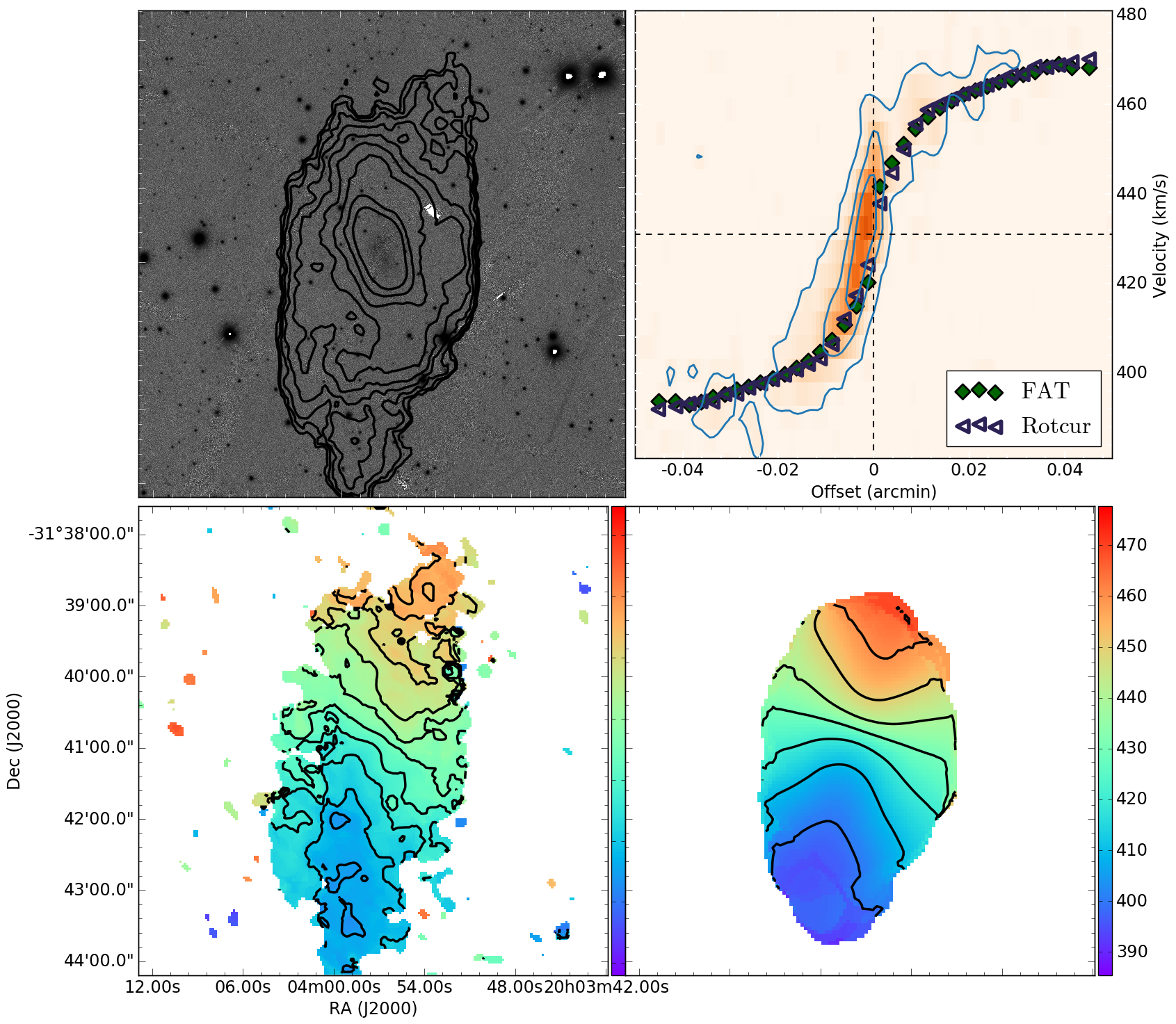}
\caption{H{\sc i} data and kinematics for the galaxy KK246. Top left: H{\sc i} contours overlaid on the SDSS $g$--band optical image, Top right: Position velocity diagram with the rotation curves overlaid on them.  Bottom left: velocity field of data, and Bottom right: velocity field of the best fitting \fat model. Velocity contours run from 380 to 470 km s$^{-1}$ with a spacing of 10 km s$^{-1}$   }
\label{fig:map1}
\end{figure*}

\begin{figure*}
\centering
\includegraphics[width = 4.75in]{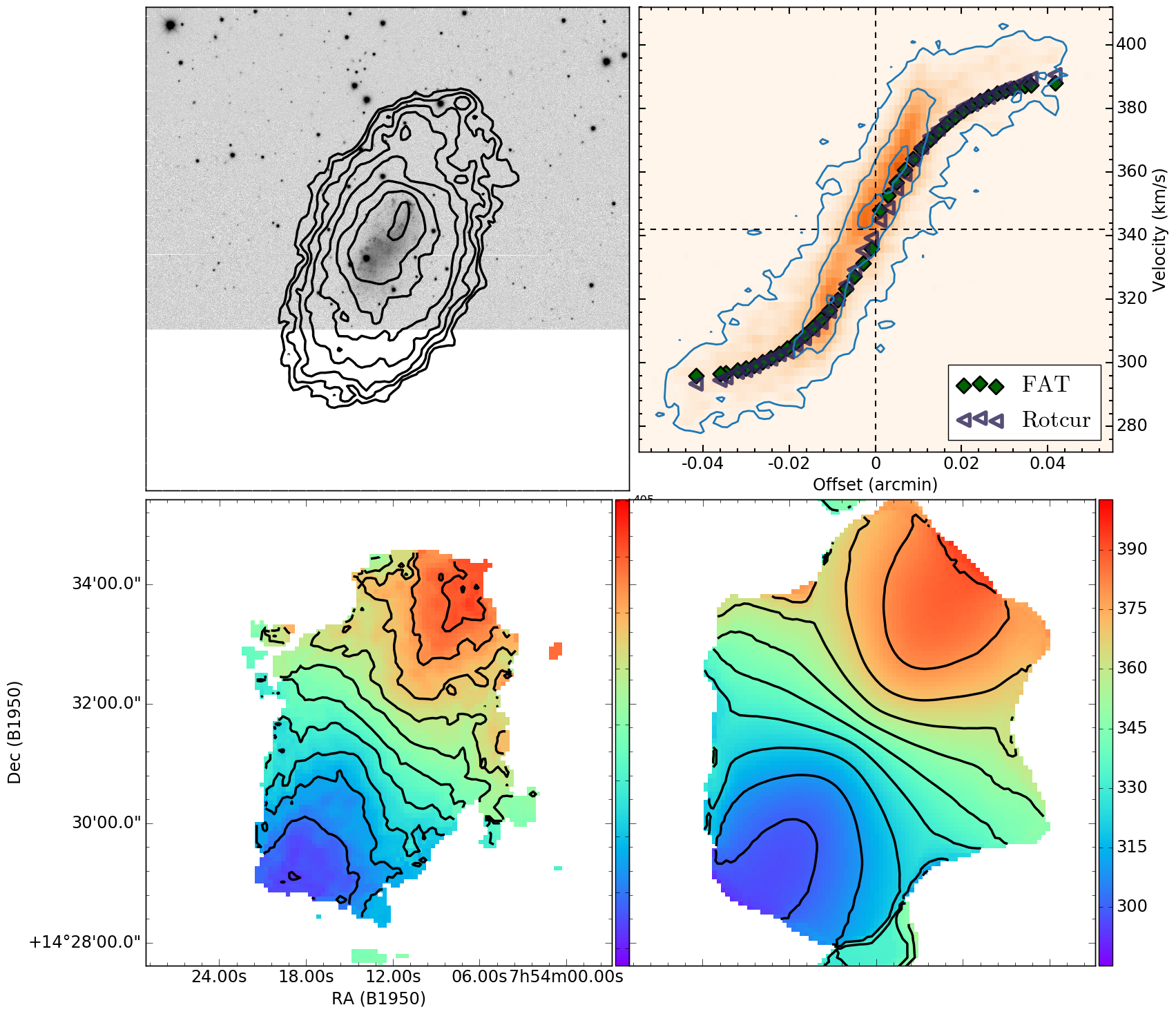}
\caption{H{\sc i} data and kinematics for the galaxy U4115. Top left: H{\sc i} distribution of the galaxy overlaid on the SDSS $g$--band data, Top right: Position velocity diagram with the rotation curves overlaid on them. Bottom left: velocity field of data, and Bottom right: velocity field of the best fitting \fat model. velocity contours run from 290 to 390 km s$^{-1}$ with a spacing of 10 km s$^{-1}$ }
\label{fig:map2}
\end{figure*}

\begin{figure*}
\centering
\includegraphics[width = 4.5in]{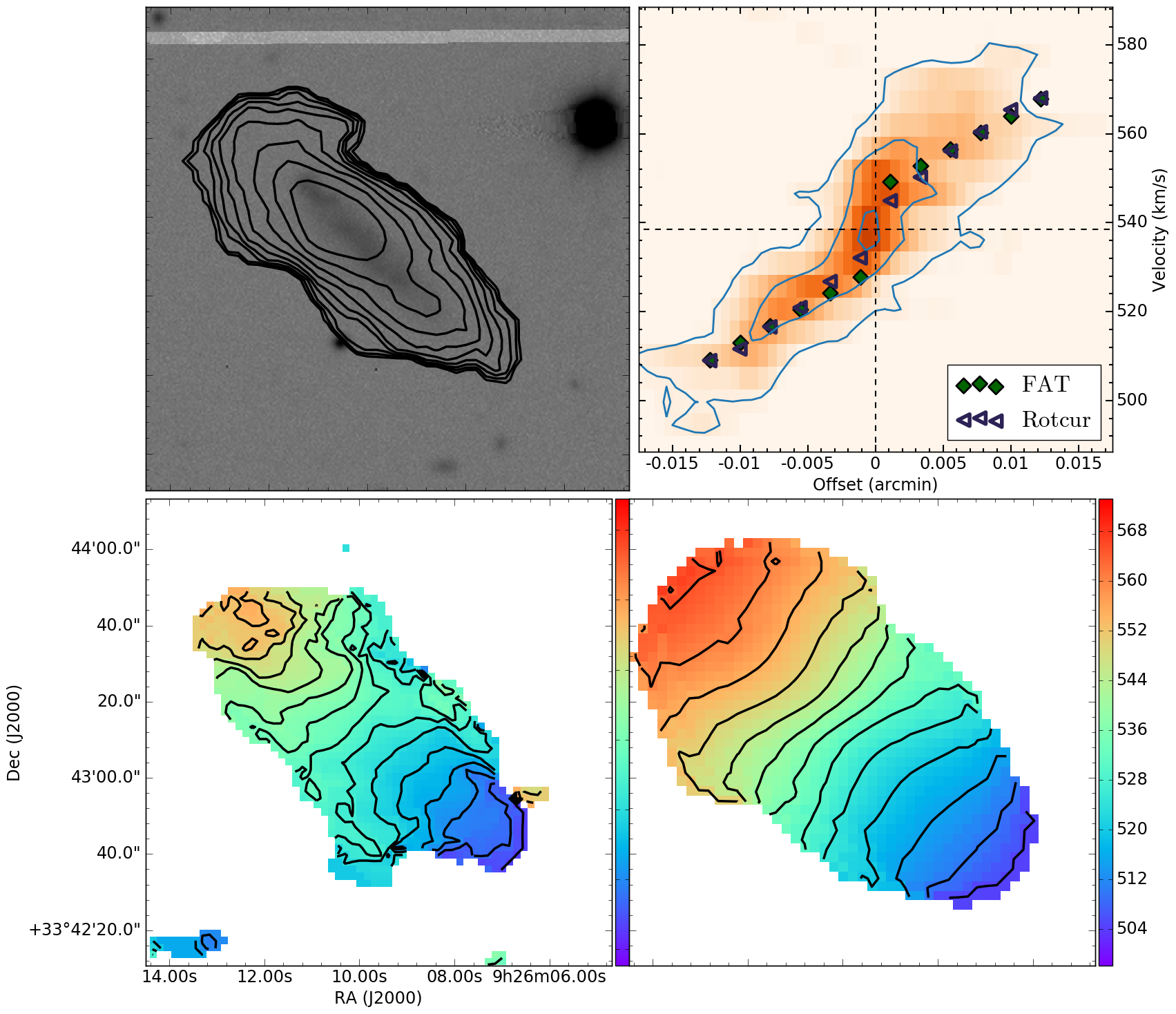}
\caption{H{\sc i} data and kinematics for the galaxy J0926+3343. Top left: H{\sc i} distribution of the galaxy overlaid on the SDSS $g$--band data, Top right: Position velocity diagram  with the rotation curves overlaid on them.  Bottom left: velocity field of data, and Bottom right: velocity field of the best fitting \fat model. velocity contours run from 505 to 565 km s$^{-1}$ with a spacing of 5 km s$^{-1}$  }
\label{fig:map3}
\end{figure*}

\begin{figure*}
\centering
\includegraphics[width = 4.5in]{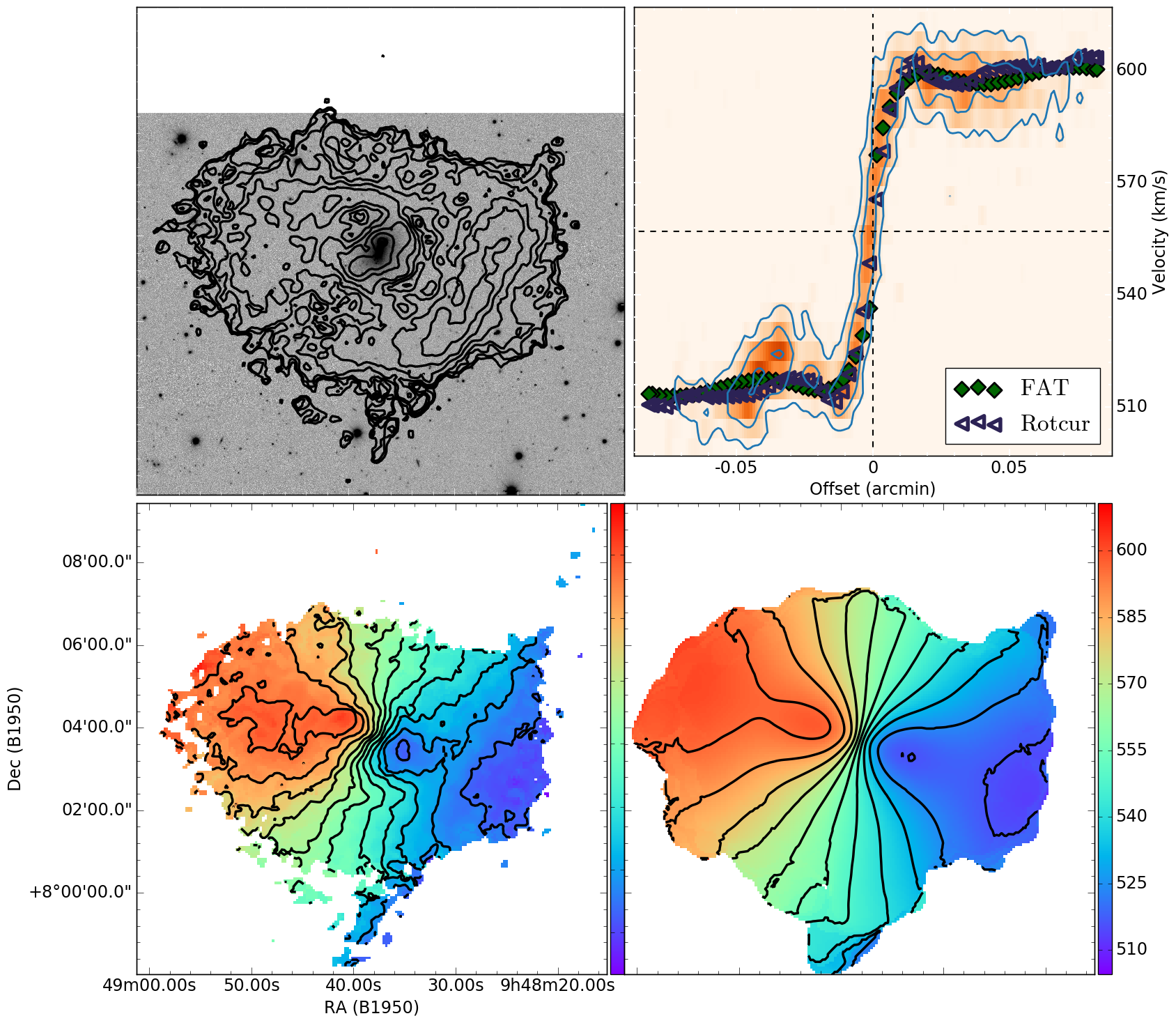}
\caption{H{\sc i} data and kinematics for the galaxy U5288. Top left panel: H{\sc i} distribution of the galaxy overlaid on the SDSS $g$--band data, Top right panel: Position velocity diagram with the rotation curves overlaid on them.  Bottom left: velocity field of data, and Bottom right: velocity field of the best fitting \fat model. velocity contours run from 516 to 596 km s$^{-1}$ with a spacing of 8 km s$^{-1}$  }
\label{fig:map4}
\end{figure*}

\begin{figure*}
\centering
\includegraphics[width = 4.5in]{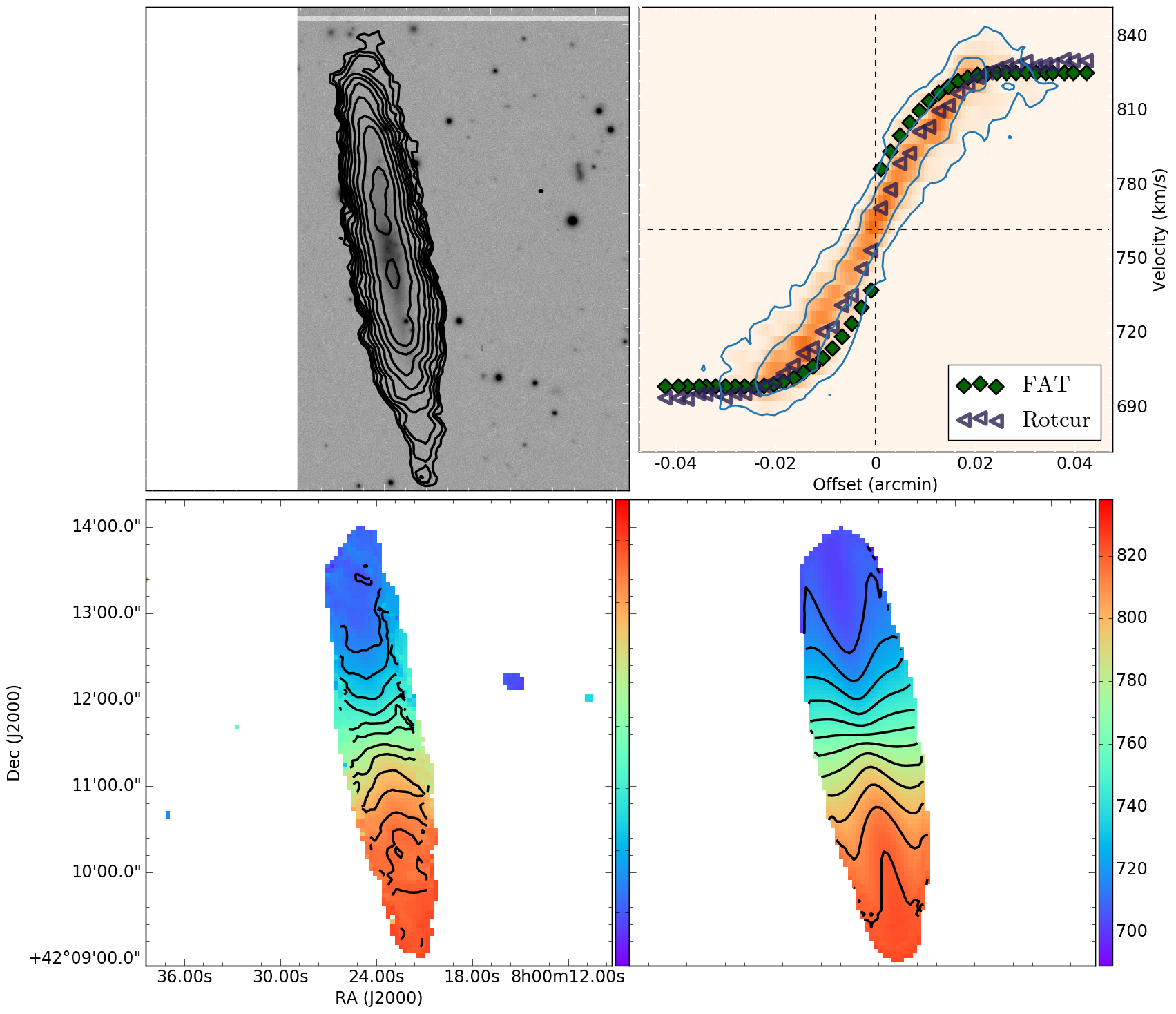}
\caption{H{\sc i} data and kinematics for the galaxy U4148. Top left: H{\sc i} distribution of the galaxy overlaid on the SDSS $g$--band data, Top right: Position velocity diagram with the rotation curves overlaid on them.  Bottom left: velocity field of data, and Bottom right: velocity field of the best fitting \fat model. velocity contours run from 700 to 820 km s$^{-1}$ with a spacing of 10 km s$^{-1}$ }
\label{fig:map5}
\end{figure*}

\begin{figure*}
\centering
\includegraphics[width = 4.5in]{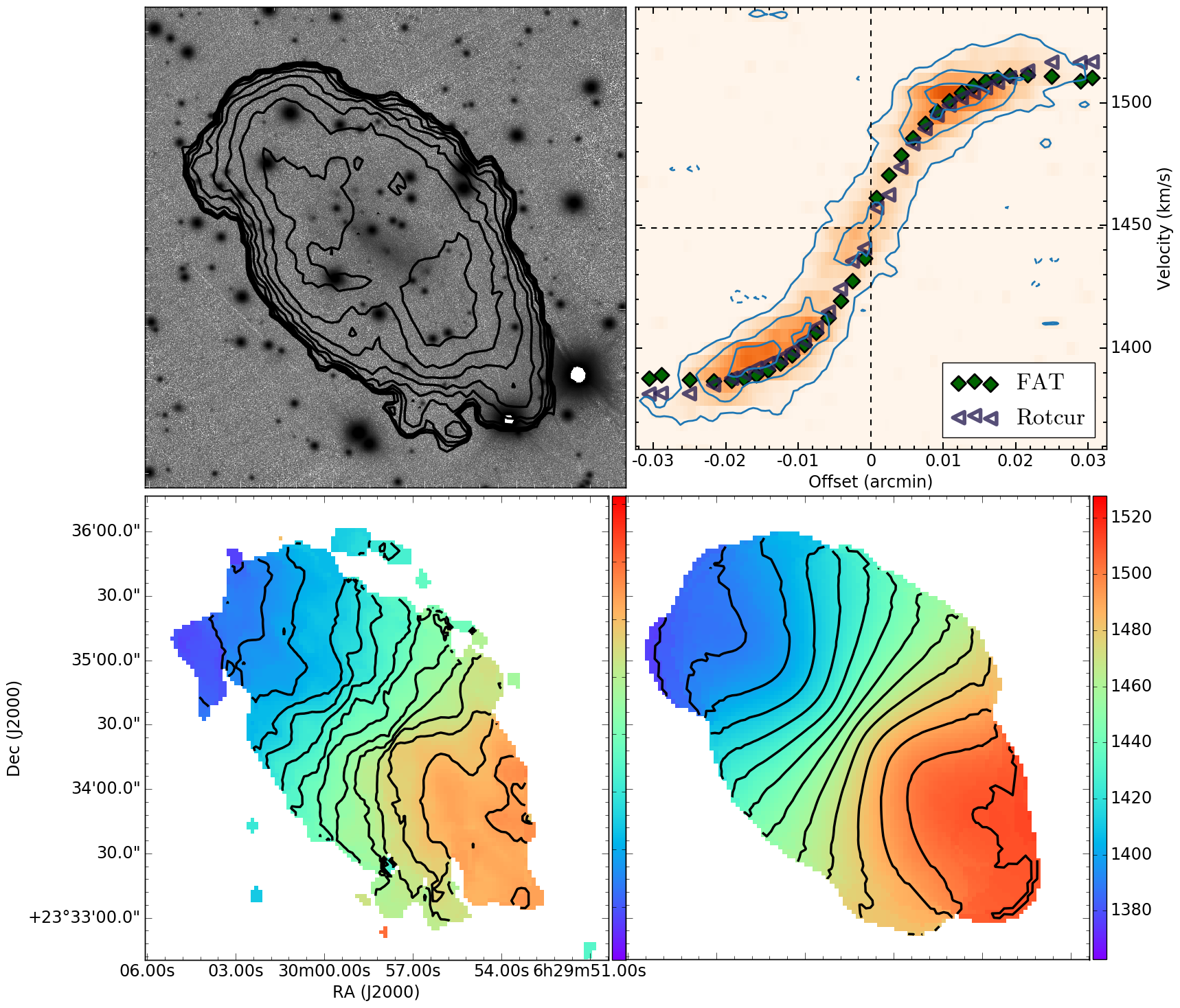}
\caption{H{\sc i} data and kinematics for the galaxy J0630+23. Top left: H{\sc i} distribution of the galaxy overlaid on the SDSS $g$--band data, Top right: Position velocity diagram with the rotation curves overlaid on them.  Bottom left: velocity field of data, and Bottom right: velocity field of the best fitting \fat model. velocity contours run from 1380 to 1510 km s$^{-1}$ with a spacing of 10 km s$^{-1}$ }
\label{fig:map6}
\end{figure*}

\begin{figure*}
\centering
\includegraphics[width = 4.5in]{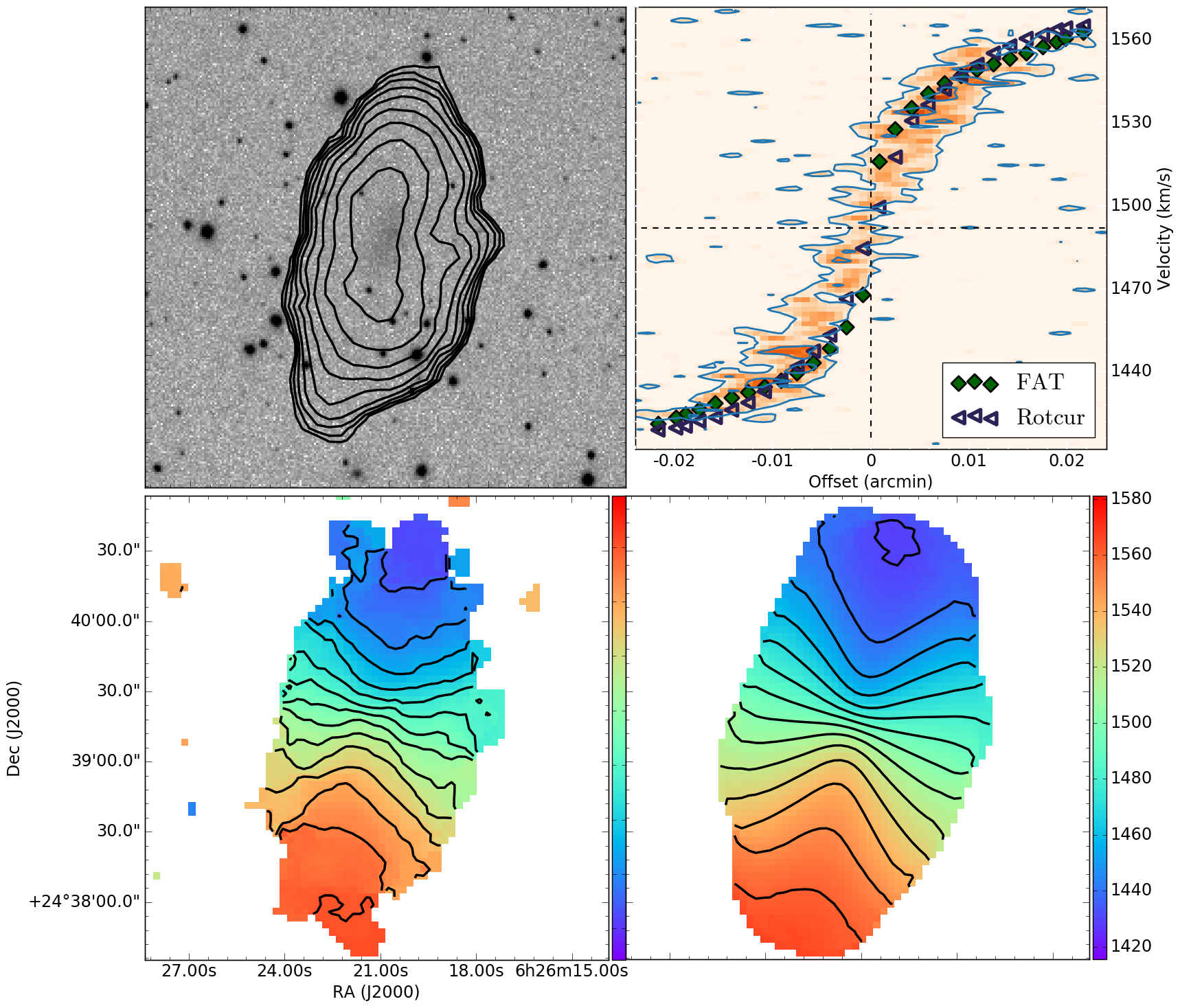}
\caption{H{\sc i} data and kinematics for the galaxy J0626+24. Top left: H{\sc i} distribution of the galaxy overlaid on the SDSS $g$--band data, Top right: Position velocity diagram with the rotation curves overlaid on them. Bottom left: velocity field of data, and Bottom right: velocity field of the best fitting \fat model. velocity contours run from 1430 to 1560 km s$^{-1}$ with a spacing of 10 km s$^{-1}$ }
\label{fig:map7}
\end{figure*}

\begin{figure*}
\centering
\includegraphics[width = 4.5in]{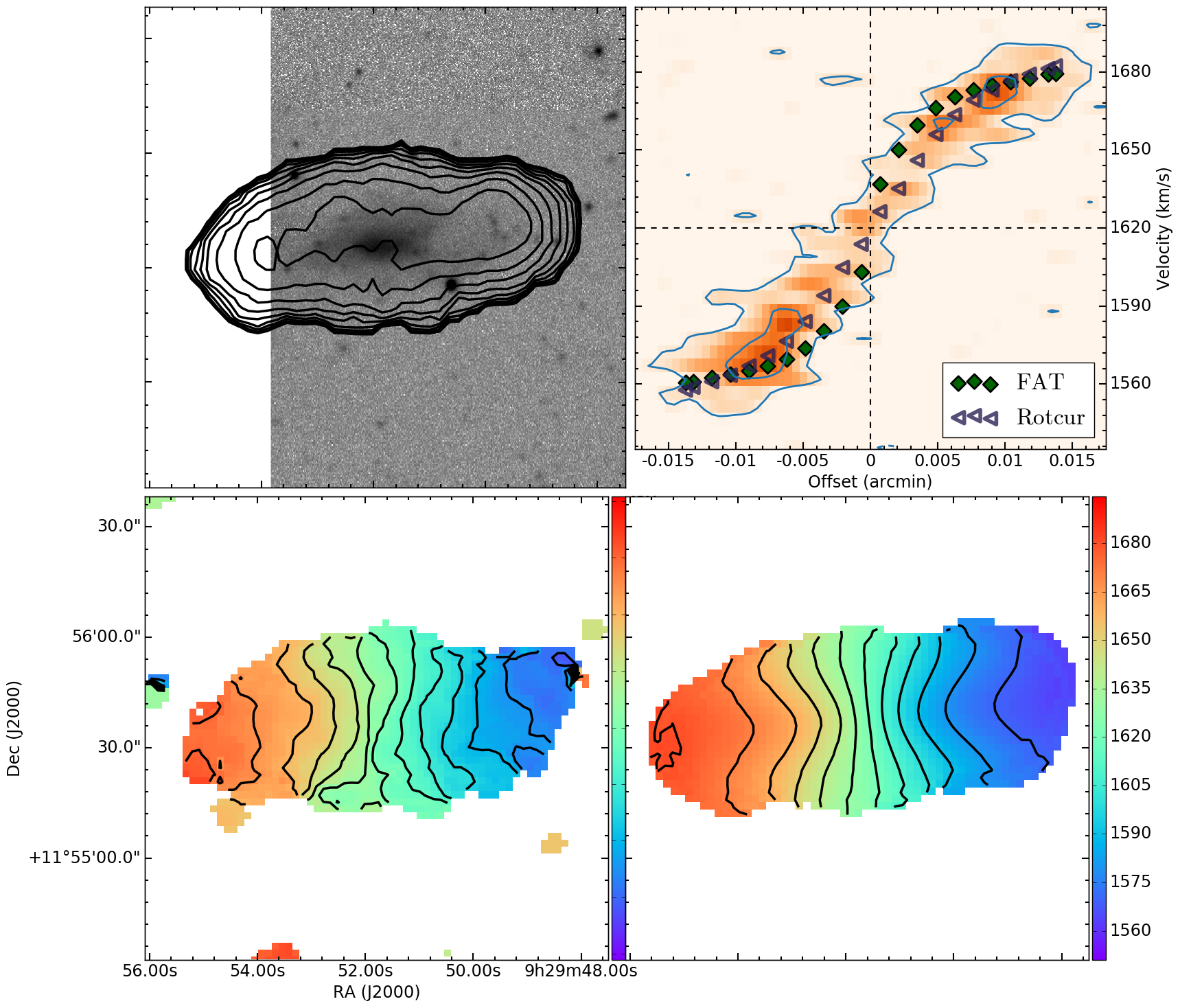}
\caption{H{\sc i} data and kinematics for the galaxy J0929+1155. Top left: H{\sc i} distribution of the galaxy overlaid on the SDSS $g$--band data, Top right: Position velocity diagram with the rotation curves overlaid on them. Bottom left: velocity field of data, and Bottom right: velocity field of the best fitting \fat model. velocity contours run from 1560 to 1680 km s$^{-1}$ with a spacing of 10 km s$^{-1}$  }
\label{fig:map8}
\end{figure*}

In this appendix, we present the data and the kinematic analysis of 8 void dwarf galaxies. In Figures \ref{fig:map1}--\ref{fig:map8}, we show (i) the integrated H{\sc i} intensity map contours overlaid on the optical image, (ii) Position-velocity diagram taken along the major axis of the galaxy with the rotation curves  overlaid on them. The dashed lines indicate the systemic velocity and kinematic center. The overplotted green diamonds represent the rotation curve derived by \fat (3D approach) and violet triangles represent the rotation curve derived by `ROTCUR' (2D approach)  (iii) the intensity weighted first moment of the galaxy, and (iv) velocity field of the best fitting \fat model. 

\section{Correction for pressure support}

\begin{figure} 
\centering
\includegraphics[width=1.0\linewidth]{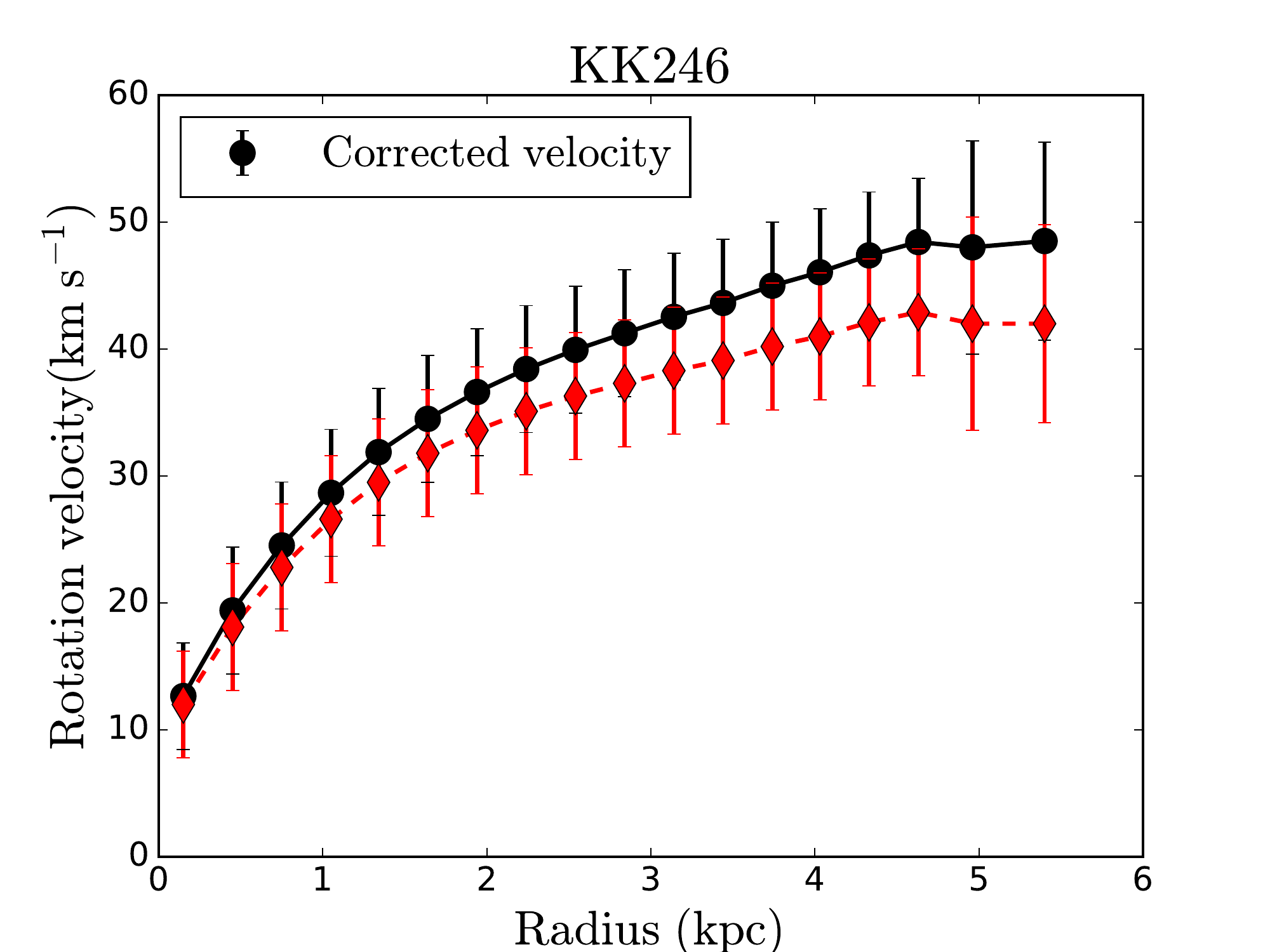}
\includegraphics[width=1.0\linewidth]{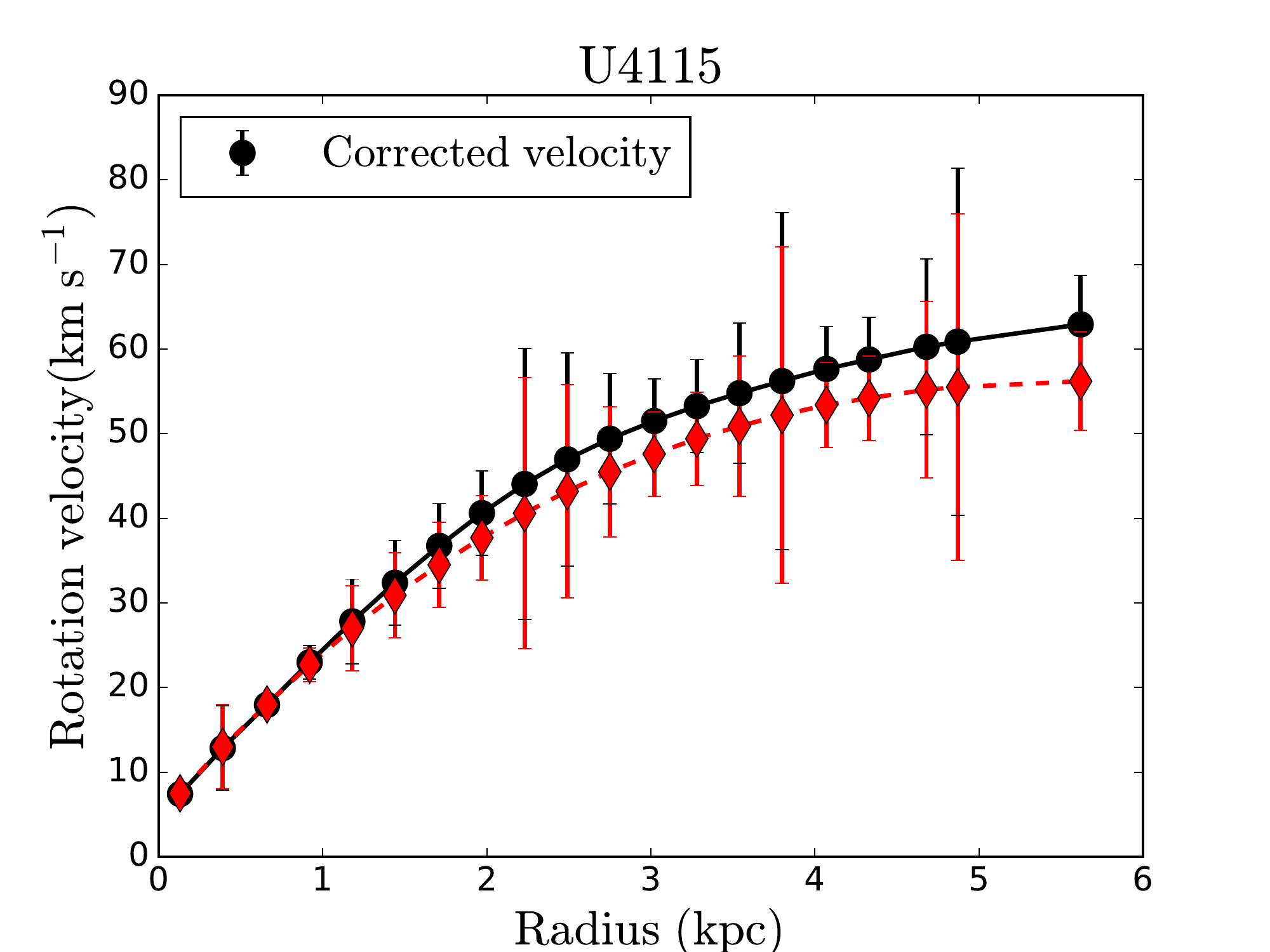}
\includegraphics[width=1.0\linewidth]{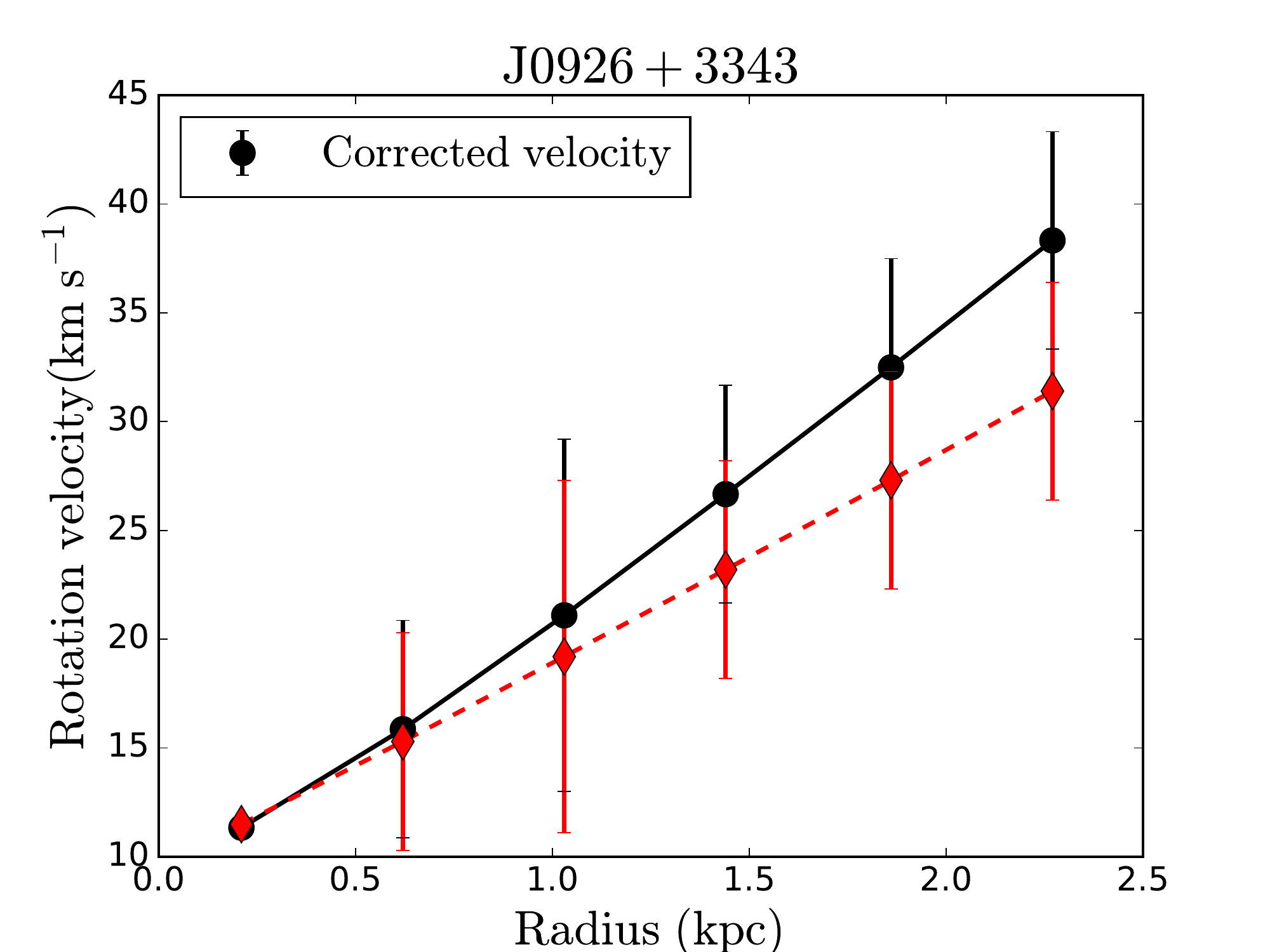}
\includegraphics[width=1.0\linewidth]{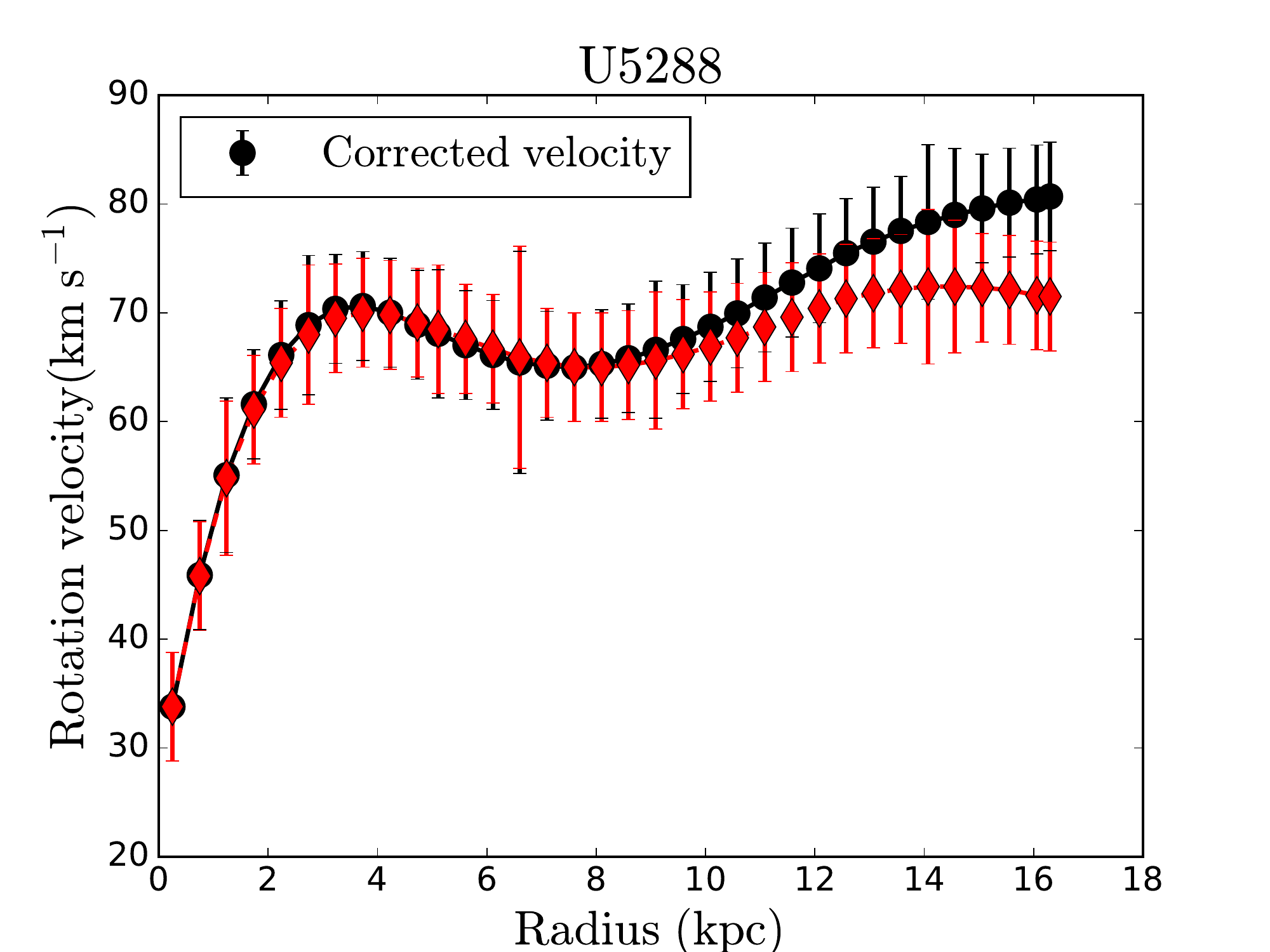}
\caption{}
\label{fig:rotfat1}
\end{figure}

\begin{figure}
\centering
\includegraphics[width=1.0\linewidth]{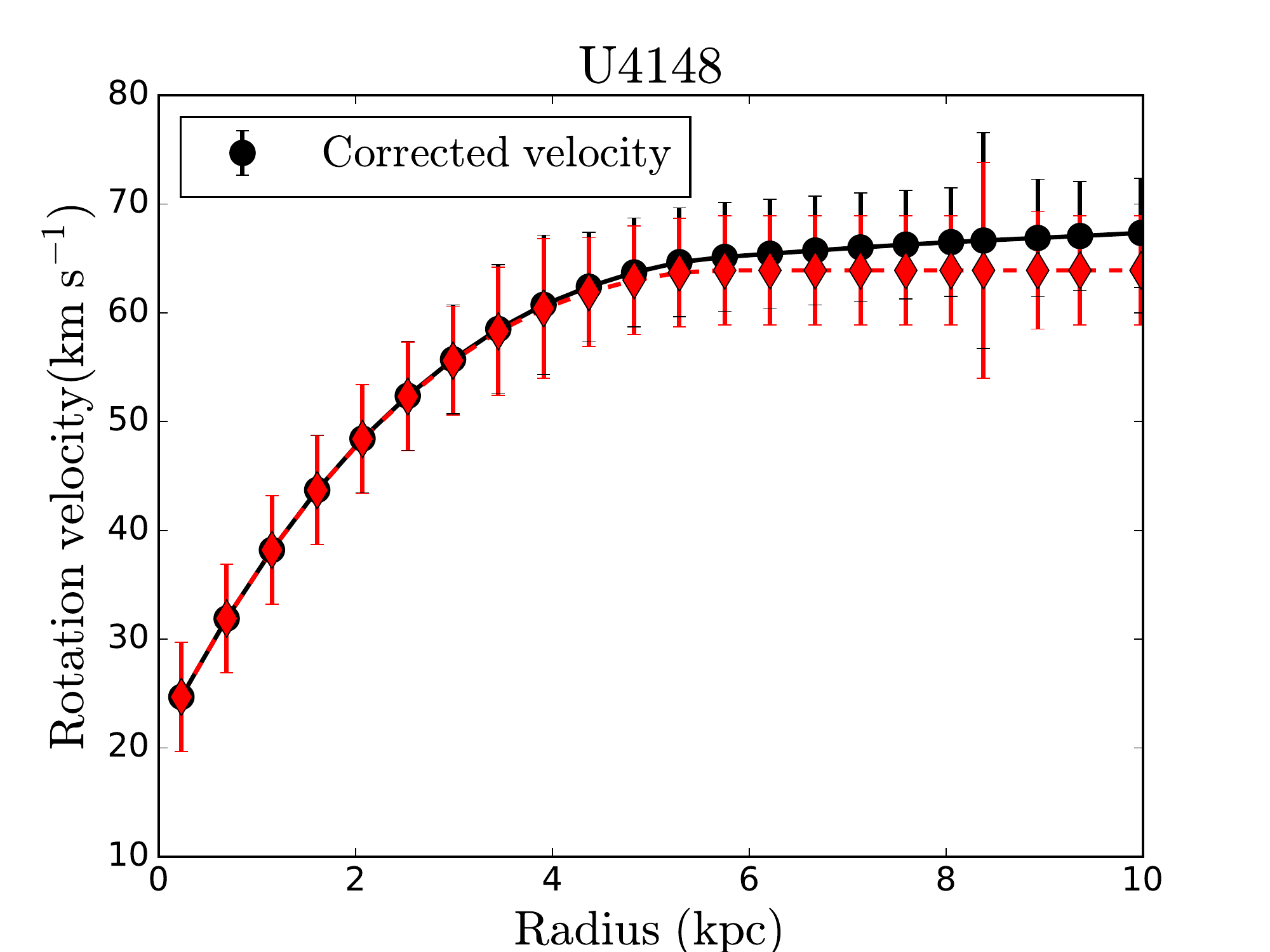}
\includegraphics[width=1.0\linewidth]{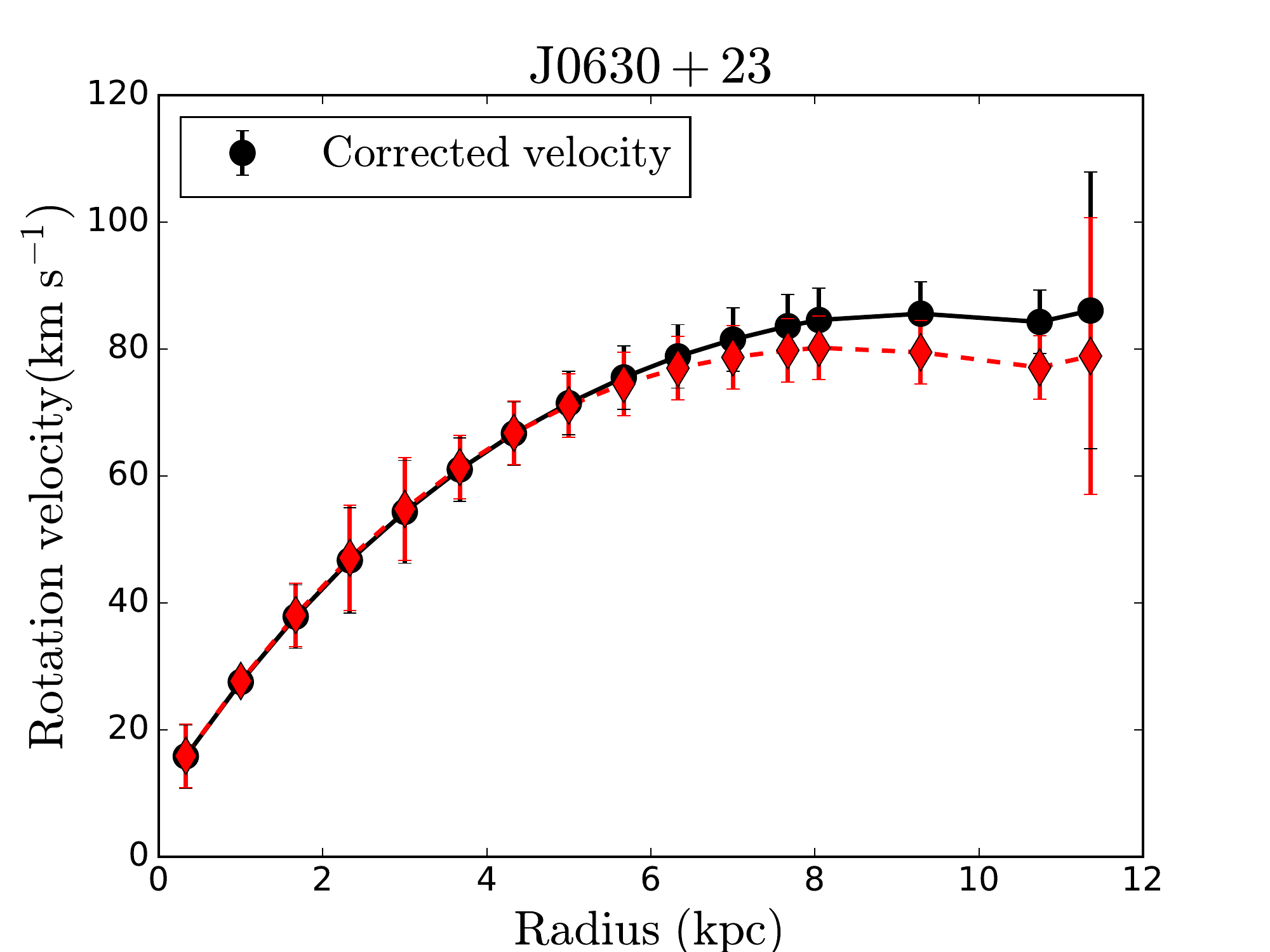}
\includegraphics[width=1.0\linewidth]{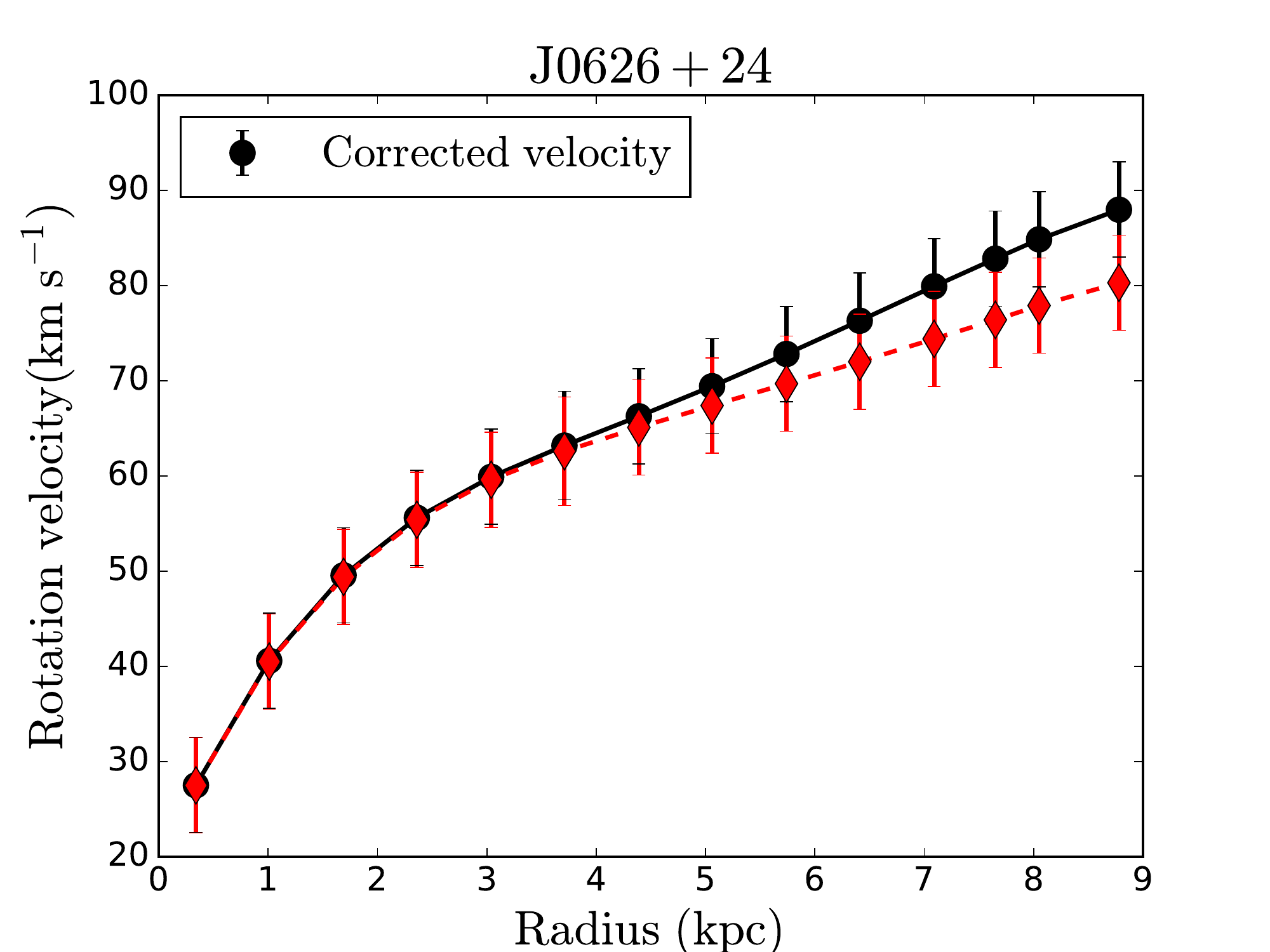}
\includegraphics[width=1.0\linewidth]{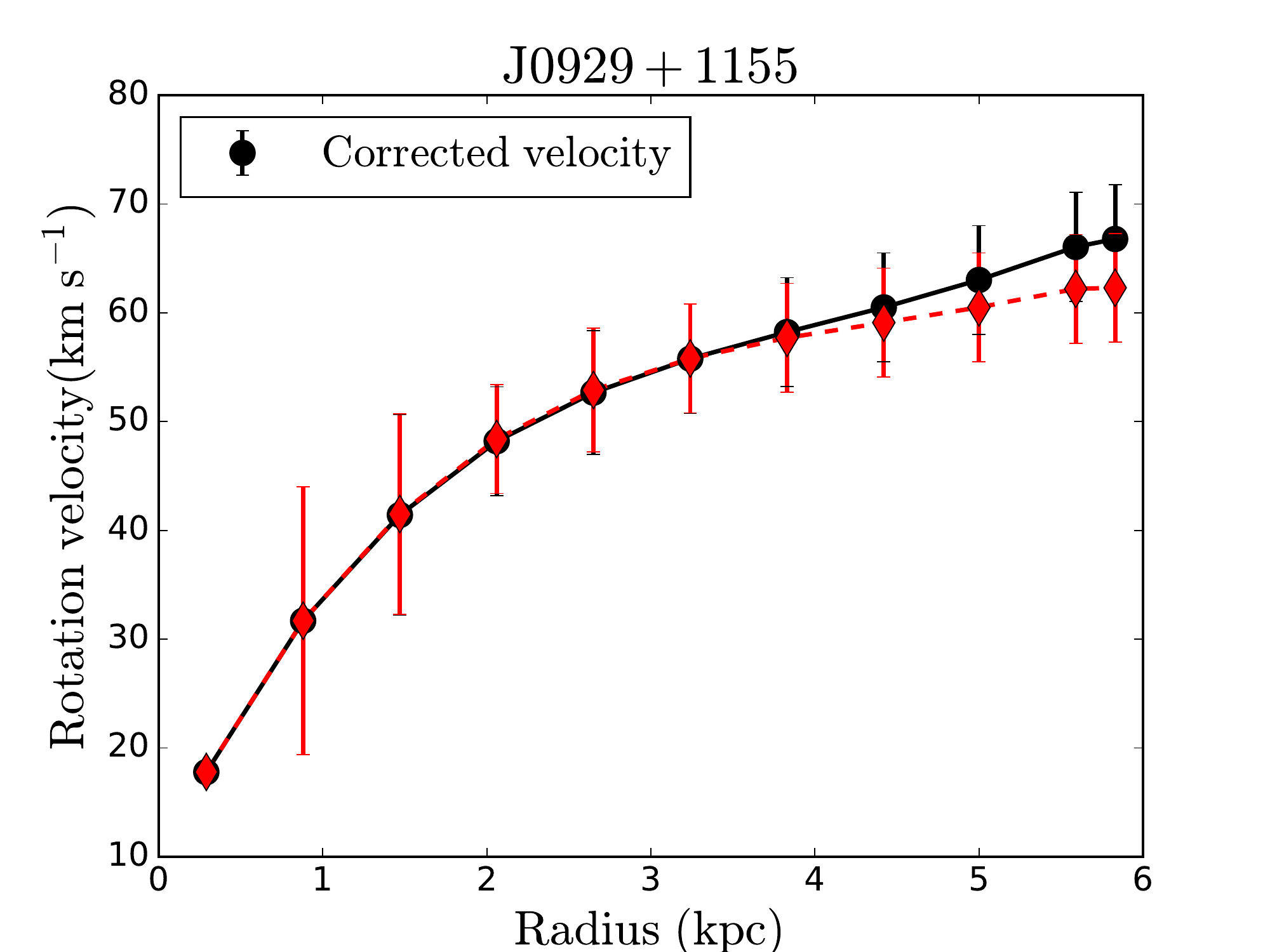}
\caption{}
\label{fig:rotfat2}
\end{figure}

In Figure \ref{fig:rotfat1} and \ref{fig:rotfat2}, we show the rotation curves of all the galaxies as derived by \fat before and after the correction for pressure support. The red diamonds indicate the derived rotation curve from the \fat software and the black circles show the rotation curve after correcting for the pressure support.

\section{Comparison of the 2D and 3D approaches}

\begin{figure*}
\centering
\subfloat{\includegraphics[width = 3.3in]{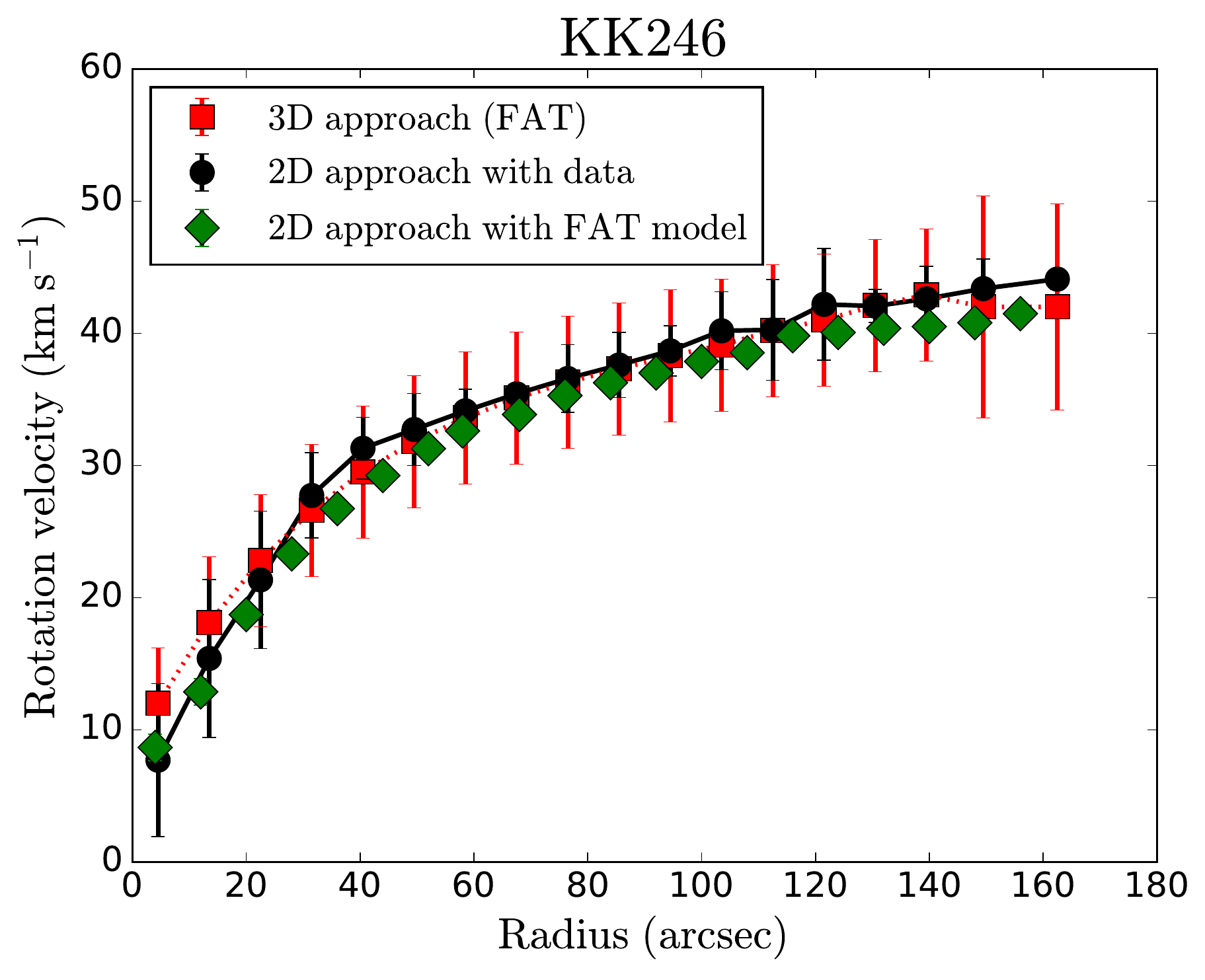}} 
\subfloat{\includegraphics[width = 3.3in]{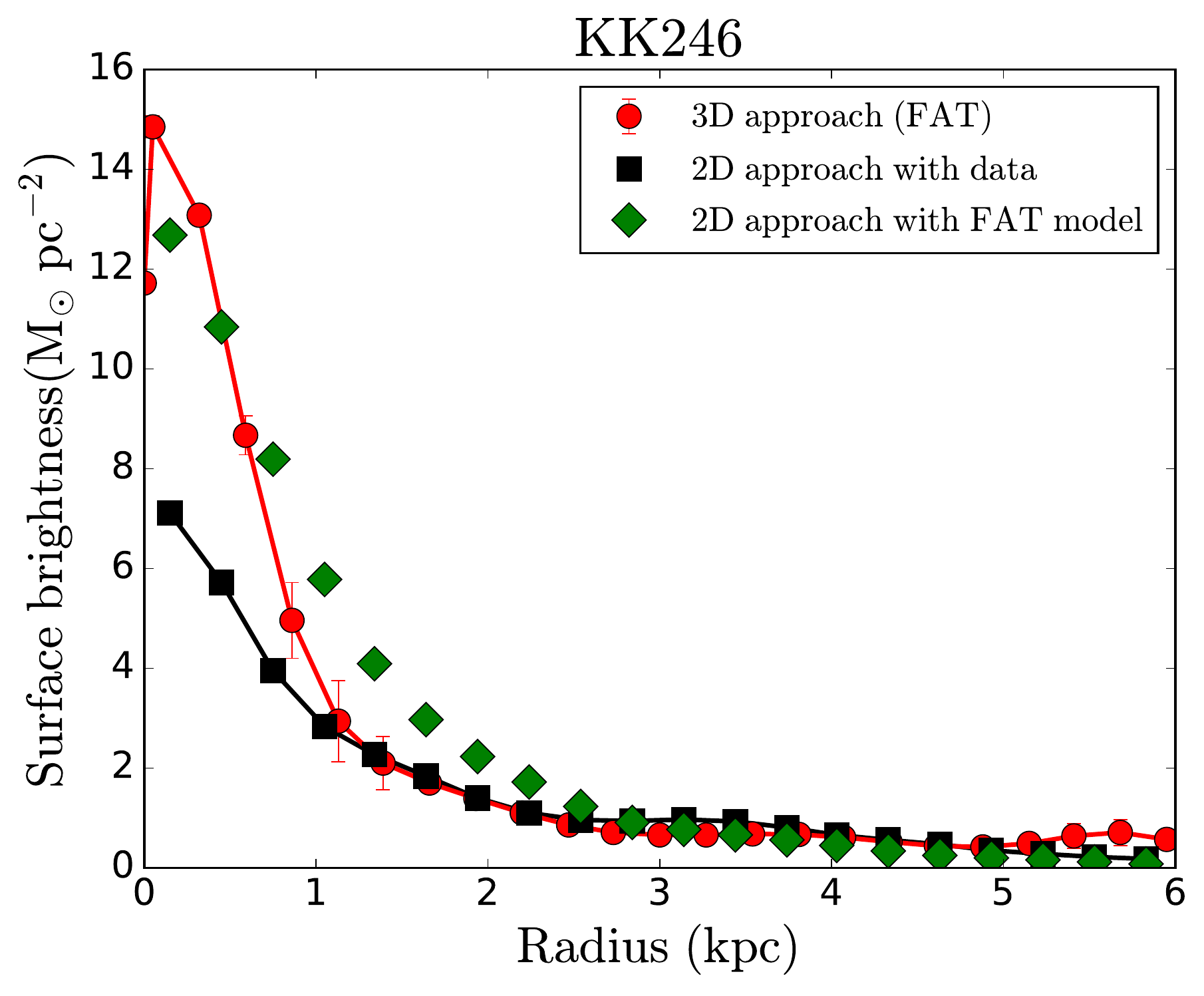}} \\
\subfloat{\includegraphics[width = 3.3in]{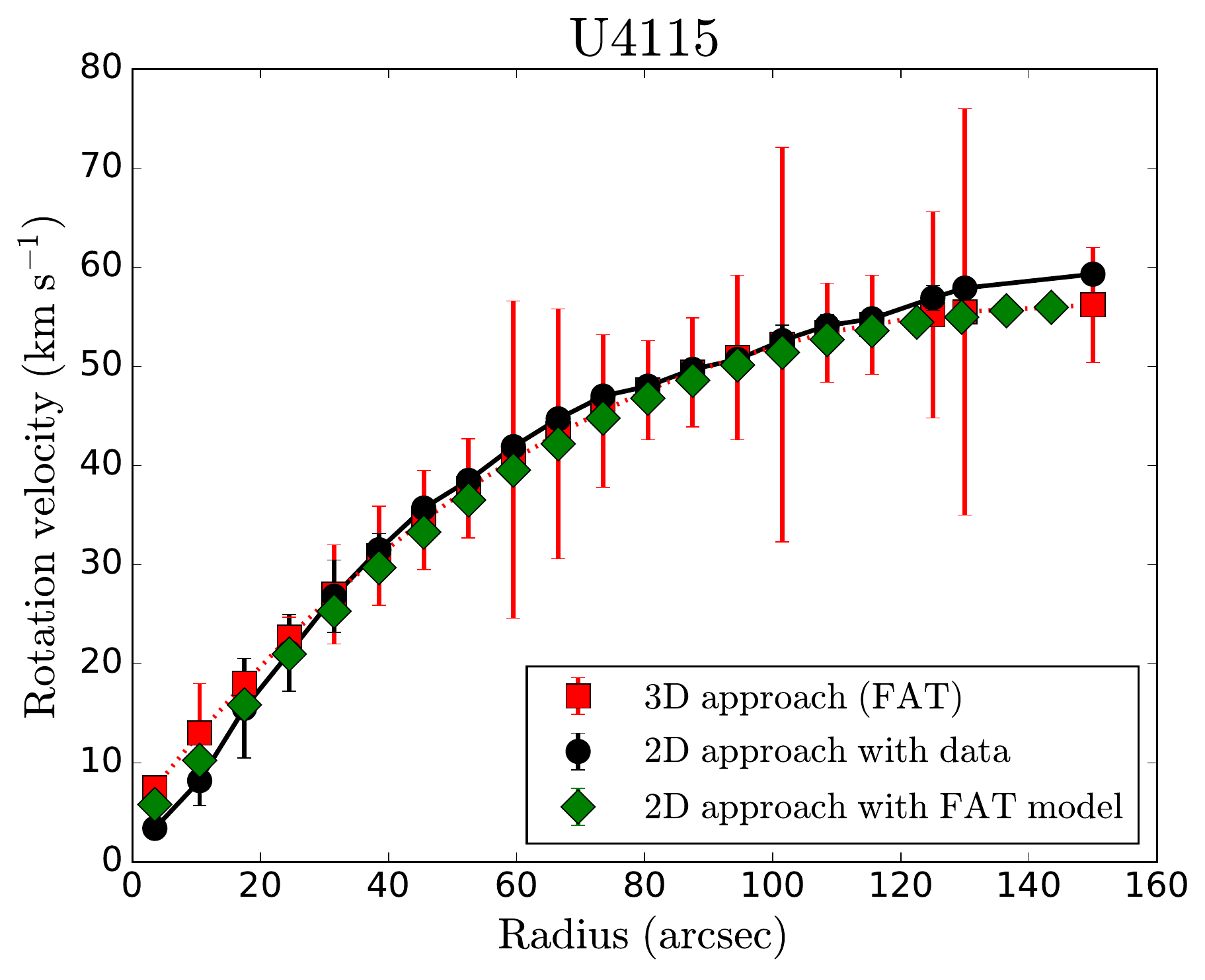}} 
\subfloat{\includegraphics[width = 3.3in]{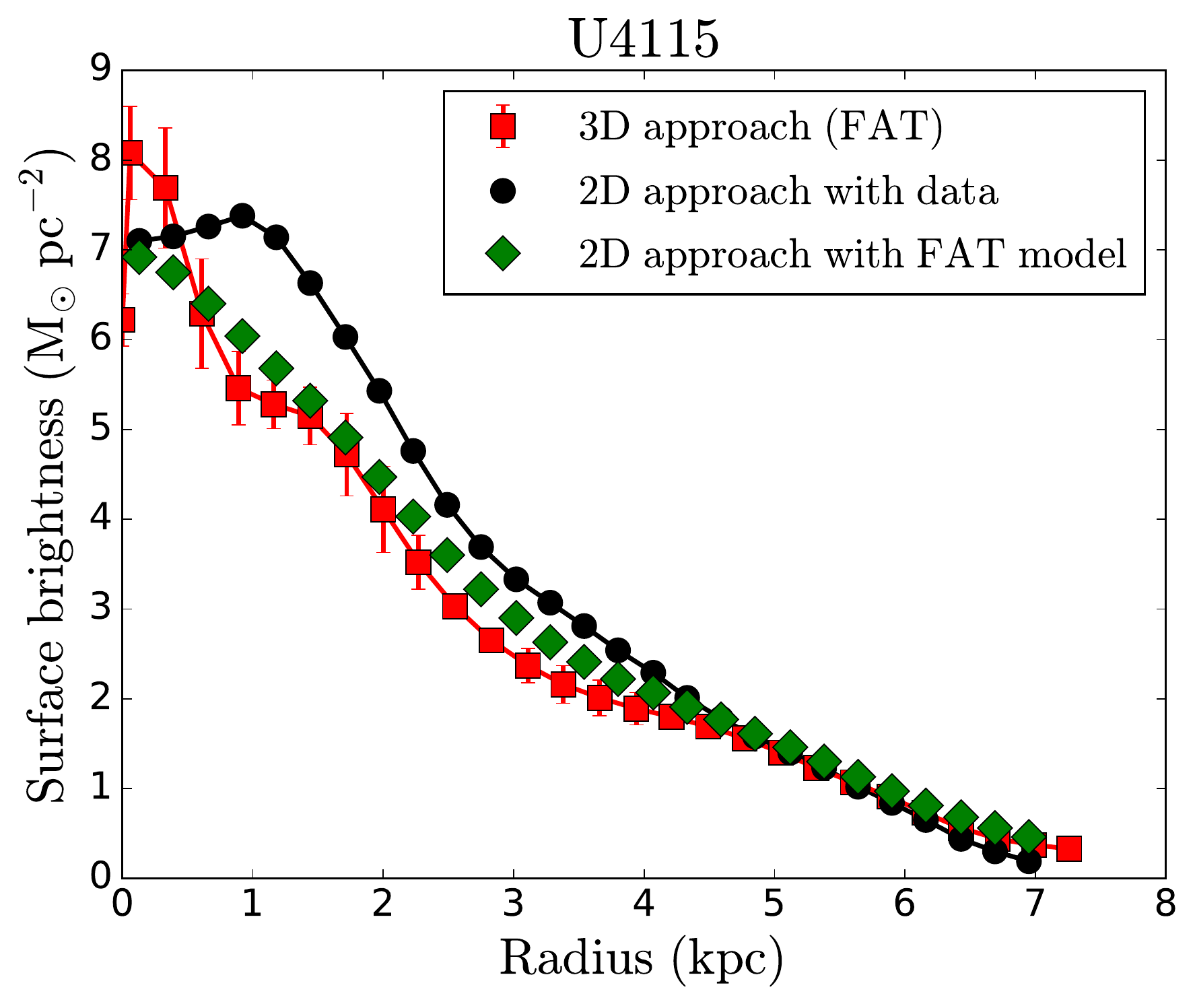}} \\
\subfloat{\includegraphics[width = 3.3in]{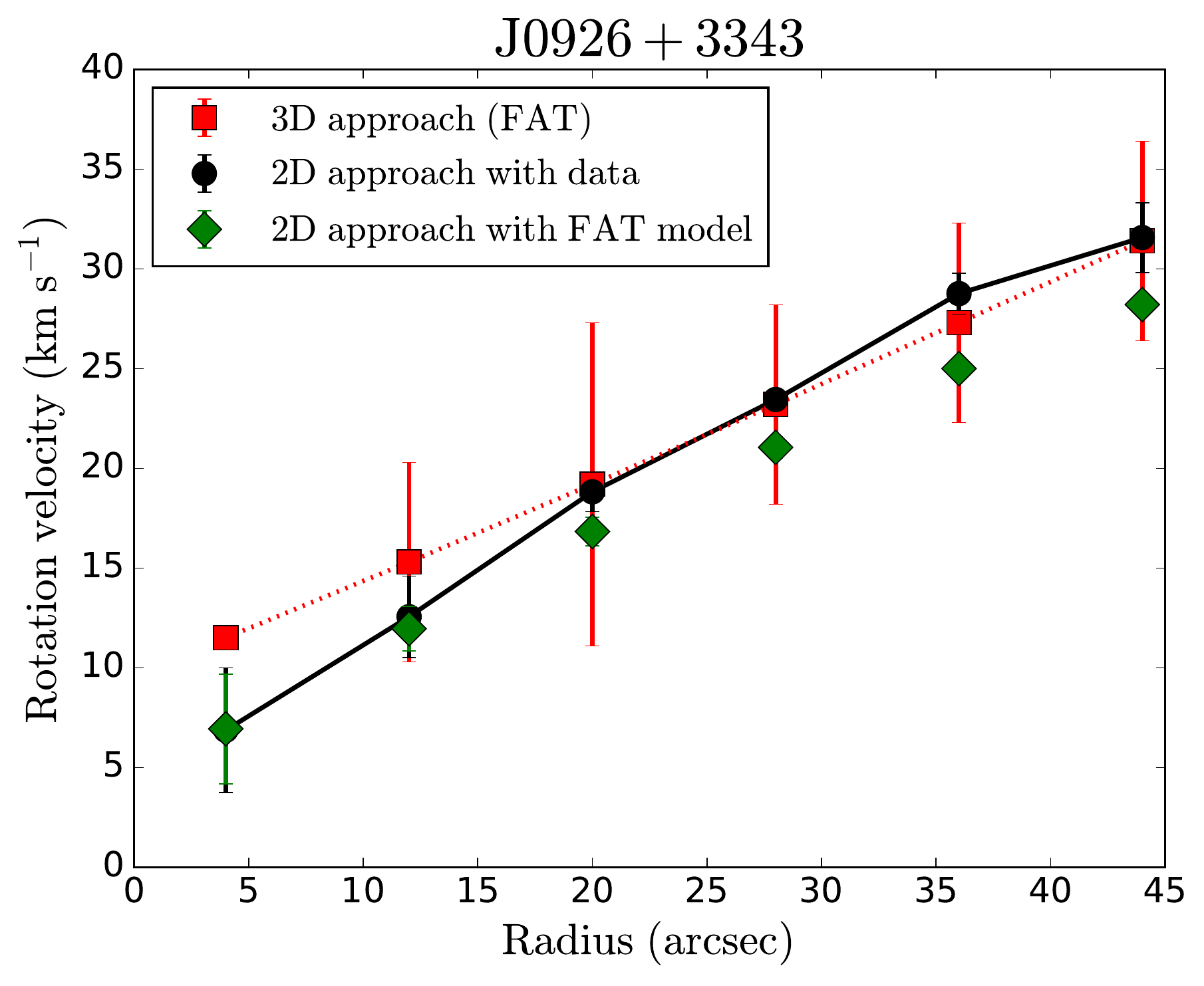}} 
\subfloat{\includegraphics[width = 3.3in]{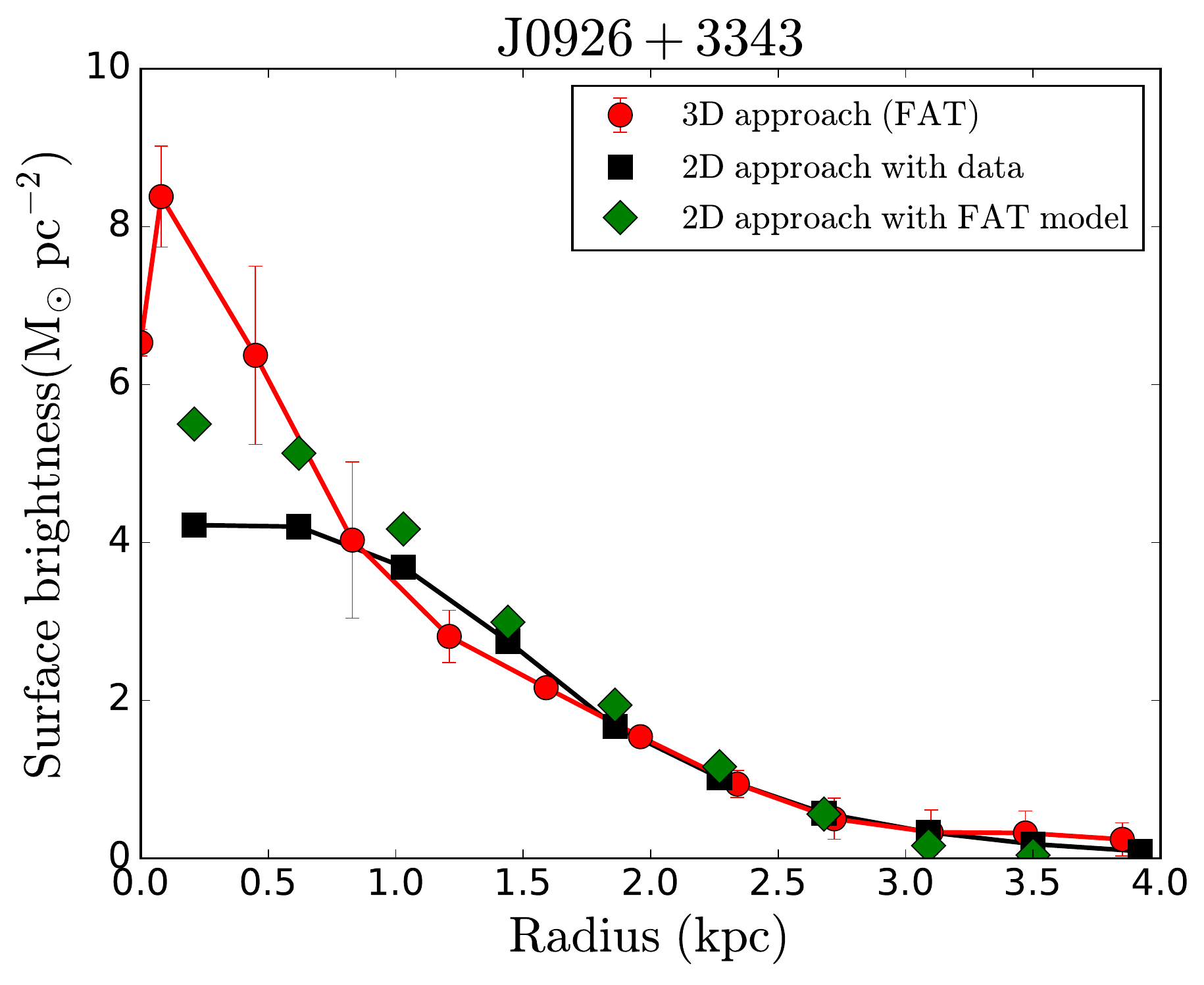}} \\
\caption{Left panels: H{\sc i} surface brightness as a function of radius for the galaxies KK246, U4115, and J0926+3343. Right panels: Rotation velocity as a function of radius for the galaxies KK246, U4115, and J0926+3343. The profile derived by fat (3D approach) are shown by red squares. The profiles derived in gipsy (2D approach) are shown by black circles. We use fat model and derive the profiles with 2D approach, which are shown by green diamonds.}
\label{fig:fatmodel1}
\end{figure*}

\begin{figure*}
\centering
\subfloat{\includegraphics[width = 3.3in]{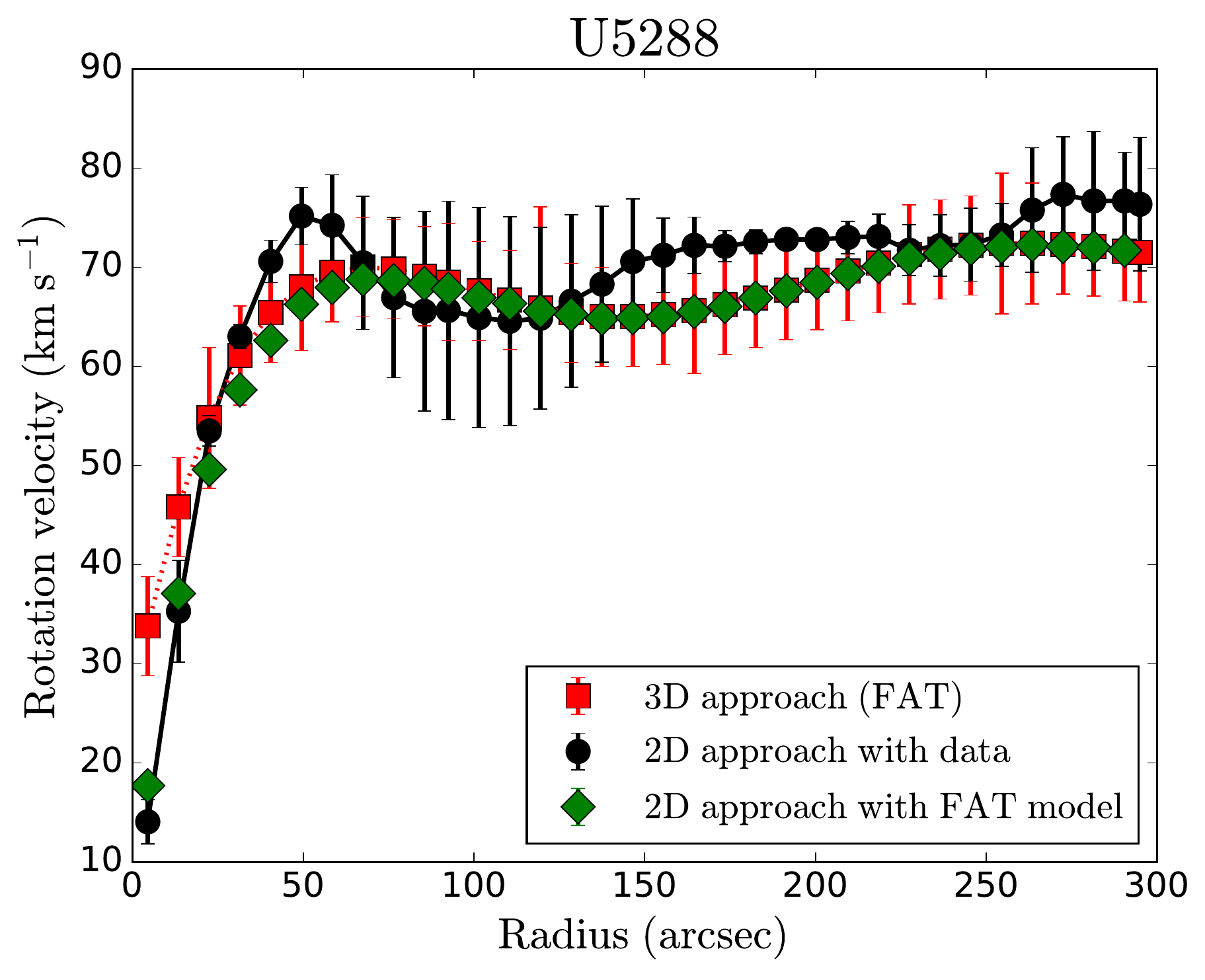}} 
\subfloat{\includegraphics[width = 3.3in]{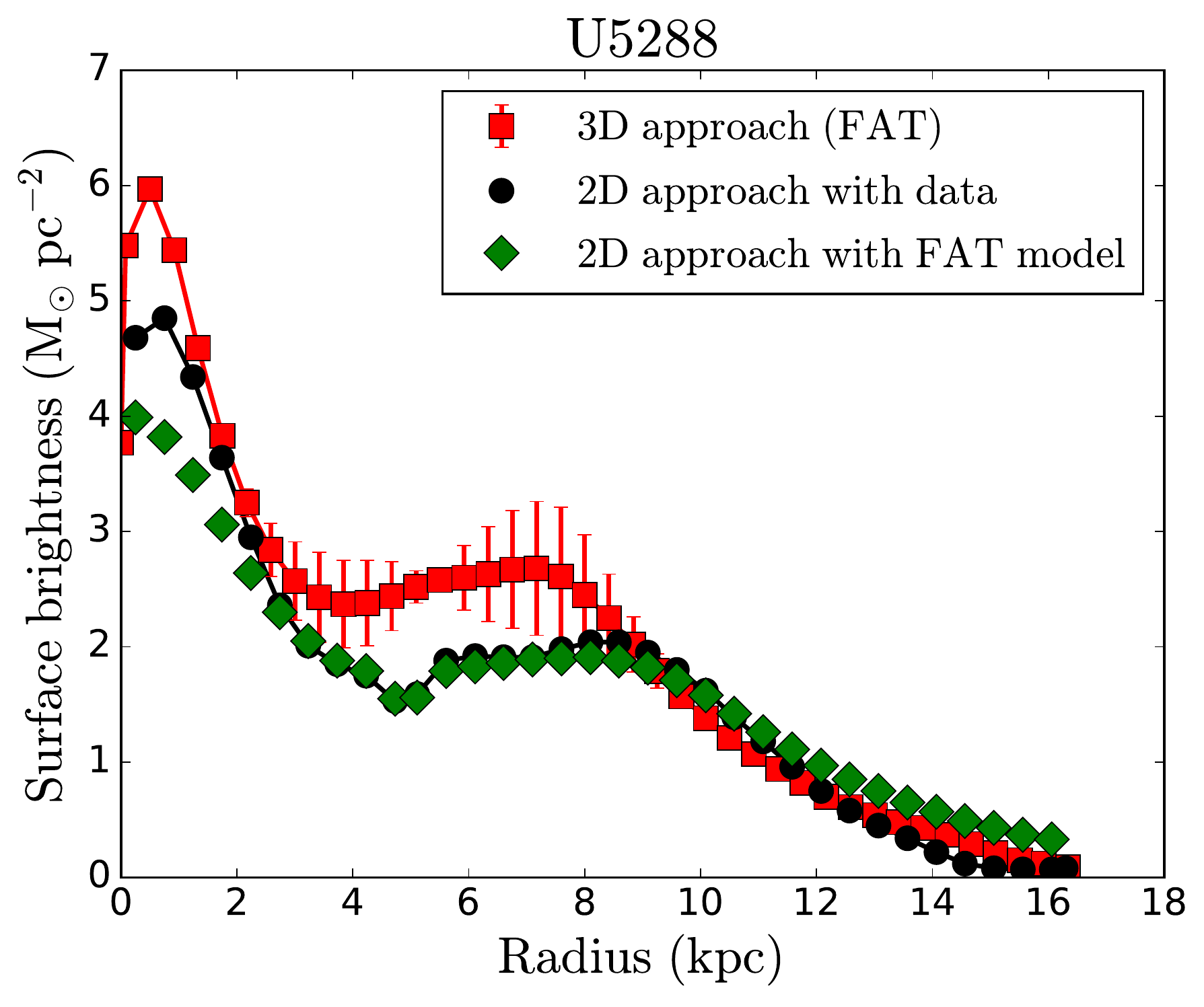}} \\
\subfloat{\includegraphics[width = 3.3in]{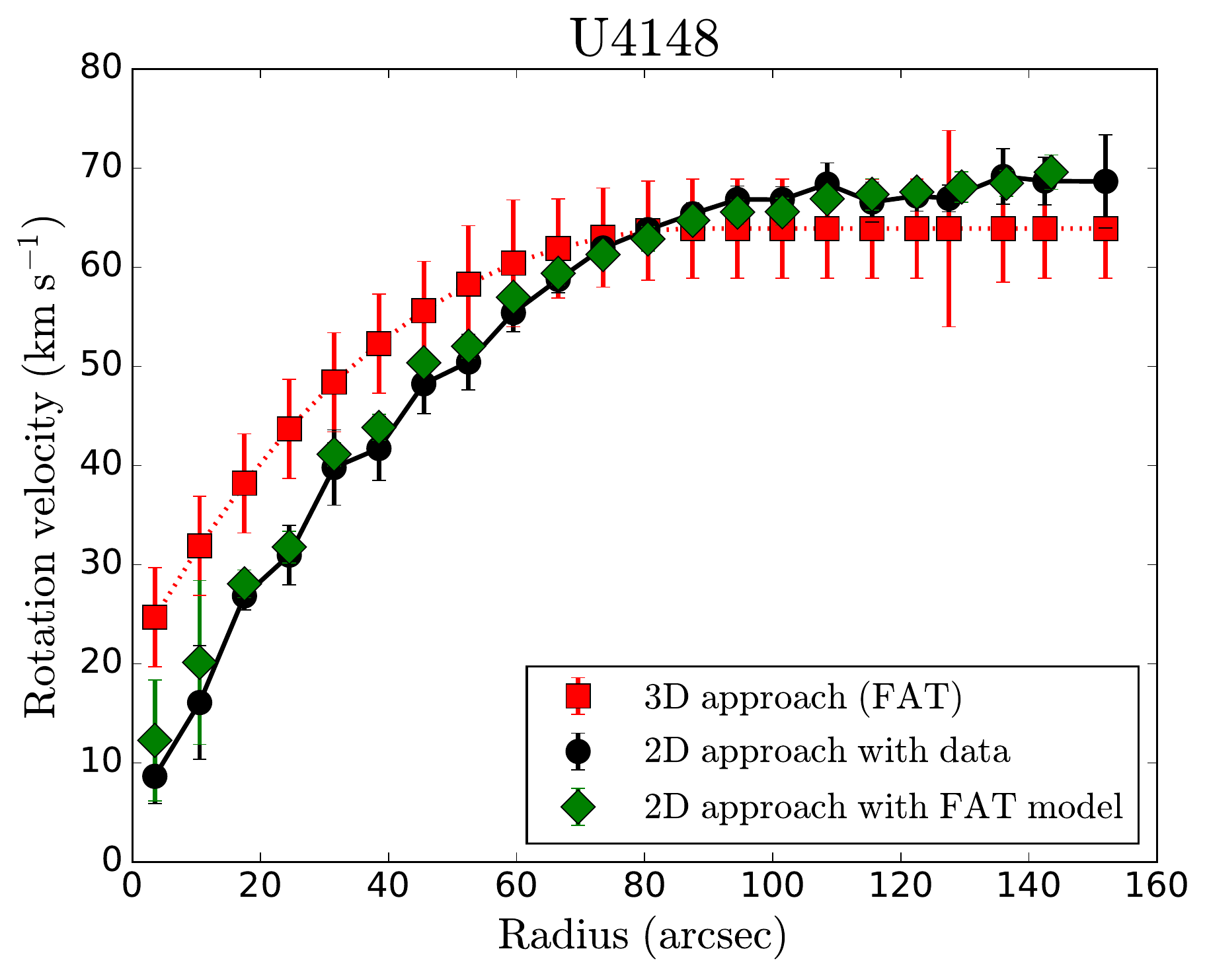}} 
\subfloat{\includegraphics[width = 3.3in]{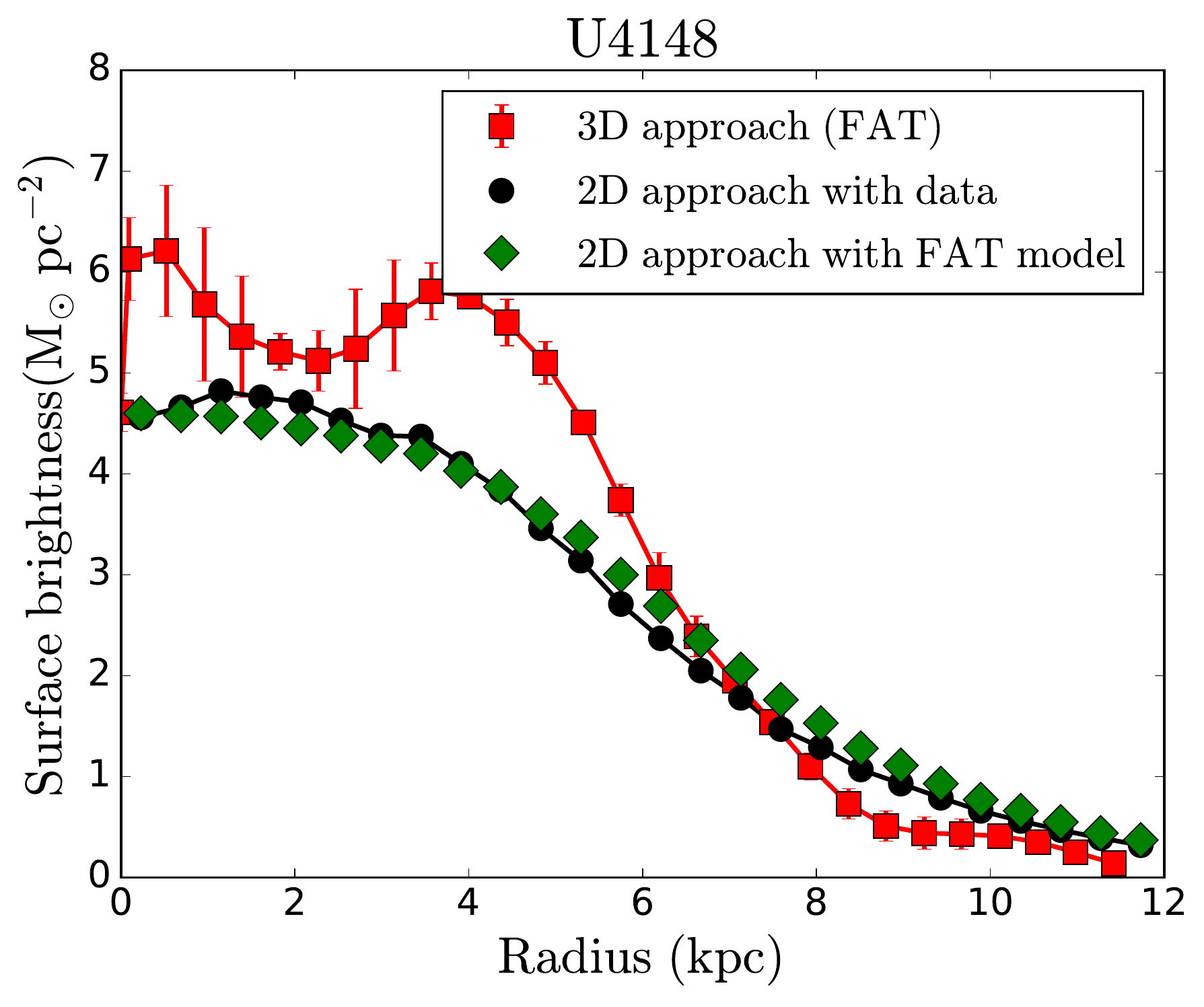}} \\
\subfloat{\includegraphics[width = 3.3in]{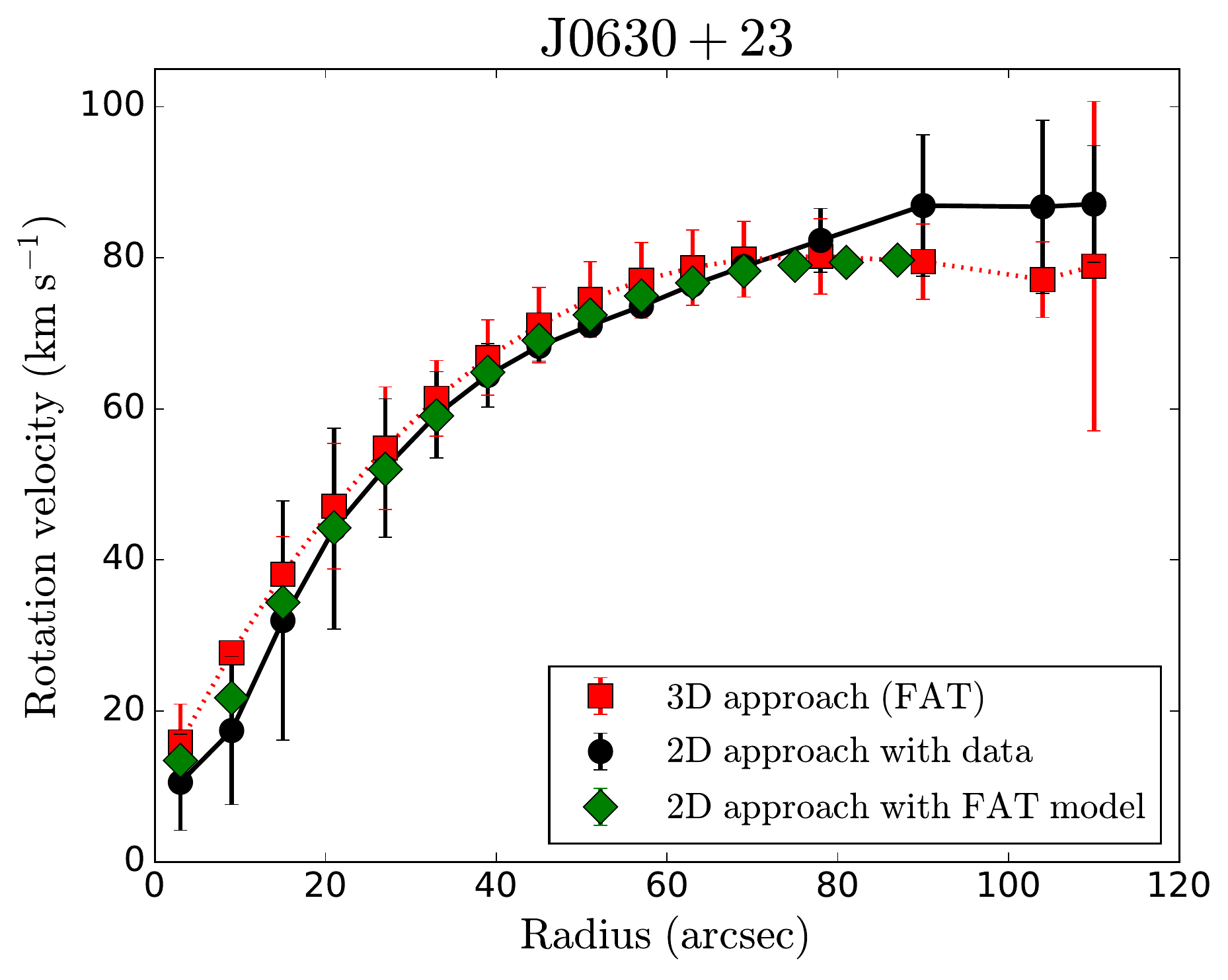}} 
\subfloat{\includegraphics[width = 3.3in]{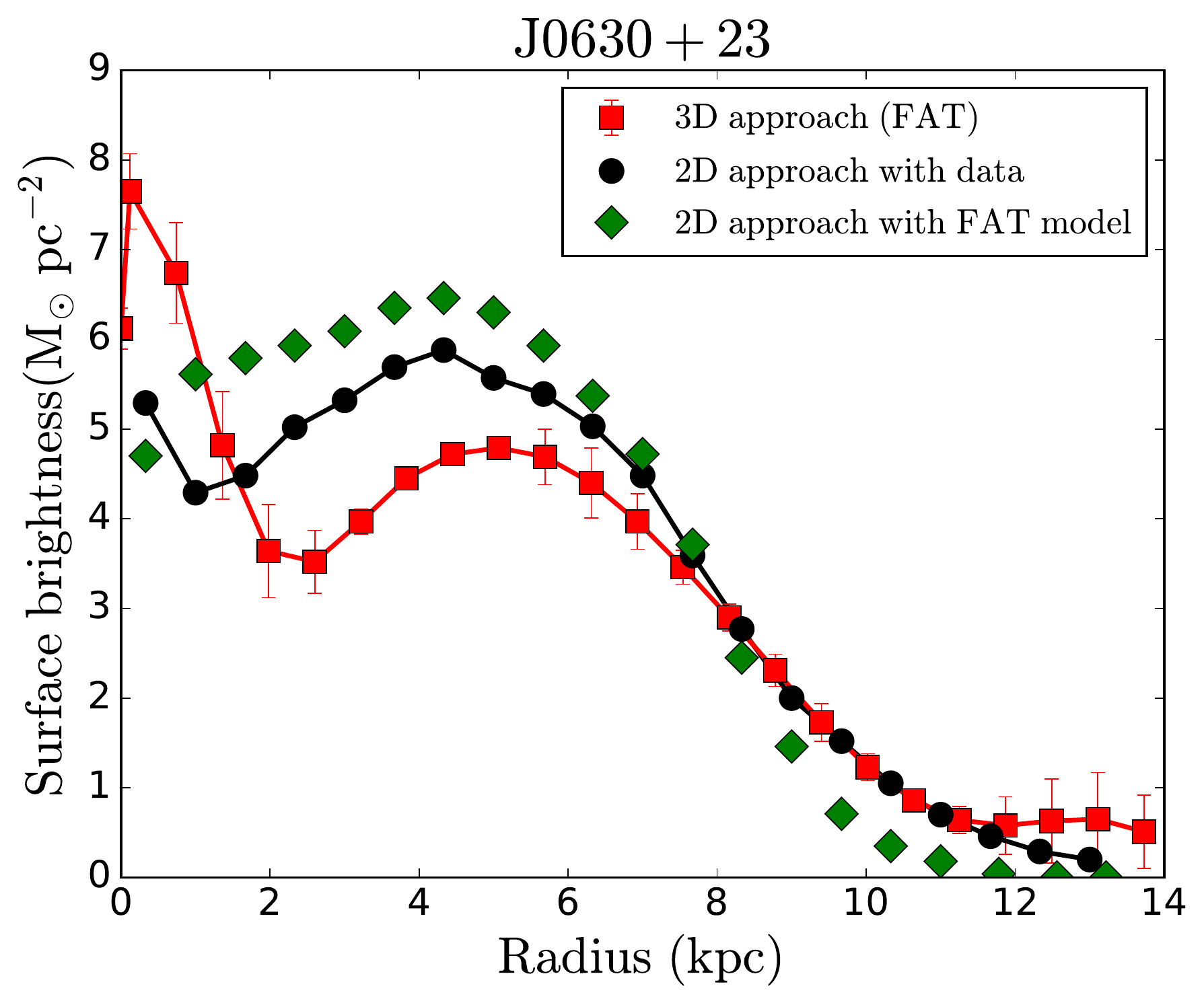}} \\
\caption{Left panels: H{\sc i} surface brightness as a function of radius for the galaxies U5288, U4148, and J0630+23. Right panels: Rotation velocity as a function of radius for the galaxies U5288, U4148, and J0630+23. The profile derived by fat (3D approach) are shown by red squares. The profiles derived in gipsy (2D approach) are shown by black circles. We use fat model and derive the profiles with 2D approach, which are shown by green diamonds. }
\label{fig:fatmodel2}
\end{figure*}

\begin{figure*}
\centering
\subfloat{\includegraphics[width = 3.3in]{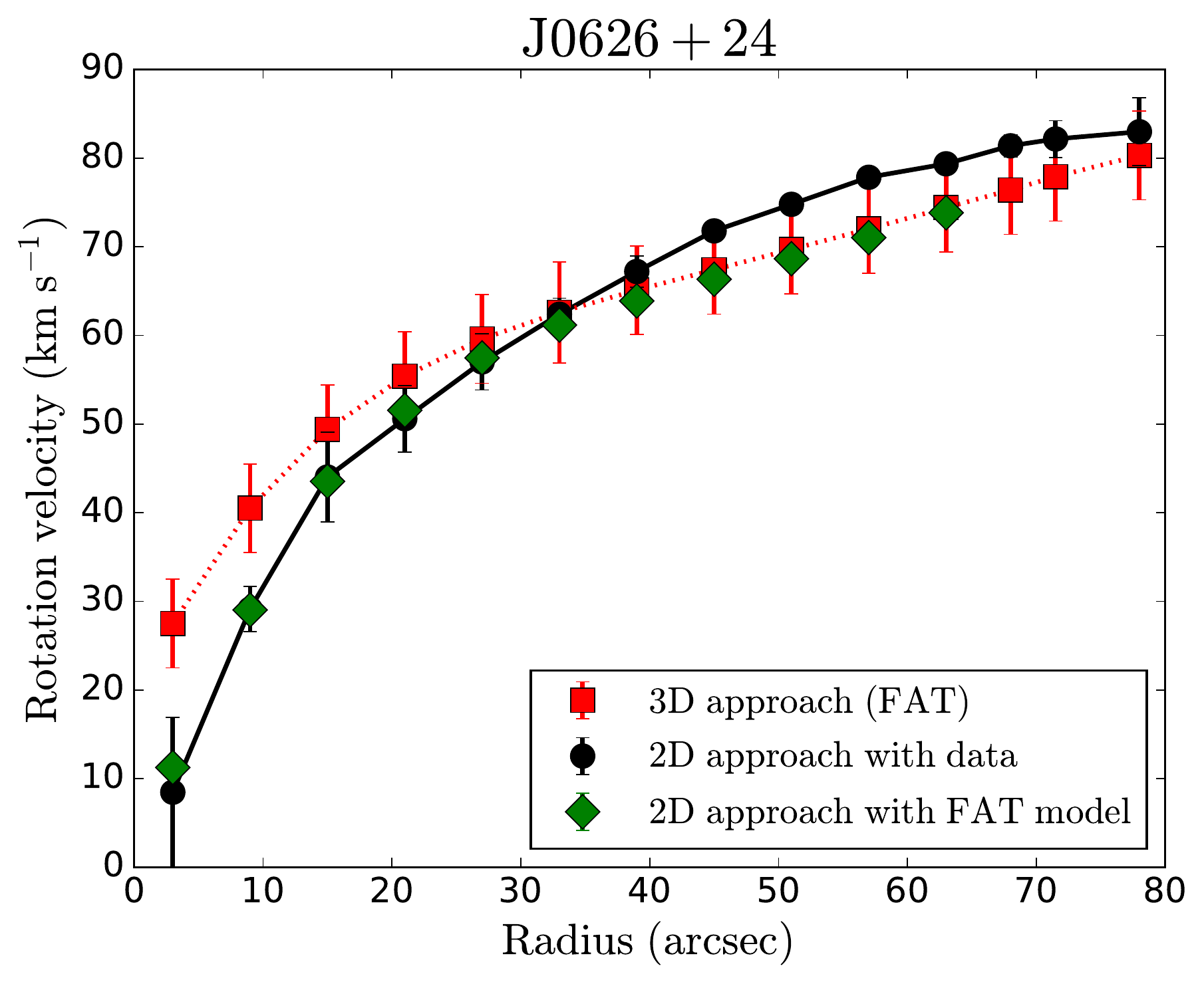}} 
\subfloat{\includegraphics[width = 3.3in]{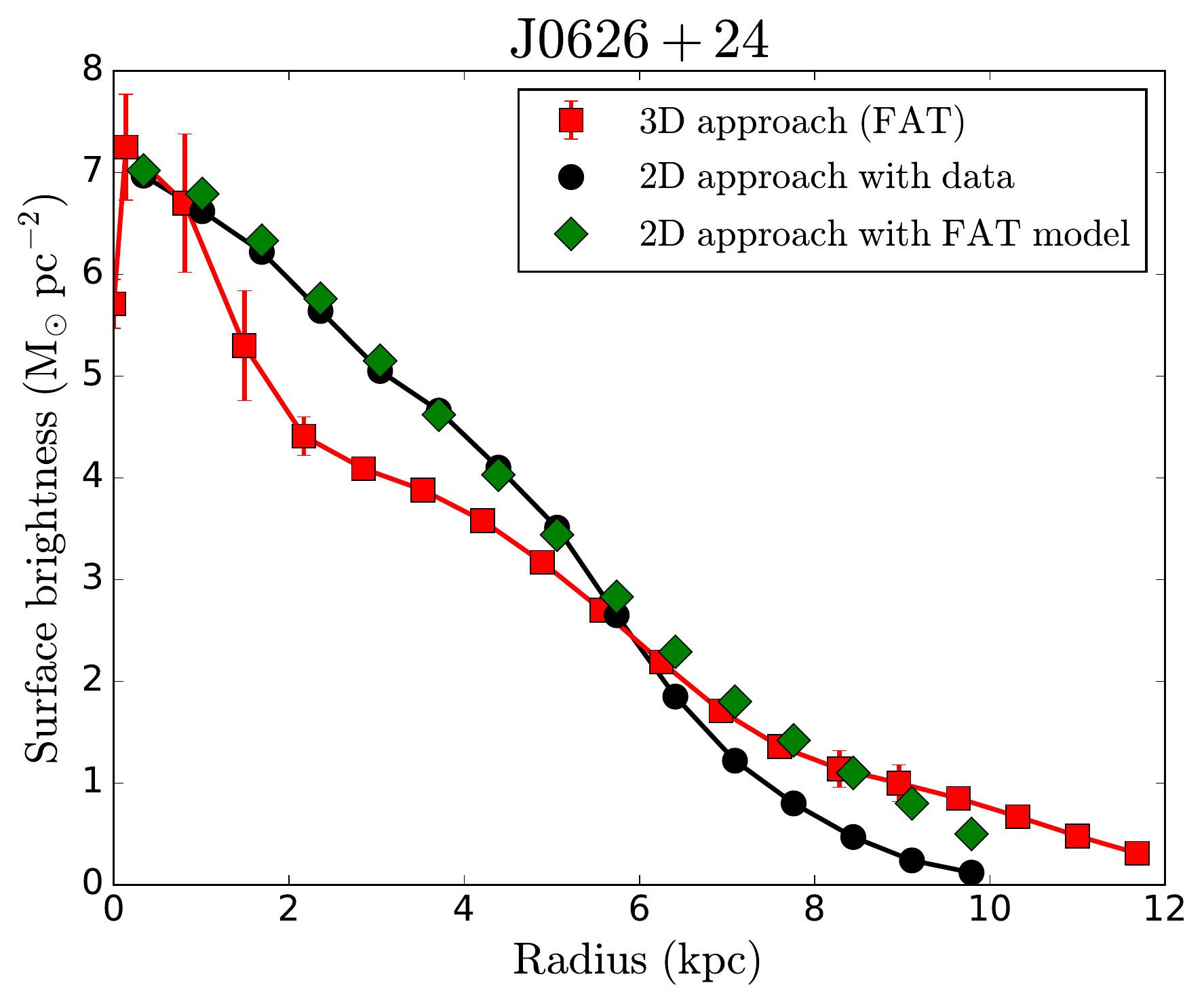}} \\
\subfloat{\includegraphics[width = 3.3in]{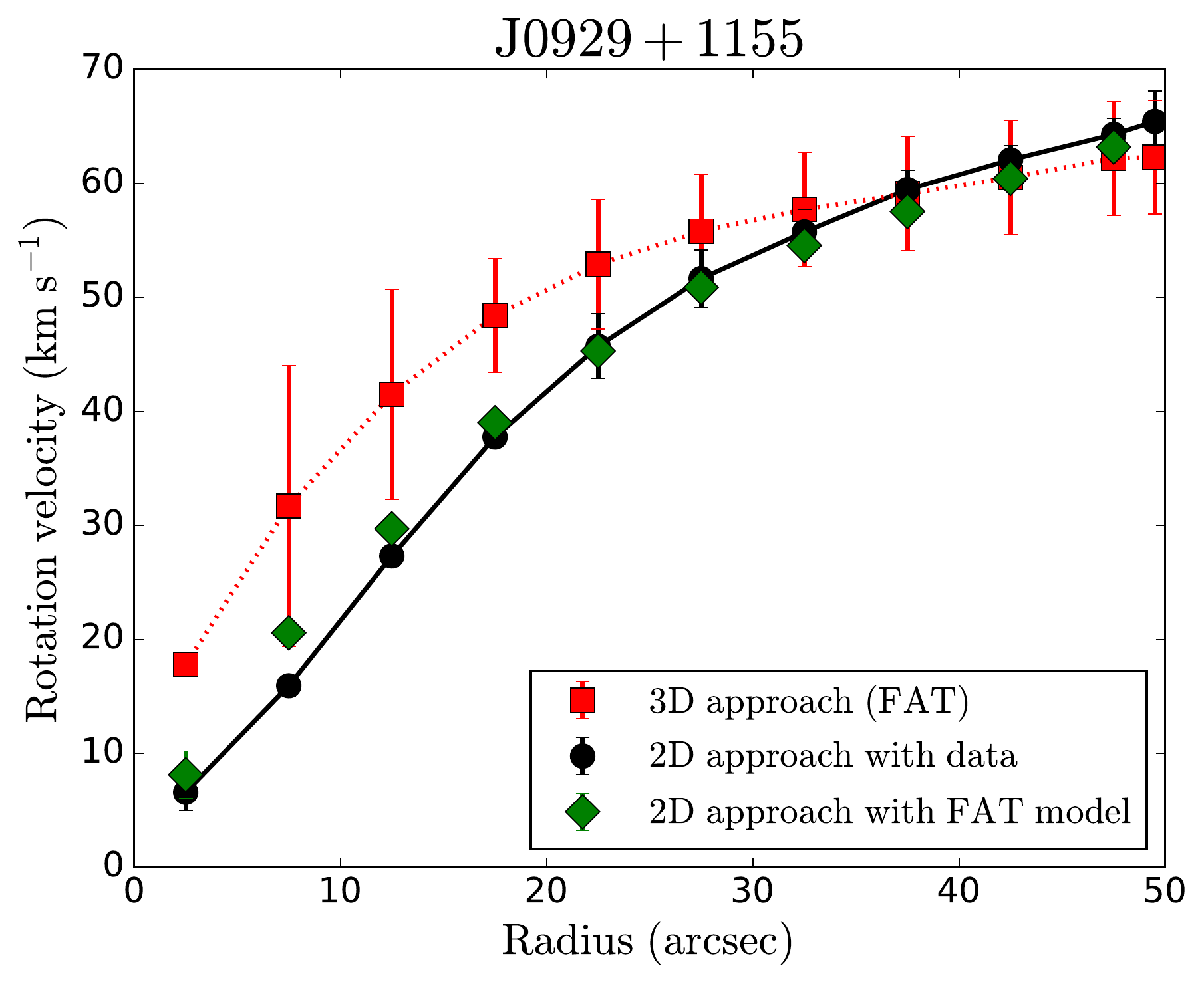}} 
\subfloat{\includegraphics[width = 3.3in]{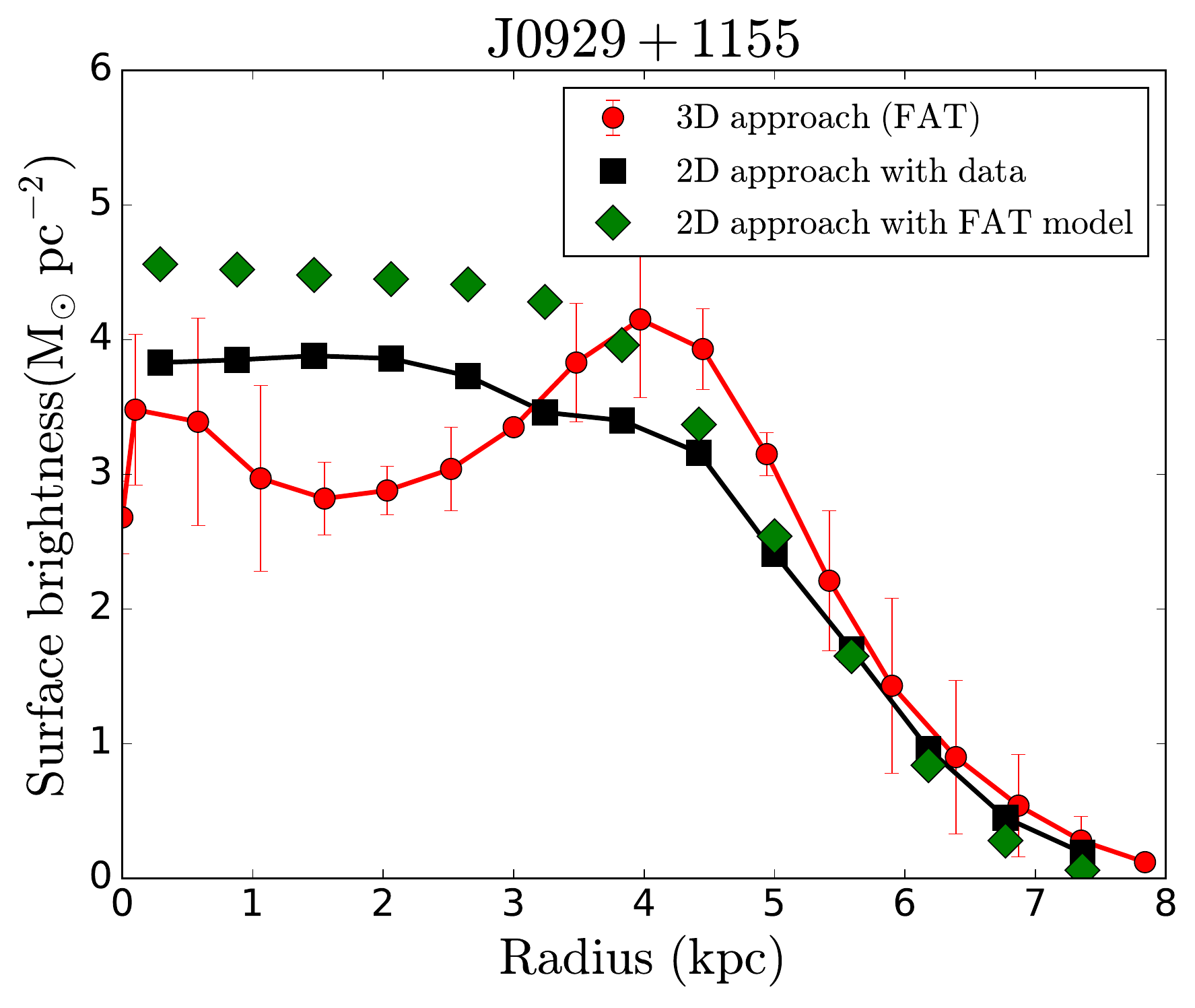}} \\

\caption{Left panels: H{\sc i} surface brightness as a function of radius for the galaxies J0626+24 and J0929+1155. Right panels: Rotation velocity as a function of radius for the galaxies J0626+24 and J0929+1155. The profile derived by fat (3D approach) are shown by red squares. The profiles derived in gipsy (2D approach) are shown by black circles. We use fat model and derive the profiles with 2D approach, which are shown by green diamonds. }
\label{fig:fatmodel3}
\end{figure*}

We derive moment maps from the best fit model data cube produced by \fat and derive rotation curves as well as surface brightness profiles from them using the 2D routines in GIPSY. For each galaxy, we present 3 sets of curves for the rotation curves and surface brightness profiles. In Figures \ref{fig:fatmodel1}, \ref{fig:fatmodel2}, and \ref{fig:fatmodel3}, we show the curves derived by fat (red squares) and GIPSY (black circles) when run on the original observed data, as well as the curves produced by the GIPSY tasks when run on the model produced by fat (green diamonds)

\begin{table*}
\begin{footnotesize}

\caption{Parameters of dark matter halos }
\label{table:gamma}
\begin{tabular}{ p{1.6cm} p{2.7cm} p{0.6cm} p{1.4cm} p{1.9cm} p{0.6cm} p{0.6cm} p{1.4cm} p{1.7cm} p{0.6cm}}
\\
\hline
\hline
Galaxy& Model &  \multicolumn{4}{c}{Isothermal} &  \multicolumn{4}{c}{NFW}   \\
\cline{3-6}
\cline{7-10}
&	&	$\gamma_{\ast}$ & r$_{c}$ & $\rho_{0}$  &$\chi^{2}_{r}$ & $\gamma_{\ast}$ & 	c   & r$_{200}$	& $\chi_{r}^{2}$ \\		
 & & & (kpc) & (M$_{\odot}$ pc$^{-3}$) & & & & (kpc) & \\
\hline
KK246 & Fixed $\gamma_{\ast}$	    & 1.07& 1.20$\pm$0.26 & 39$\pm$12& 0.30 & 1.07& 2.7$\pm$0.7 &	52$\pm$10 & 0.06 \\
& Minimum disc + gas	 		    & 0.0 & 0.69$\pm$0.16 & 93$\pm$34& 0.41 & 0.0 & 4.8$\pm$0.3 &	37$\pm$2 & 0.02 \\
& Minimum disc	 				    & 0.0 & 0.66$\pm$0.14 & 107$\pm$37& 0.42 & 0.0 & 5.1$\pm$0.2 &	38$\pm$1 & 0.01 \\
& Maximum disc	 				    & 3.10& 2.41$\pm$0.68 & 15$\pm$4 & 0.27 & ..& ... & ... & ... \\
& Free $\gamma_{\ast}$	 		    & 2.28& 1.89$\pm$0.68 & 21$\pm$9 & 0.25 & .. &  ... &	... & ... \\
\hline 
U4115 &Constant $\gamma_{\ast}$	& 0.29& 1.87$\pm$0.20 & 30$\pm$4 & 0.09 & 0.29& ... & ... & ... \\
& Minimum disc + gas	 		& 0.0 & 1.76$\pm$0.18 & 34$\pm$4 & 0.10 & 0.0 & ... & ... & ... \\
& Minimum disc	 				& 0.0 & 1.71$\pm$0.14 & 41$\pm$4 & 0.08 & 0.0 & ... & ... & ... \\
& Maximum disc	 				& 3.48& 8.45$\pm$5.62 & 9$\pm$1 & 0.07 & 3.48& ...  & ... & ... \\
& Free $\gamma_{\ast}$	 		& 2.75& 4.48$\pm$2.83 & 12$\pm$5 & 0.07 & .. &  ... &	... & ... \\

\hline 
J0926+3343 &Constant $\gamma_{\ast}$& 0.06& 0.34$\pm$0.19 & 163$\pm$143&0.52& 0.06 & 1.9$\pm$6.0 & 60$\pm$161  & 0.15 \\
& Minimum disc + gas				& 0.0 & 0.35$\pm$0.19 & 164$\pm$143 &0.52 & 0.0 & 1.9$\pm$5.6 & 60$\pm$158  & 0.15 \\
& Minimum disc	 					& 0.0 & 0.41$\pm$0.24 & 149$\pm$134 &0.64 & 0.0 & ... & ... & ... \\
& Maximum disc	 				    & 5.40& 0.28$\pm$0.27 & 138$\pm$207 &0.74 & ... & ... & ... & ... \\
& Free $\gamma_{\ast}$	 			& ... & ... 		  & ...             & ... & ... & ... &	... & ... \\
\hline 
U5288 &Constant $\gamma_{\ast}$	& 0.36& 0.25$\pm$0.05 & 1341$\pm$507&0.71 & 0.36& 11.4$\pm$1.7 & 41$\pm$2 & 1.21 \\
& Minimum disc + gas	 		& 0.0 & 0.25$\pm$0.05 & 1458$\pm$522&0.66 & 0.0 & 12.1$\pm$1.7 & 41$\pm$2 & 1.21 \\
& Minimum disc	 				& 0.0 & 0.28$\pm$0.06 & 1260$\pm$483&0.83 & 0.0 & 10.8$\pm$1.6 & 44$\pm$2 & 1.36 \\
& Maximum disc	 				& 5.40& 2.26$\pm$0.64 & 20$\pm$9    &1.22 & 5.40& 2.5$\pm$0.8  & 58$\pm$8 & 0.88 \\
& Free $\gamma_{\ast}$	 		& .. &  ... &	... & ... & .. &  ... &	... & ... \\
\hline 
U4148 &Constant $\gamma_{\ast}$		& 0.17& 0.50$\pm$0.11 & 289$\pm$118&0.69 & 0.17& 8.4$\pm$0.6 & 39$\pm$1 & 0.14\\
& Minimum disc + gas				& 0.0 & 0.51$\pm$0.11 & 289$\pm$117 &0.69 & 0.0 & 8.4$\pm$0.6 & 39$\pm$1 & 0.14 \\
& Minimum disc	 					& 0.0 & 0.73$\pm$0.16 & 172$\pm$66 &0.88 & 0.0 & 7.4$\pm$0.6 & 44$\pm$1 & 0.15 \\
& Maximum disc	 				    & 17.0& 0.14$\pm$0.08 & 1199$\pm$1270&0.95 & 17.0& 4.4$\pm$2.4 & 29$\pm$6& 1.01 \\
& Free $\gamma_{\ast}$	 		    & 2.8 & 0.45$\pm$0.13 & 316$\pm$149&0.76 & 0.51 & 8.4$\pm$0.8 & 39$\pm$2 & 0.16 \\
\hline 
J0630+23 &Constant $\gamma_{\ast}$	& 0.24& 1.92$\pm$0.24 & 42$\pm$8 &0.34 & 0.24& 3.4$\pm$0.8 & 71$\pm$10 & 0.39 \\
& Minimum disc + gas	 			& 0.0 & 1.85$\pm$0.23 & 45$\pm$8 &0.35 & 0.0 & 3.6$\pm$0.8 & 70$\pm$9 & 0.42 \\
& Minimum disc						& 0.0 & 2.17$\pm$0.29 & 41$\pm$8 &0.48 & 0.0 & 3.2$\pm$0.9 & 84$\pm$16& 0.71 \\
& Maximum disc	 				    & 2.64& 3.24$\pm$0.61 & 15$\pm$4 &0.36 & 2.64& 1.1$\pm$1.1 & 119$\pm$75& 0.34 \\
& Free $\gamma_{\ast}$	 		    & 0.73& 2.08$\pm$0.78 & 35$\pm$27&0.41 & 1.77 & 2.1$\pm$2.1 & 86$\pm$44 & 0.37 \\
\hline 
J0626+24 &Constant $\gamma_{\ast}$  & 0.27& 0.89$\pm$0.33 & 144$\pm$90 &2.14 & 0.27& 5.1$\pm$1.1 & 60$\pm$8 & 0.52 \\
& Minimum disc + gas	 			& 0.0 & 0.83$\pm$0.30 & 165 $\pm$101&2.06 & 0.0& 5.4$\pm$1.1 & 59$\pm$7 & 0.52 \\
& Minimum disc	 					& 0.0 & 1.03$\pm$0.34 & 128$\pm$69  &2.12 & 0.0& 5.1$\pm$0.9 & 64$\pm$7 & 0.35 \\
& Maximum disc	 				    & 3.24& 2.77$\pm$1.14 & 24$\pm$12   &1.98 & 3.24& 1.1$\pm$2.2 & 157$\pm$206& 0.54 \\
& Free $\gamma_{\ast}$	 			& ... & ... 		  & ...              & ... & 1.07& 4.1$\pm$3.0 & 66$\pm$26& 0.65 \\
\hline 
J0929+1155 &Constant $\gamma_{\ast}$& 0.22& 0.64$\pm$0.10 & 185$\pm$48 &0.17 & 0.22& 5.8$\pm$0.7 & 46$\pm$3 & 0.07 \\
& Minimum disc + gas	 			& 0.0 & 0.65$\pm$0.10 & 185$\pm$48 &0.17 & 0.0 & 5.8$\pm$0.7 & 47$\pm$3 & 0.07 \\
& Minimum disc	 					& 0.0 & 0.80$\pm$0.18 & 144$\pm$53 &0.38 & 0.0 & 4.7$\pm$0.5 & 57$\pm$4 & 0.04 \\
& Maximum disc	 				    & 17.6& 0.38$\pm$0.19 & 192$\pm$172&0.73 & 17.6& 6.9$\pm$4.1 & 25$\pm$6 & 0.72 \\
& Free $\gamma_{\ast}$	 		    & 1.56& 0.62$\pm$0.12 & 185$\pm$56 &0.22 & .. &  ... &	... & ... \\
\hline & 
\end{tabular}

\end{footnotesize}

\end{table*}

\begin{table*}
\begin{footnotesize}

\caption{Parameters of dark matter halos }
\label{table:rotcur}
\begin{tabular}{ p{1.6cm} p{1.5cm}  p{1.4cm} p{1.9cm} p{0.6cm} p{1.4cm} p{1.9cm} p{0.6cm}}
\\
\hline
\hline
Galaxy& Model &  \multicolumn{3}{c}{Isothermal} &  \multicolumn{3}{c}{NFW}   \\
\cline{3-5}
\cline{6-8}
&	 & r$_{c}$ & $\rho_{0}$  &$\chi^{2}_{r}$  & 	c   & r$_{200}$	& $\chi_{r}^{2}$ \\		
 &  & (kpc) & (M$_{\odot}$ pc$^{-3}$) &  & & (kpc) & \\
\hline
KK246  & \fat     & 1.20$\pm$0.26 & 39$\pm$12& 0.30 & 2.7$\pm$0.7 &	52$\pm$10 & 0.06 \\
& Rotcur	      & 1.28$\pm$0.06 & 37$\pm$2 & 0.08 & 3.3$\pm$0.4 &	46$\pm$4 & 0.13 \\
\hline 
U4155 & \fat      & 1.87$\pm$0.20 & 30$\pm$4 &0.09 & ... & ... & ... \\
& Rotcur          & 2.06$\pm$0.10 & 29$\pm$1 &0.17 & ... & ... & ... \\
\hline
J0926+3343 & \fat & 0.34$\pm$0.19 & 163$\pm$143 &0.52 & 1.9$\pm$6.0 & 60$\pm$161  & 0.15 \\
& Rotcur          & 2.23$\pm$1.08 & 21$\pm$5  &0.46 & ... & ... & ... \\
\hline 
U5288 & \fat      & 0.25$\pm$0.05 & 1341$\pm$507&0.71 & 11.4$\pm$1.7 & 41$\pm$2 & 1.21 \\
& Rotcur	      & 0.58$\pm$0.07 & 314$\pm$68 &1.90 & 10.4$\pm$1.3 & 44$\pm$2 & 4.32 \\
\hline 
U4148 & \fat      & 0.50$\pm$0.11 & 289$\pm$118&0.69 & 8.4$\pm$0.6 & 39$\pm$1 & 0.14 \\
& Rotcur		  & 1.74$\pm$0.11 & 37$\pm$3 &0.49 & 3.2$\pm$0.7 & 61$\pm$9 & 2.43 \\
\hline
J0630+23 & \fat   & 1.92$\pm$0.24 & 42$\pm$8 &0.34 & 3.4$\pm$0.8 & 71$\pm$10 & 0.39 \\
& Rotcur 		  & 2.97$\pm$0.38 & 22$\pm$3 &0.21 & 1.5$\pm$1.0 & 125 $\pm$60 & 0.29 \\
\hline 
J0626+24 & \fat   & 0.89$\pm$0.33 & 144$\pm$90&2.14 & 5.1$\pm$1.1 & 60$\pm$8 & 0.52 \\
& Rotcur	 	  & 3.04$\pm$0.35 & 26$\pm$4 &0.85 & 1.0$\pm$1.3 & 193$\pm$153 & 0.71 \\
\hline 
J0929+1155 & \fat & 0.64$\pm$0.10 & 185$\pm$48 &0.17 & 5.8$\pm$0.7 & 46$\pm$3 & 0.07 \\
& Rotcur  		  & 2.71$\pm$0.14 & 22$\pm$1 &0.15 & ... & ... & ... \\
\hline & 
\end{tabular}

\end{footnotesize}

\end{table*}

\begin{table*}
\begin{footnotesize}

\caption{Parameters of dark matter halos }
\label{table:rotcur}
\begin{tabular}{ p{1.6cm} p{1.5cm}  p{1.4cm} p{1.9cm} p{0.6cm} p{1.4cm} p{1.9cm} p{0.6cm}}
\\
\hline
\hline
Galaxy& Model &  \multicolumn{3}{c}{Isothermal} &  \multicolumn{3}{c}{NFW}   \\
\cline{3-5}
\cline{6-8}
&	 & r$_{c}$ & $\rho_{0}$  &$\chi^{2}_{r}$  & 	c   & r$_{200}$	& $\chi_{r}^{2}$ \\		
 &  & (kpc) & (M$_{\odot}$ pc$^{-3}$) &  & & (kpc) & \\
\hline
KK246  & 2D SBR   & 1.20$\pm$0.26 & 39$\pm$12& 0.30 & 2.7$\pm$0.7 &	52$\pm$10 & 0.06 \\
& 3D SBR	      & 1.28$\pm$0.27 & 36$\pm$10& 0.29 & 2.4$\pm$0.8 &	57$\pm$14 & 0.08 \\
\hline 
U4155 & 2D SBR    & 1.87$\pm$0.20 & 30$\pm$4 &0.09 & ... & ... & ... \\
& 3D SBR          & 1.83$\pm$0.16 & 34$\pm$3 &0.07 & ... & ... & ... \\
\hline
J0926+3343&2D SBR & 0.34$\pm$0.19 & 163$\pm$143 &0.52 & 1.9$\pm$6.1 & 60$\pm$161  & 0.15 \\
& 3D SBR          & 0.41$\pm$0.28 & 131$\pm$134 &0.65 & ... & ... & ... \\
\hline 
U5288 & 2D SBR    & 0.25$\pm$0.05 & 1341$\pm$507&0.71 & 11.4$\pm$1.7 & 41$\pm$2 & 1.21 \\
& 3D SBR          & 0.26$\pm$0.05 & 1339$\pm$532&0.78 & 11.3$\pm$1.7 & 41$\pm$2 & 1.30 \\
\hline 
U4148 & 2D SBR    & 0.50$\pm$0.11 & 289$\pm$118&0.69 & 8.4$\pm$0.6 & 39$\pm$1 & 0.14 \\
& 3D SBR		  & 0.49$\pm$0.11 & 304$\pm$123&0.66 & 8.5$\pm$0.6 & 39$\pm$1 & 0.14 \\
\hline
J0630+23 & 2D SBR & 1.92$\pm$0.24 & 42$\pm$8 &0.34 & 3.4$\pm$0.8 & 71$\pm$10 & 0.39 \\
& 3D SBR 		  & 2.01$\pm$0.28 & 40$\pm$8 &0.43 & 3.4$\pm$0.9 & 74$\pm$13 & 0.59 \\
\hline 
J0626+24 & 2D SBR & 0.89$\pm$0.33 & 144$\pm$89&2.14 & 5.1$\pm$1.1 & 60$\pm$8 & 0.52 \\
& 3D SBR	 	  & 0.98$\pm$0.34 & 128$\pm$74&2.03 & 5.1$\pm$0.9 & 62$\pm$7 & 0.37 \\
\hline 
J0929+1155&2D SBR & 0.64$\pm$0.10 & 185$\pm$48 &0.17 & 5.8$\pm$0.7 & 46$\pm$3 & 0.07 \\
& 3D SBR  		  & 0.70$\pm$0.12 & 168$\pm$50 &0.23 & 5.4$\pm$0.6 & 50$\pm$3& 0.06 \\
\hline & 
\end{tabular}

\end{footnotesize}

\end{table*}

\begin{table*}
\begin{footnotesize}

\caption{Inner density slopes of dark matter halos }
\label{table:alpha}
\begin{tabular}{ p{1.6cm} p{1.2cm} p{1.9cm} p{2.8cm}   p{2.0cm} p{2.0cm} p{2.0cm} }
\\
\hline
Galaxy      &R$_{\rm min}$  & R$_{\rm max}$    &$\alpha$  &  $\alpha$ & $\alpha$  & $\alpha$     \\
 & & H{\sc i}$_{\rm beam}^{-1}$ & (3D: minimum disc) & (3D: DM only) & (2D: Moment) & (2D: Hermite) \\
\hline
KK246		& 0.15 & 20.3 & -1.41 $\pm$ 0.12 & -1.72 $\pm$ 0.37 & -0.92 $\pm$0.13 & -1.12 $\pm$0.03 \\
U4115		& 0.25 & 32.7 & -1.09 $\pm$ 0.16 & -1.13 $\pm$ 0.20 & 0.04 $\pm$0.24 & -0.35 $\pm$ 0.05\\
J0926+3343	& 0.21 & 5.9  & -1.46 $\pm$ 0.38 & -1.64 $\pm$ 0.37 & -0.34 $\pm$0.21 & -0.64 $\pm$ 0.6\\
U5288		& 0.23 & 23.4 & -1.73 $\pm$ 0.15 & -1.77 $\pm$ 0.19 & -0.38 $\pm$0.3 & -0.94 $\pm$0.1 \\
U4148		& 0.13 & 21.4 & -1.76 $\pm$ 0.23 & -1.78 $\pm$ 0.23 & -0.52 $\pm$0.02 & -0.16$\pm$0.14 \\			
J0630+23	& 0.33 & 18.3 & -1.03 $\pm$ 0.14 & -1.06 $\pm$ 0.15 & -0.64 $\pm$0.49 & -0.80 $\pm$0.46 \\	
J0626+24	& 0.34 & 19.5 & -1.54 $\pm$ 0.08 & -1.59 $\pm$ 0.08 & -1.21 $\pm$0.02 & -1.02 $\pm$0.40 \\	
J0929+1155	& 0.29 & 9.0  & -1.16 $\pm$ 0.01 & -1.17 $\pm$ 0.01 & -0.07 $\pm$0.19 & . . .\\	
\hline & 
\end{tabular}

\end{footnotesize}

\end{table*}




\bibliographystyle{mnras}
\bibliography{masmodel} 





\appendix


\bsp	
\label{lastpage}
\end{document}